\documentclass[11pt]{article}
\usepackage{amsmath}

\newcommand{\be}{\begin{equation}}
\newcommand{\ee}{\end{equation}}
\newcommand{\bea}{\begin{eqnarray}}
\newcommand{\eea}{\end{eqnarray}}
\newcommand{\te}[1]{{\text{#1}}}
\newcommand{\N}{{\rm NS}}

\DeclareMathOperator{\R}{Re}
\DeclareMathOperator{\I}{Im}
\DeclareMathOperator{\Tr}{Tr}
\DeclareMathOperator{\sign}{sign}

\begin{document}

\author{Matthew Headrick \\ Martin Fisher School of Physics \\ Brandeis University \\ {\tt mph@brandeis.edu}}
\title{A solution manual for Polchinski's \emph{String Theory}}
\date{}

\maketitle

\begin{abstract}
We present detailed solutions to 81 of the 202 problems in J. Polchinski's two-volume textbook \emph{String Theory}.
\end{abstract}

\vspace{5in}

\begin{flushleft}BRX-TH-604\end{flushleft}

\thispagestyle{empty}

\newpage

\setcounter{page}{1}

\tableofcontents

\newpage

\setcounter{section}{-1}

\section{Preface}

The following pages contain detailed solutions to 81 of the 202 problems in J. Polchinski's two-volume textbook \emph{String Theory} \cite{polchinski1,polchinski2}. I originally wrote up these solutions while teaching myself the subject, and then later decided that they may be of some use to others doing the same. These solutions are the work of myself alone, and carry no endorsement from Polchinski.

I would like to thank R. Britto, S. Minwalla, D. Podolsky, and M. Spradlin for help on these problems. This work was done while I was a graduate student at Harvard University, and was supported by an NSF Graduate Research Fellowship.

\newpage

\section{Chapter 1}

\subsection{Problem 1.1}

\paragraph{(a)}
We work in the gauge where $\tau=X^0$. Non-relativistic motion means
$\dot X^i\equiv v^i\ll1$. Then
\bea
S_{\rm pp} &=& -m\int d\tau\,\sqrt{-\dot X^\mu\dot X_\mu}\nonumber \\
&=& -m\int dt\,\sqrt{1-v^2}\nonumber \\
&\approx& \int dt\,(\frac{1}{2}mv^2-m).
\eea

\paragraph{(b)}
Again, we work in the gauge $\tau=X^0$, and assume $\dot X^i\equiv
v^i\ll1$. Defining $u^i\equiv\partial_\sigma X^i$, the induced metric
$h_{ab}=\partial_aX^\mu\partial_bX_\mu$ becomes:
\be
\{h_{ab}\} = \left(\begin{array}{cc} -1+v^2 & \mathbf{u\cdot v} \\ \mathbf{u\cdot v} & u^2
\end{array}\right).
\ee
Using the fact that the transverse velocity of the string is
\be
\mathbf v_{\rm T} = \mathbf{v-\frac{u\cdot v}{\mathnormal{u^2}}u},
\ee
the Nambu-Goto Lagrangian can be written:
\bea
L &=& -\frac{1}{2\pi\alpha'}\int
d\sigma\,\left(-\det\{h_{ab}\}\right)^{1/2}\nonumber \\
&=& -\frac{1}{2\pi\alpha'}\int d\sigma\,\left(u^2(1-v^2)+(\mathbf{u\cdot
v})^2\right)^{1/2}\nonumber \\
&\approx& -\frac{1}{2\pi\alpha'}\int d\sigma\,|\mathbf
u|\left(1-\frac{1}{2}v^2+\frac{\mathbf{u\cdot v}^2}{2u^2}\right)\nonumber
\\
&=& \int d\sigma\,|\mathbf u|\frac{1}{2}\rho v_{\rm T}^2-L_{\rm s}T,
\eea
where
\be
\rho = \frac{1}{2\pi\alpha'}
\ee
is the mass per unit length of the string, 
\be
L_{\rm s} = \int d\sigma\,|\mathbf u|
\ee
is its physical length, and
\be
T = \frac{1}{2\pi\alpha'} = \rho
\ee
is its tension.

\subsection{Problem 1.3}
It is well known that $\chi$, the Euler characteristic of the surface, is a
topological invariant, i.e.\ does not depend on the metric. We will prove
by explicit computation that, in particular, $\chi$ is invariant under Weyl
transformations,
\be
\gamma'_{ab} = e^{2\omega(\sigma,\tau)}\gamma_{ab}.
\ee
For this we will need the transformation law for the connection
coefficients,
\be
\Gamma'^a_{bc} =
\Gamma^a_{bc}+\partial_b\omega\delta^a_c+\partial_c\omega\delta^a_b-\partial_d\omega\gamma^{ad}\gamma_{bc},
\ee
and for the curvature scalar,
\be
R' = e^{-2\omega}(R-2\nabla_a\partial^a\omega).
\ee
Since the tangent and normal vectors at the boundary are normalized, they
transform as
\bea
t'^a &=& e^{-\omega}t^a, \\
n'_a &=& e^\omega n_a.
\eea
The curvature of the boundary thus transforms as follows:
\bea
k' &=& \pm t'^an'_b(\partial_at'^b+\Gamma'^b_{ac}t'^c)\nonumber \\
&=& e^{-\omega}(k\mp t^at_an^b\partial_d),
\eea
where we have used (9), (11), (12), and the fact that $n$ and $t$ are
orthogonal. If the boundary is timelike then $t^at_a=-1$ and we must use the
upper sign, whereas if it is spacelike then $t^at_a=1$ and we must use the
lower sign. Hence
\be
k' = e^{-\omega}(k+n^a\partial_a\omega).
\ee
Finally, since $ds=(-\gamma_{\tau\tau})^{1/2}d\tau$ for a timelike boundary
and $ds=\gamma_{\sigma\sigma}^{1/2}d\sigma$ for a spacelike bounday,
\be
ds' = ds\,e^\omega.
\ee
Putting all of this together, and applying Stokes theorem, which says that
for any vector $v^a$,
\be
\int_Md\tau\,d\sigma\,(-\gamma)^{1/2}\nabla_av^a = \int_{\partial M}ds\,n^av_a,
\ee
we find the transformation law for $\chi$:
\bea
\chi' &=&
\frac{1}{4\pi}\int_Md\tau\,d\sigma\,(-\gamma')^{1/2}R'+\frac{1}{2\pi}\int_{\partial
M}ds'\,k'\nonumber \\
&=&
\frac{1}{4\pi}\int_Md\tau\,d\sigma\,(-\gamma)^{1/2}(R-2\nabla_a\partial^a\omega)+\frac{1}{2\pi}\int_{\partial
M}ds\,(k+n^a\partial_a\omega)\nonumber \\
&=& \chi.
\eea

\subsection{Problem 1.5}
For simplicity, let us define $a\equiv(\pi/2p^+\alpha'l)^{1/2}$. Then we
wish to evaluate
\bea
\lefteqn{\sum_{n=1}^\infty(n-\theta)\exp[-(n-\theta)\epsilon a]}\nonumber \\
&&\qquad\qquad = -\frac{d}{d(\epsilon a)}\sum_{n=1}^\infty\exp[-(n-\theta)\epsilon
a]\nonumber \\
&&\qquad\qquad = -\frac{d}{d(\epsilon a)}\frac{e^{\theta\epsilon a}}{e^{\epsilon
a}-1}\nonumber \\
&&\qquad\qquad = -\frac{d}{d(\epsilon a)}\left(\frac{1}{\epsilon
a}+\theta-\frac{1}{2}+\left(\frac{1}{12}-\frac{\theta}{2}+\frac{\theta^2}{2}\right)\epsilon
a+\mathcal{O}(\epsilon a)^2\right)\nonumber \\
&&\qquad\qquad = \frac{1}{(\epsilon
a)^2}-\frac{1}{2}\left(\frac{1}{6}-\theta+\theta^2\right)+\mathcal{O}(\epsilon a).
\eea
As expected, the cutoff dependent term is independent of $\theta$; the
finite result is
\be
-\frac{1}{2}\left(\frac{1}{6}-\theta+\theta^2\right).
\ee

\subsection{Problem 1.7}
The mode expansion satisfying the boundary conditions is
\be
X^{25}(\tau,\sigma) =
\sqrt{2\alpha'}\sum_n\frac{1}{n}\alpha_n^{25}\exp\left[-\frac{i\pi
nc\tau}{l}\right]\sin\frac{\pi n\sigma}{l},
\ee
where the sum runs over the half-odd-integers,
$n=1/2,-1/2,3/2,-3/2,\dots$. Note that there is no $p^{25}$. Again,
Hermiticity of $X^{25}$ implies
$\alpha^{25}_{-n}=(\alpha^{25}_n)^\dagger$. Using (1.3.18), 
\be
\Pi^{25}(\tau,\sigma) =
-\frac{i}{\sqrt{2\alpha'}l}\sum_n\alpha^{25}_n\exp\left[-\frac{i\pi
nc\tau}{l}\right]\sin\frac{\pi n\sigma}{l}.
\ee

We will now determine the commutation relations among the $\alpha^{25}_n$
from the equal time commutation relations (1.3.24b). Not surprisingly, they
will come out the same as for the free string (1.3.25b). We have:
\bea
i\delta(\sigma-\sigma') &=&
[X^{25}(\tau,\sigma),\Pi^{25}(\tau,\sigma)]\\
&=&
-\frac{i}{l}\sum_{n,n'}\frac{1}{n}[\alpha_n^{25},\alpha_{n'}^{25}]\exp\left[-\frac{i\pi(n+n')c\tau}{l}\right]\sin\frac{\pi
n\sigma}{l}\sin\frac{\pi n'\sigma'}{l}.\nonumber
\eea
Since the LHS does not depend on $\tau$, the coefficient of $\exp[-i\pi
mc\tau/l]$ on the RHS must vanish for $m\neq0$:
\be
\frac{1}{l}\sum_n\frac{1}{n}[\alpha^{25}_n,\alpha^{25}_{m-n}]\sin\frac{\pi
n\sigma}{l}\sin\frac{\pi(n-m)\sigma'}{l} =
\delta(\sigma-\sigma')\delta_{m,0}.
\ee
Multiplying both sides by $\sin[\pi n'\sigma/l]$ and integrating over
$\sigma$ now yields,
\bea
\frac{1}{2n}\left([\alpha^{25}_n,\alpha^{25}_{m-n}]\sin\frac{\pi(n-m)\sigma'}{l}+[\alpha^{25}_{n+m},\alpha^{25}_{-n}]\sin\frac{\pi(n+m)\sigma'}{l}\right)\qquad\qquad&& \\
 = \sin\frac{\pi n\sigma'}{l}\delta_{m,0},&&\nonumber
\eea
or,
\be
[\alpha^{25}_n,\alpha^{25}_{m-n}] = n\delta_{m,0},
\ee
as advertised.

The part of the Hamiltonian (1.3.19) contributed by the $X^{25}$
oscillators is 
\bea
\lefteqn{\frac{l}{4\pi\alpha'p^+}\int_0^ld\sigma\,\left(2\pi\alpha'\left(\Pi^{25}\right)^2+\frac{1}{2\pi\alpha'}\left(\partial_\sigma
X^{25}\right)^2\right)}\qquad\qquad\qquad\qquad\nonumber \\
&=&
\frac{1}{4\alpha'p^+l}\sum_{n,n'}\alpha^{25}_n\alpha^{25}_{n'}\exp\left[-\frac{i\pi(n+n')c\tau}{l}\right]\nonumber
\\
&&\qquad\times\int_0^ld\sigma\,\left(-\sin\frac{\pi
n\sigma}{l}\sin\frac{\pi n'\sigma}{l}+\cos\frac{\pi
n\sigma}{l}\cos\frac{\pi n'\sigma}{l}\right)\nonumber \\
&=& \frac{1}{4\alpha'p^+}\sum_n\alpha^{25}_n\alpha^{25}_{-n}\nonumber \\
&=&
\frac{1}{4\alpha'p^+}\sum_{n=1/2}^\infty\left(\alpha^{25}_n\alpha^{25}_{-n}+\alpha^{25}_{-n}\alpha^{25}_n\right)\nonumber
\\
&=&
\frac{1}{2\alpha'p^+}\sum_{n=1/2}^\infty\left(\alpha^{25}_{-n}\alpha^{25}_n+\frac{n}{2}\right)\nonumber
\\
&=& \frac{1}{2\alpha'p^+}\left(\sum_{n=1/2}^\infty\alpha^{25}_{-n}\alpha^{25}_n+\frac{1}{48}\right),
\eea
where we have used (19) and (25). Thus the mass spectrum (1.3.36) becomes
\bea
m^2 &=& 2p^+H-p^ip^i\qquad(i=2,\dots,24)\nonumber \\
&=& \frac{1}{\alpha'}\left(N-\frac{15}{16}\right),
\eea
where the level spectrum is given in terms of the occupation numbers by
\be
N = \sum_{i=2}^{24}\sum_{n=1}^\infty nN_{in}+\sum_{n=1/2}^\infty nN_{25,n}.
\ee
The ground state is still a tachyon,
\be
m^2 = -\frac{15}{16\alpha'}.
\ee
The first excited state has the lowest $X^{25}$ oscillator excited
($N_{25,1/2}=1$), and is also tachyonic:
\be
m^2 = -\frac{7}{16\alpha'}.
\ee
There are no massless states, as the second excited state is already massive:
\be
m^2 = \frac{1}{16\alpha'}.
\ee
This state is 24-fold degenerate, as it can be reached either by
$N_{i,1}=1$ or by $N_{25,1/2}=2$. Thus it is a massive vector with respect to
the SO(24,1) Lorentz symmetry preserved by the D-brane. The third excited
state, with
\be
m^2 = \frac{9}{16\alpha'},
\ee
is 25-fold degenerate and corresponds to a vector plus a scalar on the
D-brane---it can be reached by $N_{25,1/2}=1$, by $N_{25,1/2}=3$, or by
$N_{i,1}=1$, $N_{25,1/2}=1$.

\subsection{Problem 1.9}
The mode expansion for $X^{25}$ respecting the boundary conditions is
essentially the same as the mode expansion (1.4.4), the only differences
being that the first two terms are no longer allowed, and the oscillator
label $n$, rather than running over the non-zero integers, must now run
over the half-odd-integers as it did in Problem 1.7:
\bea
\lefteqn{X^{25}(\tau,\sigma) =}\qquad\qquad \\
&&
i\sqrt{\frac{\alpha'}{2}}\sum_n\left(\frac{\alpha^{25}_n}{n}\exp\left[-\frac{2\pi
in(\sigma+c\tau)}{l}\right]+\frac{\tilde{\alpha}^{25}_n}{n}\exp\left[\frac{2\pi
in(\sigma-c\tau)}{l}\right]\right).\nonumber
\eea
The canonical commutators are the same as for the untwisted closed string,
(1.4.6c) and (1.4.6d),
\bea
[\alpha^{25}_m,\alpha^{25}_n] &=& m\delta_{m,-n}, \\ 
{}[\tilde{\alpha}^{25}_m,\tilde{\alpha}^{25}_n] &=& m\delta_{m,-n},
\eea
as are the mass formula (1.4.8),
\be
m^2 = \frac{2}{\alpha'}(N+\tilde{N}+A+\tilde{A}),
\ee
the generator of $\sigma$-translations (1.4.10),
\be
P = -\frac{2\pi}{l}(N-\tilde{N}),
\ee
and (therefore) the level-matching condition (1.4.11),
\be
N = \tilde{N}.
\ee
However, the level operator $N$ is now slightly different,
\be
N = \sum_{i=2}^{24}\sum_{n=1}^\infty\alpha^i_{-n}\alpha^i_n+\sum_{n=1/2}^\infty\alpha^{25}_{-n}\alpha^{25}_n;
\ee
in fact, it is the same as the level operator for the open string on a
D24-brane of Problem 1.7. The left-moving level spectrum is therefore given
by (28), and similarly for the right-moving level operator $\tilde{N}$. The
zero-point constants are also the same as in Problem 1.7:
\bea
A = \tilde{A} &=& \frac{1}{2}\left(\sum_{i=2}^{24}\sum_{n=1}^\infty
n+\sum_{n=1/2}^\infty n\right)\nonumber \\
&=& -\frac{15}{16}.
\eea
At a given level $N=\tilde{N}$, the occupation numbers $N_{in}$ and
$\tilde{N}_{in}$ may be chosen independently, so long as both sets satisfy
(28). Therefore the spectrum at that level will consist of the product of
two copies of the D-brane open string spectrum, and
the mass-squared of that level (36) will be 4 times the
open string mass-squared (27). We will have tachyons at levels $N=0$ and
$N=1/2$, with
\be
m^2 = -\frac{15}{4\alpha'}
\ee
and
\be
m^2 = -\frac{7}{4\alpha'},
\ee
respectively. The lowest non-tachyonic states will again be at level $N=1$:
a second rank SO(24) tensor with
\be
m^2 = \frac{1}{4\alpha'},
\ee
which can be decomposed into a scalar, an antisymmetric tensor, and a
traceless symmetric tensor.

\setcounter{equation}{0}

\newpage

\section{Chapter 2}

\subsection{Problem 2.1}

\paragraph{(a)}
For holomorphic test functions $f(z)$,
\bea
\int_Rd^2\!z\,\partial\bar\partial\ln|z|^2f(z) &=&
\int_Rd^2\!z\,\bar\partial\frac{1}{z}f(z)\nonumber \\
&=& -i\oint_{\partial R}dz\,\frac{1}{z}f(z)\nonumber \\
&=& 2\pi f(0).
\eea
For antiholomorphic test functions $f(\bar z)$,
\bea
\int_Rd^2\!z\,\partial\bar\partial\ln|z|^2f(\bar z) &=&
\int_Rd^2\!z\,\partial\frac{1}{\bar z}f(\bar z)\nonumber \\
&=& i\oint_{\partial R}d\bar z\,\frac{1}{\bar z}f(\bar z)\nonumber \\
&=& 2\pi f(0).
\eea

\paragraph{(b)}
We regulate $\ln|z|^2$ by replacing it with $\ln(|z|^2+\epsilon)$. This
lead to regularizations also of $1/\bar z$ and $1/z$:
\be
\partial\bar\partial\ln(|z|^2+\epsilon) = \partial\frac{z}{|z|^2+\epsilon}
= \bar\partial\frac{\bar z}{|z|^2+\epsilon} =
\frac{\epsilon}{(|z|^2+\epsilon)^2}.
\ee
Working in polar coordinates, consider a general test function
$f(r,\theta)$, and define $g(r^2)\equiv\int d\theta\,f(r,\theta)$. Then,
assuming that $g$ is sufficiently well behaved at zero and infinity,
\bea
\lefteqn{\int d^2\!z\,\frac{\epsilon}{(|z|^2+\epsilon)^2}f(z,\bar
z)}\qquad\qquad\qquad\nonumber \\
&=& \int_0^\infty du\,\frac{\epsilon}{(u+\epsilon)^2}g(u)\nonumber \\
&=&
\left(-\frac{\epsilon}{u+\epsilon}g(u)+\epsilon\ln(u+\epsilon)g'(u)\right)_0^\infty-\int_0^\infty
du\,\epsilon\ln(u+\epsilon)g''(u).\nonumber \\
&=& g(0)\nonumber \\
&=& 2\pi f(0).
\eea

\subsection{Problem 2.3}

\paragraph{(a)}
The leading behavior of the expectation value as $z_1\rightarrow z_2$ is
\bea
\lefteqn{\left\langle\prod_{i=1}^n:e^{ik_i\cdot X(z_i,\bar
z_i)}:\right\rangle}\nonumber \\
&=& iC^X(2\pi)^D\delta^D\left(\sum_{i=1}^nk_i\right)\prod_{i,j=1}^n|z_{ij}|^{\alpha'k_i\cdot
k_j}\nonumber \\
&=& |z_{12}|^{\alpha'k_1\cdot
k_2}iC^X(2\pi)^D\delta^D(k_1+k_2+\sum_{i=3}^nk_i)\nonumber \\
&&\qquad\qquad\quad\qquad\qquad\qquad\qquad\times\prod_{i=3}^n\left(|z_{1i}|^{\alpha'k_1\cdot
k_i}|z_{2i}|^{\alpha'k_2\cdot
k_i}\right)\prod_{i,j=3}^n|z_{ij}|^{\alpha'k_i\cdot k_j}\nonumber \\
&\approx& |z_{12}|^{\alpha'k_1\cdot
k_2}iC^X(2\pi)^D\delta^D(k_1+k_2+\sum_{i=3}^nk_i)\prod_{i=3}^n|z_{2i}|^{\alpha'(k_1+k_2)\cdot
k_i}\prod_{i,j=3}^n|z_{ij}|^{\alpha'k_i\cdot k_j}\nonumber \\
&=& |z_{12}|^{\alpha'k_1\cdot k_2}\left\langle:e^{i(k_1+k_2)\cdot X(z_2,\bar
z_2)}:\prod_{i=3}^n:e^{ik_i\cdot X(z_i,\bar z_i)}:\right\rangle,
\eea
in agreement with (2.2.14).

\paragraph{(b)}
The $z_i$-dependence of the expectation value is given by
\bea
\lefteqn{|z_{23}|^{\alpha'k_2\cdot k_3}|z_{12}|^{\alpha'k_1\cdot
k_2}|z_{13}|^{\alpha'k_1\cdot k_3}}\qquad\quad\nonumber \\
&=& |z_{23}|^{\alpha'k_2\cdot k_3}|z_{12}|^{\alpha'k_1\cdot
k_2}|z_{23}|^{\alpha'k_1\cdot
k_3}\left|1+\frac{z_{12}}{z_{23}}\right|^{\alpha'k_1\cdot k_3} \\
&=& |z_{23}|^{\alpha'(k_1+k_2)\cdot k_3}|z_{12}|^{\alpha'k_1\cdot
k_2}\left(\sum_{k=0}^\infty\frac{\Gamma(\frac{1}{2}\alpha'k_1\cdot
k_3+1)}{k!\,\Gamma(\frac{1}{2}\alpha'k_1\cdot
k_3-k+1)}\left(\frac{z_{12}}{z_{23}}\right)^k\right)\nonumber \\
&& \qquad\qquad\qquad\qquad\qquad\quad\times\left(\sum_{k=0}^\infty\frac{\Gamma(\frac{1}{2}\alpha'k_1\cdot
k_3+1)}{k!\,\Gamma(\frac{1}{2}\alpha'k_1\cdot k_3-k+1)}\left(\frac{\bar
z_{12}}{\bar z_{23}}\right)^k\right).\nonumber
\eea
The radius of convergence of a power series is given by the limit as
$k\rightarrow\infty$ of $|a_k/a_{k+1}|$, where the $a_k$ are the
coefficients of the series. In this case, for both of the above power
series,
\bea
R &=&
\lim_{k\rightarrow\infty}\left|\frac{(k+1)!\,\Gamma(\frac{1}{2}\alpha'k_1\cdot
k_3-k)}{k!\,\Gamma(\frac{1}{2}\alpha'k_1\cdot
k_3-k+1)}z_{23}\right|\nonumber \\
&=& |z_{23}|.
\eea

\paragraph{(c)}
Consider the interior of the dashed line in figure 2.1, that is, the set of
points $z_1$ satisfying
\be
|z_{12}| < |z_{23}|.
\ee
By equation (2.1.23), the expectation value
\be
\left\langle:X^\mu(z_1,\bar z_1)X^\nu(z_2,\bar z_2):\mathcal A(z_3,\bar
z_3)\mathcal B(z_4,\bar z_4)\right\rangle
\ee
is a harmonic function of $z_1$ within this region. It can therefore be
written as the sum of a holomorphic and an antiholomorphic function (this
statement is true in any simply connected region). The Taylor expansion of
a function that is holomorphic on an open disk (about the center of the
disk), converges on the disk; similarly for an antiholomorphic
function. Hence the two Taylor series on the RHS of (2.2.4) must converge
on the disk.

\subsection{Problem 2.5}
Under the variation of the fields
$\phi_\alpha(\sigma)\rightarrow\phi_\alpha(\sigma)+\delta\phi_\alpha(\sigma)$,
the variation of the Lagrangian is
\be
\delta\mathcal L = \frac{\partial\mathcal
L}{\partial\phi_\alpha}\delta\phi_\alpha+\frac{\partial\mathcal
L}{\partial(\partial_a\phi_\alpha)}\partial_a\delta\phi_\alpha.
\ee
The Lagrangian equations of motion (Euler-Lagrange equations) are derived
by assuming that the action is stationary under an arbitrary variation
$\delta\phi_\alpha(\sigma)$ that vanishes at infinity:
\bea
0 &=& \delta S\nonumber \\
&=& \int d^d\!\sigma\,\delta\mathcal L\nonumber \\
&=& \int d^d\!\sigma\,\left(\frac{\partial\mathcal
L}{\partial\phi_\alpha}\delta\phi_\alpha+\frac{\partial\mathcal
L}{\partial(\partial_a\phi_\alpha)}\partial_a\delta\phi_\alpha\right)\nonumber
\\
&=& \int d^d\!\sigma\,\left(\frac{\partial\mathcal
L}{\partial\phi_\alpha}-\partial_a\frac{\partial\mathcal
L}{\partial(\partial_a\phi_\alpha)}\right)\delta\phi_\alpha
\eea
implies
\be
\frac{\partial\mathcal
L}{\partial\phi_\alpha}-\partial_a\frac{\partial\mathcal
L}{\partial(\partial_a\phi_\alpha)} = 0.
\ee
Instead of assuming that $\delta\phi_\alpha$ vanishes at infinity, let us
assume that it is a symmetry. In this case, the variation of the Lagrangian
(10) must be a total derivative to insure that the action on bounded regions
varies only by a surface term, thereby not affecting the equations of motion:
\be
\delta\mathcal L = \epsilon\partial_a\mathcal K^a;
\ee
$\mathcal K^a$ is assumed to be a local function of the
fields and their derivatives, although it is not obvious how to prove that
this can always be arranged. Using (10), (12), and (13),
\bea
\partial_aj^a &=& 2\pi i\partial_a\left(\frac{\partial\mathcal
L}{\partial(\partial_a\phi_\alpha)}\epsilon^{-1}\delta\phi_\alpha-\mathcal
K^a\right)\nonumber \\
&=& \frac{2\pi i}{\epsilon}\left(\frac{\partial\mathcal
L}{\partial\phi_\alpha}\delta\phi_\alpha+\frac{\partial\mathcal
L}{\partial(\partial_a\phi_\alpha)}\partial_a\delta\phi_\alpha-\delta\mathcal
L\right)\nonumber \\
&=& 0.
\eea

If we now vary the fields by $\rho(\sigma)\delta\phi_\alpha(\sigma)$, where
$\delta\phi_\alpha$ is a symmetry as before but $\rho$ is
an arbitrary function, then the variation of the action will be
\bea
\delta\mathcal L &=& \frac{\partial\mathcal
L}{\partial(\partial_a\phi_\alpha)}\partial_a(\delta\phi_\alpha\rho)+\frac{\partial\mathcal
L}{\partial\phi_\alpha}\delta\phi_\alpha\rho\nonumber \\
&=& \left(\frac{\partial\mathcal
L}{\partial(\partial_a\phi_\alpha)}\partial_a\delta\phi_\alpha+\frac{\partial\mathcal
L}{\partial\phi_\alpha}\delta\phi_\alpha\right)\rho+\frac{\partial\mathcal
L}{\partial(\partial_a\phi_\alpha)}\delta\phi_\alpha\partial_a\rho.
\eea
Equation (13) must be satisfied in the case $\rho(\sigma)$ is identically
1, so the factor in parentheses must equal $\epsilon\partial_a\mathcal
K^a$:
\bea
\delta S &=& \int
d^d\!\sigma\,\left(\epsilon\partial_a\mathcal K^a\rho+\frac{\partial\mathcal
L}{\partial(\partial_a\phi_\alpha)}\delta\phi_\alpha\partial_a\rho\right)\nonumber
\\
&=& \int d^d\!\sigma\,\left(-\epsilon\mathcal K^a+\frac{\partial\mathcal
L}{\partial(\partial_a\phi_\alpha)}\delta\phi_\alpha\right)\partial_a\rho\nonumber
\\
&=& \frac{\epsilon}{2\pi i}\int d^d\!\sigma\,j^a\partial_a\rho,
\eea
where we have integrated by parts, assuming that $\rho$ falls off at
infinity. Since $\delta\exp(-S) = -\exp(-S)\delta S$, this agrees with
(2.3.4) for the case of flat space, ignoring the transformation of the
measure.

\subsection{Problem 2.7}

\paragraph{(a)}

$X^\mu$:
\bea
&& T(z)X^\mu(0,0) = -\frac{1}{\alpha'}:\partial
X^\nu(z)\partial X_\nu(z):X^\mu(0,0) \sim \frac{1}{z}\partial X^\mu(z) \sim
\frac{1}{z}\partial X^\mu(0)\nonumber \\
&& \tilde T(\bar z)X^\mu(0,0) = -\frac{1}{\alpha'}:\bar\partial
X^\nu(\bar z)\bar\partial X_\nu(\bar z):X^\mu(0,0) \sim
\frac{1}{\bar z}\bar\partial X^\mu(\bar z) \sim \frac{1}{\bar z}\bar\partial
X^\mu(0)\qquad
\eea
$\partial X^\mu$:
\bea
&& T(z)\partial X^\mu(0) \sim \frac{1}{z^2}\partial
X^\mu(z) \sim \frac{1}{z^2}\partial
X^\mu(0)+\frac{1}{z}\partial^2X^\mu(0)\nonumber \\
&& \tilde T(\bar z)\partial X^\mu(0) \sim 0
\eea
$\bar\partial X^\mu$:
\bea
&& T(z)\bar\partial X^\mu(0) \sim 0\nonumber \\
&& \tilde T(\bar z)\bar\partial X^\mu(0) \sim \frac{1}{\bar z^2}\bar\partial
X^\mu(\bar z) \sim \frac{1}{\bar z^2}\bar\partial X^\mu(0)+\frac{1}{\bar
z}\bar\partial^2X^\mu(0)
\eea
$\partial^2X^\mu$:
\bea
&& T(z)\partial^2X^\mu(0) \sim \frac{2}{z^3}\partial X^\mu(z) \sim
\frac{2}{z^3}\partial
X^\mu(0)+\frac{2}{z^2}\partial^2X^\mu(0)+\frac{1}{z}\partial^3X^\mu(0)\nonumber
\\
&& \tilde T(\bar z)\partial^2X^\mu(0) \sim 0
\eea
$:e^{ik\cdot X}:$:
\bea
T(z):e^{ik\cdot X(0,0)}: &\sim& \frac{\alpha'k^2}{4z^2}:e^{ik\cdot
X(0,0)}:+\frac{1}{z}ik_\mu:\partial X^\mu(z)e^{ik\cdot X(0,0)}:\nonumber \\
&\sim& \frac{\alpha'k^2}{4z^2}:e^{ik\cdot
X(0,0)}:+\frac{1}{z}ik_\mu:\partial X^\mu(0)e^{ik\cdot X(0,0)}:\nonumber \\
\tilde T(\bar z):e^{ik\cdot X(0,0)}: &\sim& \frac{\alpha'k^2}{4\bar z^2}:e^{ik\cdot
X(0,0)}:+\frac{1}{\bar z}ik_\mu:\bar\partial X^\mu(\bar z)e^{ik\cdot X(0,0)}:\nonumber \\
&\sim& \frac{\alpha'k^2}{4\bar z^2}:e^{ik\cdot
X(0,0)}:+\frac{1}{\bar z}ik_\mu:\partial X^\mu(0)e^{ik\cdot X(0,0)}:
\eea

\paragraph{(b)}
In the linear dilaton theory, the energy-momentum tensor is
\bea
T &=& -\frac{1}{\alpha'}:\partial X^\mu\partial X_\mu:+V_\mu\partial^2X^\mu,\nonumber
\\
\tilde T &=& -\frac{1}{\alpha'}:\bar\partial X^\mu\bar\partial
X_\mu:+V_\mu\bar\partial^2X^\mu,
\eea
so it suffices to calculate the OPEs of the various operators with the terms
$V_\mu\partial^2X^\mu$ and $V_\mu\bar\partial^2X^\mu$ and add them to the
results found in part (a).

$X^\mu$:
\bea
&& V_\nu\partial^2X^\nu(z)X^\mu(0,0) \sim
\frac{\alpha'V^\mu}{2z^2}\nonumber \\
&& V_\nu\bar\partial^2X^\nu(\bar z)X^\mu(0,0) \sim
\frac{\alpha'V^\mu}{2\bar z^2}
\eea
Not only is $X^\mu$ is not a tensor anymore, but it does not even have
well-defined weights, because it is not an eigenstate of rigid
transformations.

$\partial X^\mu$:
\bea
&& V_\nu\partial^2X^\nu(z)\partial X^\mu(0) \sim
\frac{\alpha'V^\mu}{z^3}\nonumber \\
&& V_\nu\bar\partial^2X^\nu(\bar z)\partial X^\mu(0) \sim 0
\eea
So $\partial X^\mu$ still has weights (1,0), but it is no longer a
tensor operator.

$\bar\partial X^\mu$:
\bea
&& V_\nu\partial^2X^\nu(z)\bar\partial X^\mu(0) \sim 0\nonumber \\
&& V_\nu\bar\partial^2X^\nu(\bar z)\bar\partial X^\mu(0) \sim
\frac{\alpha'V^\mu}{\bar z^3}
\eea
Similarly, $\bar\partial X^\mu$ still has weights (0,1), but is no longer a
tensor.

$\partial^2X^\mu$:
\bea
&& V_\nu\partial^2X^\nu(z)\partial^2X^\mu(0) \sim
\frac{3\alpha'V^\mu}{z^4}\nonumber \\
&& V_\nu\bar\partial^2X^\nu(\bar z)\partial^2X^\mu(0) \sim 0
\eea
Nothing changes from the scalar theory: the weights are still (2,0), and
$\partial^2X^\mu$ is still not a tensor.

$:e^{ik\cdot X}:$:
\bea
&&
V_\nu\partial^2X^\nu(z):e^{ik\cdot X(0,0)}: \sim \frac{i\alpha'V\cdot
k}{2z^2}:e^{ik\cdot X(0,0)}:\nonumber \\
&& V_\nu\bar\partial^2X^\nu(\bar z):e^{ik\cdot X(0,0)}: \sim
\frac{i\alpha'V\cdot k}{2\bar z^2}:e^{ik\cdot X(0,0)}:
\eea
Thus $:e^{ik\cdot X}:$ is still a tensor, but, curiously, its weights are
now complex:
\be
\left(\frac{\alpha'}{4}(k^2+2iV\cdot k),\frac{\alpha'}{4}(k^2+2iV\cdot
k)\right).
\ee

\subsection{Problem 2.9}
Since we are interested in finding the central charges of these theories,
it is only necessary to calculate the $1/z^4$ terms in the $TT$ OPEs, the
rest of the OPE being determined by general considerations as in equation
(2.4.25). In the following, we will therefore drop all terms less singular
than $1/z^4$. For the linear dilaton CFT,
\bea
T(z)T(0) &=& \frac{1}{\alpha'^2}:\partial X^\mu(z)\partial
X_\mu(z)::\partial X^\nu(0)\partial
X_\nu(0):\nonumber \\
&& \qquad{}-\frac{2V_\nu}{\alpha'}:\partial X^\mu(z)\partial
X_\mu(z):\partial^2X^\nu(0)\nonumber \\
&& \qquad{}-\frac{2V_\mu}{\alpha'}\partial^2X^\mu(z):\partial
X^\nu(0)\partial
X_\nu(0):+V_\mu V_\nu\partial^2X^\mu(z)\partial^2X^\nu(0)\nonumber \\
&\sim& \frac{D}{2z^4}+\frac{3\alpha'V^2}{z^4}+\mathcal
O\left(\frac{1}{z^2}\right),
\eea
so
\be
c = D+6\alpha'V^2.
\ee
Similarly,
\bea
\tilde T(\bar z)\tilde T(0) &=& \frac{1}{\alpha'^2}:\bar\partial X^\mu(\bar
z)\bar\partial
X_\mu(\bar z)::\bar\partial X^\nu(0)\bar\partial
X_\nu(0):\nonumber \\
&& \qquad{}-\frac{2V_\nu}{\alpha'}:\bar\partial X^\mu(\bar z)\bar\partial
X_\mu(\bar z):\bar\partial^2X^\nu(0)\nonumber \\
&& \qquad{}-\frac{2V_\mu}{\alpha'}\bar\partial^2X^\mu(\bar z):\bar\partial
X^\nu(0)\bar\partial
X_\nu(0):+V_\mu V_\nu\bar\partial^2X^\mu(\bar z)\bar\partial^2X^\nu(0)\nonumber \\
&\sim& \frac{D}{2\bar z^4}+\frac{3\alpha'V^2}{\bar z^4}+\mathcal
O\left(\frac{1}{\bar z^2}\right),
\eea
so
\be
\tilde c = D+6\alpha'V^2.
\ee

For the $bc$ system,
\bea
T(z)T(0) &=& (1-\lambda)^2:\partial b(z)c(z)::\partial
b(0)c(0):\nonumber \\
&& \quad{}-\lambda(1-\lambda):\partial b(z)c(z)::b(0)\partial
c(0):\nonumber \\
&& \quad{}-\lambda(1-\lambda):b(z)\partial c(z)::\partial
b(0)c(0):\nonumber \\
&& \quad{}+\lambda^2:b(z)\partial c(z)::b(0)\partial
c(0):\nonumber \\
&\sim& \frac{-6\lambda^2+6\lambda-1}{z^4}+\mathcal
O\left(\frac{1}{z^2}\right),
\eea
so
\be
c = -12\lambda^2+12\lambda-2.
\ee
Of course $\tilde T(\bar z)\tilde T(0) = 0$, so $\tilde c=0$.

The $\beta\gamma$ system has the same energy-momentum tensor and almost
the same OPEs as the $bc$ system. While $\gamma(z)\beta(0)\sim1/z$ as in
the $bc$ system, now $\beta(z)\gamma(0)\sim-1/z$. Each term in (33)
involved one $b(z)c(0)$ contraction and one $c(z)b(0)$ contraction, so the
central charge of the $\beta\gamma$ system is minus that of the $bc$
system:
\be
c = 12\lambda^2-12\lambda+2.
\ee
Of course $\tilde c=0$ still.

\subsection{Problem 2.11}

Assume without loss of generality that $m>1$; for $m=0$ and $m=\pm1$ the
central charge term in (2.6.19) vanishes, while $m<-1$ is equivalent to
$m>1$. Then $L_m$ annihilates $|0;0\rangle$, as do all but $m-1$ of the
terms in the mode expansion (2.7.6) of $L_{-m}$:
\be
L_{-m}|0;0\rangle =
\frac{1}{2}\sum_{n=1}^{m-1}\alpha^\mu_{n-m}\alpha_{\mu(-n)}|0;0\rangle.
\ee
Hence the LHS of (2.6.19), when applied to $|0;0\rangle$, yields,
\bea
\lefteqn{[L_m,L_{-m}]|0;0\rangle}\qquad\quad\nonumber \\
&=& L_mL_{-m}|0;0\rangle-L_{-m}L_m|0;0\rangle\nonumber \\
&=&
\frac{1}{4}\sum_{n'=-\infty}^\infty\sum_{n=1}^{m-1}\alpha^\nu_{m-n'}\alpha_{\nu
n'}\alpha^\mu_{n-m}\alpha_{\mu(-n)}|0;0\rangle\nonumber \\
&=&
\frac{1}{4}\sum_{n=1}^{m-1}\sum_{n'=-\infty}^\infty\left((m-n')n'\eta^{\nu\mu}\eta_{\nu\mu}\delta_{n'n}+(m-n')n'\delta^\nu_\mu\delta^\mu_\nu\delta_{m-n',n}\right)|0;0\rangle\nonumber
\\
&=& \frac{D}{2}\sum_{n=1}^{m-1}n(m-n)|0;0\rangle\nonumber \\
&=& \frac{D}{12}m(m^2-1)|0;0\rangle.
\eea
Meanwhile, the RHS of (2.6.19) applied to the same state yields,
\be
\left(2mL_0+\frac{c}{12}(m^3-m)\right)|0;0\rangle =
\frac{c}{12}(m^3-m)|0;0\rangle,
\ee
so
\be
c = D.
\ee

\newcommand{\no}{{}^{{}_\circ}_{{}^\circ}}

\subsection{Problem 2.13}

\paragraph{(a)}
Using (2.7.16) and (2.7.17),
\bea
\no b(z)c(z')\no &=&\sum_{m,m'=-\infty}^\infty\frac{\no
b_mc_{m'}\no}{z^{m+\lambda}z'^{m'+1-\lambda}}\nonumber \\
&=&
\sum_{m,m'=-\infty}^\infty\frac{b_mc_{m'}}{z^{m+\lambda}z'^{m'+1-\lambda}}-\sum_{m=0}^\infty\frac{1}{z^{m+\lambda}z'^{-m+1-\lambda}}\nonumber
\\
&=& b(z)c(z')-\left(\frac{z}{z'}\right)^{1-\lambda}\frac{1}{z-z'}.
\eea
With (2.5.7),
\be
:b(z)c(z'):-\no b(z)c(z')\no =
\frac{1}{z-z'}\left(\left(\frac{z}{z'}\right)^{1-\lambda}-1\right).
\ee

\paragraph{(b)}
By taking the limit of (41) as $z'\rightarrow z$, we find,
\be
:b(z)c(z):-\no b(z)c(z)\no = \frac{1-\lambda}{z}.
\ee
Using (2.8.14) we have,
\bea
N^{\rm g} &=& Q^{\rm g}-\lambda+\frac{1}{2}\nonumber \\
&=& \frac{1}{2\pi i}\oint dz\,j_z-\lambda+\frac{1}{2}\nonumber \\
&=& -\frac{1}{2\pi i}\oint dz\,:b(z)c(z):-\lambda+\frac{1}{2}\nonumber \\
&=& -\frac{1}{2\pi i}\oint dz\,\no b(z)c(z)\no-\frac{1}{2}.
\eea

\paragraph{(c)}
If we re-write the expansion (2.7.16) of $b(z)$ in the $w$-frame using the
tensor transformation law (2.4.15), we find,
\bea
b(w) &=& (\partial_zw)^{-\lambda}b(z)\nonumber \\
&=& (-iz)^\lambda\sum_{m=-\infty}^\infty\frac{b_m}{z^{m+\lambda}}\nonumber
\\
&=& e^{-\pi i\lambda/2}\sum_{m=-\infty}^\infty e^{imw}b_m.
\eea
Similarly,
\be
c(w) = e^{-\pi i(1-\lambda)/2}\sum_{m=-\infty}^\infty e^{imw}c_m.
\ee
Hence, ignoring ordering,
\bea
j_w(w) &=& -b(w)c(w)\nonumber \\
&=& i\sum_{m,m'=-\infty}^\infty e^{i(m+m')w}b_mc_{m'},
\eea
and
\bea
N^{\rm g} &=& -\frac{1}{2\pi i}\int_0^{2\pi}dw\,j_w\nonumber \\
&=& -\sum_{m=-\infty}^\infty b_mc_{-m}\nonumber \\
&=& -\sum_{m=-\infty}^\infty\no b_mc_{-m}\no-\sum_{m=0}^\infty1.
\eea
The ordering constant is thus determined by the value of the second
infinite sum. If we write, more generally, $\sum_{m=0}^\infty a$, then we
must regulate the sum in such a way that the divergent part is independent
of $a$. For instance,
\bea
\sum_{m=0}^\infty ae^{-\epsilon a} &=& \frac{a}{1-e^{-\epsilon a}}\nonumber
\\
&=& \frac{1}{\epsilon}+\frac{a}{2}+\mathcal O(\epsilon);
\eea
the $\epsilon$-independent part is $a/2$, so the ordering constant in (47)
equals $-1/2$.

\subsection{Problem 2.15}

To apply the doubling trick to the field $X^\mu(z,\bar z)$, define for $\Im
z<0$,
\be
X^\mu(z,\bar z) \equiv X^\mu(z^*,\bar z^*).
\ee
Then
\be
\partial^mX^\mu(z) = \bar\partial^mX^\mu(\bar z^*),
\ee
so that in particular for $z$ on the real line,
\be
\partial^mX^\mu(z) = \bar\partial^mX^\mu(\bar z),
\ee
as can also be seen from the mode expansion (2.7.26). The modes
$\alpha^\mu_m$ are defined as integrals over a semi-circle of $\partial
X^\mu(z)+\bar\partial X^\mu(\bar z)$, but with the doubling trick the
integral can be extended to the full circle:
\be
\alpha^\mu_m =
\sqrt{\frac{2}{\alpha'}}\oint\frac{dz}{2\pi}z^m\partial X^\mu(z) = 
-\sqrt{\frac{2}{\alpha'}}\oint\frac{d\bar z}{2\pi}\bar z^m\bar\partial
X^\mu(\bar z).
\ee

At this point the derivation proceeds in exactly the same manner as for the
closed string treated in the text. With no operator at the origin, the
fields are holomorphic inside the contour, so with $m$ positive, the
contour integrals (52) vanish, and the state corresponding to the unit
operator ``inserted'' at the origin must be the ground state $|0;0\rangle$:
\be
1(0,0) \cong |0;0\rangle.
\ee
The state $\alpha^\mu_{-m}|0;0\rangle$ ($m$ positive) is given by
evaluating the integrals (52), with the fields holomorphic inside the
contours:
\be
\alpha^\mu_{-m}|0;0\rangle \cong
\left(\frac{2}{\alpha'}\right)^{1/2}\frac{i}{(m-1)!}\partial^mX^\mu(0) =
\left(\frac{2}{\alpha'}\right)^{1/2}\frac{i}{(m-1)!}\bar\partial^mX^\mu(0).
\ee
Similarly, using the mode expansion (2.7.26), we see that
$X^\mu(0,0)|0;0\rangle = x^\mu|0;0\rangle$, so
\be
x^\mu|0;0\rangle \cong X^\mu(0,0).
\ee
As in the closed string case, the same correspondence applies when these
operators act on states other than the ground state, as long as we normal
order the resulting local operator. The result is therefore exactly the
same as (2.8.7a) and (2.8.8) in the text; for example, (2.8.9) continues to
hold.

\subsection{Problem 2.17}

Take the matrix element of (2.6.19) between $\langle1|$ and $|1\rangle$,
with $n=-m$ and $m>1$. The LHS yields,
\bea
\langle1|[L_m,L_{-m}]|1\rangle &=&
\langle1|L_{-m}^\dagger L_{-m}|1\rangle\nonumber \\
&=& \|L_{-m}|1\rangle\|^2,
\eea
using (2.9.9). Also by (2.9.9), $L_0|1\rangle=0$, so on the RHS we are left
with the term
\be
\frac{c}{12}(m^3-m)\langle1|1\rangle.
\ee
Hence
\be
c = \frac{12}{m^3-m}\frac{\|L_{-m}|1\rangle\|^2}{\langle1|1\rangle} \ge 0.
\ee

\setcounter{equation}{0}

\newpage

\section{Chapter 3}

\subsection{Problem 3.1}

\paragraph{(a)}
The definition of the geodesic curvature $k$ of a boundary given in Problem
1.3 is
\be
k = -n_bt^a\nabla_at^b,
\ee
where $t^a$ is the unit tangent vector to the boundary and $n^b$ is the
outward directed unit normal. For a flat unit disk, $R$ vanishes, while the
geodesic curvature of the boundary is 1 (since
$t^a\nabla_at^b=-n^b$). Hence
\be
\chi = \frac{1}{2\pi}\int_0^{2\pi}d\theta = 1.
\ee
For the unit hemisphere, on the other hand, the boundary is a geodesic,
while $R=2$. Hence
\be
\chi = \frac{1}{4\pi}\int d^2\!\sigma\,g^{1/2}2 = 1,
\ee
in agreement with (2).

\paragraph{(b)}
If we cut a surface along a closed curve, the two new boundaries will have
oppositely directed normals, so their contributions to the Euler number of
the surface will cancel, leaving it unchanged. The Euler number of the unit
sphere is
\be
\chi = \frac{1}{4\pi}\int d^2\!\sigma\,g^{1/2}2 = 2.
\ee
If we cut the sphere along $b$ small circles, we will be left with $b$
disks and a sphere with $b$ holes. The Euler number of the disks is $b$
(from part (a)), so the Euler number of the sphere with $b$ holes is
\be
\chi = 2-b.
\ee

\paragraph{(c)}
A finite cylinder has Euler number 0, since we can put on it a globally
flat metric for which the boundaries are geodesics. If we remove from a
sphere $b+2g$ holes, and then join to $2g$ of the holes $g$ cylinders, the
result will be a sphere with $b$ holes and $g$ handles; its Euler
number will be
\be
\chi = 2-b-2g.
\ee

\subsection{Problem 3.2}

\paragraph{(a)}
This is easiest to show in complex coordinates, where $g^{zz}=g^{\bar z\bar
z}=0$. Contracting two indices of a symmetric tensor with lower indices by
$g^{ab}$ will pick out the components where one of the indices is $z$ and
the other $\bar z$. If the tensor is traceless then all such components
must vanish. The only non-vanishing components are therefore the one with
all $z$ indices and the one with all $\bar z$ indices.

\paragraph{(b)}
Let $v_{a_1\cdots a_n}$ be a traceless symmetric tensor. Define $P_n$ by
\be
(P_nv)_{a_1\cdots a_{n+1}} \equiv \nabla_{(a_1}v_{a_2\cdots
a_{n+1})}-\frac{n}{n+1}g_{(a_1a_2}\nabla^{}_{|b|}v^b_{\hspace{0.4em}a_3\cdots a_{n+1})}.
\ee
This tensor is symmetric by construction, and it is easy to see that it is
also traceless. Indeed, contracting with $g^{a_1a_2}$, the first term
becomes
\be
g^{a_1a_2}\nabla_{(a_1}v_{a_2\cdots a_{n+1})} =
\frac{2}{n+1}\nabla^{}_bv^b_{\hspace{0.4em}a_3\cdots a_{n+1}},
\ee
where we have used the symmetry and tracelessness of $v$, and the second
cancels the first:
\bea
\lefteqn{g^{a_1a_2}g_{(a_1a_2}\nabla^{}_{|b|}v^b_{\hspace{0.4em}a_3\cdots
a_{n+1})}}\qquad\quad\nonumber \\
&=&
\frac{2}{n(n+1)}g^{a_1a_2}g_{a_1a_2}\nabla^{}_bv^b_{\hspace{0.4em}a_3\cdots
a_{n+1}}+\frac{2(n-1)}{n(n+1)}g^{a_1a_2}g_{a_1a_3}\nabla^{}_bv^b_{\hspace{0.4em}a_2a_4\cdots
a_{n+1}}\nonumber \\
&=& \frac{2}{n}\nabla^{}_bv^b_{\hspace{0.4em}a_3\cdots a_{n+1}}.
\eea

\paragraph{(c)}
For $u_{a_1\cdots a_{n+1}}$ a traceless symmetric tensor, define $P^T_n$ by
\be
(P^T_nu)_{a_1\cdots a_n} \equiv -\nabla^{}_bu^b_{\hspace{0.4em}a_1\cdots
a_n}.
\ee
This inherits the symmetry and tracelessness of $u$.

\paragraph{(d)}
\bea
(u,P_nv) &=& \int d^2\!\sigma\,g^{1/2}u^{a_1\cdots
a_{n+1}}(P_nv)_{a_1\cdots a_{n+1}}\nonumber \\
&=& \int d^2\!\sigma\,g^{1/2}u^{a_1\cdots
a_{n+1}}\left(\nabla_{a_1}v_{a_2\cdots
a_{n+1}}-\frac{n}{n+1}g_{a_1a_2}\nabla^{}_{b}v^b_{\hspace{0.4em}a_3\cdots
a_{n+1}}\right)\nonumber \\
&=& -\int d^2\!\sigma\,g^{1/2}\nabla_{a_1}u^{a_1\cdots a_{n+1}}v_{a_2\cdots
a_{n+1}}\nonumber \\
&=& \int d^2\!\sigma\,g^{1/2}(P^T_nu)^{a_2\cdots a_{n+1}}v_{a_2\cdots
a_{n+1}}\nonumber \\
&=& (P^T_nu,v)
\eea

\subsection{Problem 3.3}

\paragraph{(a)}
The conformal gauge metric in complex coordinates is $g_{z\bar z}=g_{\bar
zz} =e^{2\omega}/2$, $g_{zz}=g_{\bar z\bar z}=0$. Connection coefficients
are quickly calculated:
\bea
\Gamma^z_{zz} &=& \frac{1}{2}g^{z\bar z}(\partial_zg_{z\bar
z}+\partial_zg_{\bar zz}-\partial_{\bar z}g_{zz})\nonumber \\
&=& 2\partial\omega, \\
\Gamma^{\bar z}_{\bar z\bar z} &=& 2\bar\partial\omega,
\eea
all other coefficients vanishing.

This leads to the following simplification in the formula for the covariant
derivative:
\bea
\nabla^{}_zT^{a_1\cdots a_m}_{b_1\cdots b_n} &=& \partial_zT^{a_1\cdots
a_m}_{b_1\cdots b_n}+\sum_{i=1}^m\Gamma^{a_i}_{zc}T^{a_1\cdots c\cdots
a_m}_{b_1\cdots b_n}-\sum_{j=1}^n\Gamma^c_{zb_j}T^{a_1\cdots
a_m}_{b_1\cdots c\cdots b_n}\nonumber \\
&=&
\left(\partial+2\partial\omega\sum_{i=1}^m\delta_z^{a_i}-2\partial\omega\sum_{j=1}^n\delta_{b_j}^z\right)T^{a_1\cdots
a_m}_{b_1\cdots b_n};
\eea
in other words, it counts the difference between the number of upper $z$
indices and lower $z$ indices, while $\bar z$ indices do not
enter. Similarly,
\be
\nabla^{}_{\bar z}T^{a_1\cdots a_m}_{b_1\cdots b_n} =
\left(\bar\partial+2\bar\partial\omega\sum_{i=1}^m\delta_{\bar
z}^{a_i}-2\bar\partial\omega\sum_{j=1}^n\delta_{b_j}^{\bar
z}\right)T^{a_1\cdots a_m}_{b_1\cdots b_n}.
\ee
In particular, the covariant derivative with respect to $z$ of a tensor
with only $\bar z$ indices is equal to its regular derivative, and vice
versa:
\bea
\nabla_zT^{\bar z\cdots\bar z}_{\bar z\cdots\bar z} &=& \partial T^{\bar
z\cdots\bar z}_{\bar z\cdots\bar z},\nonumber \\
\nabla_{\bar z}T^{z\cdots z}_{z\cdots z} &=& \bar\partial T^{z\cdots
z}_{z\cdots z}.
\eea

\paragraph{(b)}
As shown in problem 3.2(a), the only non-vanishing components of a
traceless symmetric tensor with lowered indices have all them $z$ or all of
them $\bar z$. If $v$ is an $n$-index traceless symmetric tensor, then
$P_nv$ will be an $(n+1)$-index traceless symmetric tensor, and will
therefore have only two non-zero components:
\bea
(P_nv)_{z\cdots z} &=& \nabla_zv_{z\cdots z}\nonumber \\
&=& \left(\frac{1}{2}e^{2\omega}\right)^n\nabla_zv^{\bar z\cdots\bar
z}\nonumber \\
&=& \left(\frac{1}{2}e^{2\omega}\right)^n\partial v^{\bar z\cdots\bar
z}\nonumber \\
&=& (\partial-2n\partial\omega)v_{z\cdots z}; \\
(P_nv)_{\bar z\cdots\bar z} &=& (\bar\partial-2n\bar\partial\omega)v_{\bar
z\cdots\bar z}.
\eea
Similarly, if $u$ is an $(n+1)$-index traceless symmetric tensor, then
$P_n^Tu$ will be an $n$-index traceless symmetric tensor, and will have
only two non-zero components:
\bea
(P^T_nu)_{z\cdots z} &=&
-\nabla^{}_bu^b_{\hspace{0.4em}z\cdots z}\nonumber \\
&=& -2e^{-2\omega}\nabla^{}_zu_{\bar zz\cdots
z}-2e^{-2\omega}\nabla^{}_{\bar z}u_{zz\cdots z}\nonumber \\
&=& -\left(\frac{1}{2}e^{2\omega}\right)^{n-1}\partial u_{\bar
z}^{\hspace{0.4em}\bar z\cdots\bar z}-2e^{-2\omega}\bar\partial u_{z\cdots
z}\nonumber \\
&=& -2e^{-2\omega}\bar\partial u_{z\cdots z}; \\
(P^T_nu)_{\bar z\cdots\bar z} &=& -2e^{-2\omega}\partial u_{\bar
z\cdots\bar z}.
\eea

\subsection{Problem 3.4}
The Faddeev-Popov determinant is defined by,
\be
\Delta_{\rm FP}(\phi) \equiv
\left(\int[d\zeta]\,\delta\left(F^A(\phi^\zeta)\right)\right)^{-1}.
\ee
By the gauge invariance of the measure $[d\zeta]$ on the gauge group, this
is a gauge-invariant function. It can be used to re-express the
gauge-invariant formulation of the path integral, with arbitrary
gauge-invariant insertions $f(\phi)$, in a gauge-fixed way:
\bea
\frac{1}{V}\int[d\phi]\,e^{-S_1(\phi)}f(\phi) &=&
\frac{1}{V}\int[d\phi]\,e^{-S_1(\phi)}\Delta_{\rm
FP}(\phi)\int[d\zeta]\,\delta\left(F^A(\phi^\zeta)\right)f(\phi)\nonumber
\\
&=&
\frac{1}{V}\int[d\zeta\,d\phi^\zeta]\,e^{-S_1(\phi^\zeta)}\Delta_{\rm
FP}(\phi^\zeta)\delta\left(F^A(\phi^\zeta)\right)f(\phi^\zeta)\nonumber \\
&=& \int[d\phi]\,e^{-S_1(\phi)}\Delta_{\rm FP}(\phi)\delta\left(F^A(\phi)\right)f(\phi).
\eea
In the second equality we used the gauge invariance of
$[d\phi]\,e^{-S_1(\phi)}$ and $f(\phi)$, and in the third line we renamed
the variable of integration, $\phi^\zeta\to\phi$.

In the last line of (22), $\Delta_{\rm FP}$ is evaluated only for $\phi$ on
the gauge slice, so it suffices to find an expression for it that is valid
there. Let $\hat\phi$ be on the gauge slice (so $F^A(\hat\phi)=0$),
parametrize the gauge group near the identity by coordinates $\epsilon^B$,
and define
\be
\delta_BF^A(\hat\phi) \equiv \left.\frac{\partial F^A(\hat\phi^\zeta)}{\partial\epsilon^B}\right|_{\epsilon=0}
= \left.\frac{\partial
F^A}{\partial\phi_i}\frac{\partial\hat\phi_i^\zeta}{\partial\epsilon^B}\right|_{\epsilon=0}.
\ee
If the $F^A$ are properly behaved (i.e.\ if they have non-zero and linearly
independent gradients at $\hat\phi$), and if there are no gauge
transformations that leave $\hat\phi$ fixed, then $\delta_BF^A$ will be a
non-singular square matrix. If we choose the coordinates $\epsilon^B$ such
that $[d\zeta]=[d\epsilon^B]$ locally, then the Faddeev-Popov determinant
is precisely the determinant of $\delta_BF^A$, and can be represented as a
path integral over ghost fields:
\bea
\Delta_{\rm FP}(\hat\phi) &=&
\left(\int[d\epsilon^B]\,\delta\left(F^A(\hat\phi^\zeta)\right)\right)^{-1}\nonumber
\\
&=&
\left(\int[d\epsilon^B]\,\delta\left(\delta_BF^A(\hat\phi)\epsilon^B\right)\right)^{-1}\nonumber
\\
&=& \det\left(\delta_BF^A(\hat\phi)\right)\nonumber \\
&=& \int[db_A\,dc^B]\,e^{-b_A\delta_BF^A(\hat\phi)c^B}.
\eea
Finally, we can express the delta function appearing in the gauge-fixed
path integral (22)
as a path integral itself:
\be
\delta\left(F^A(\phi)\right) = \int[dB_A]e^{iB_AF^A(\phi)}.
\ee
Putting it all together, we obtain (4.2.3):
\be
\int[d\phi\,db_A\,dc^B\,dB_A]\,e^{-S_1(\phi)-b_A\delta_BF^A(\phi)c^B+iB_AF^A(\phi)}f(\phi).
\ee

\subsection{Problem 3.5}
For each field configuration $\phi$, there is a unique gauge-equivalent
configuration $\hat\phi_F$ in the gauge slice defined by the $F^A$, and a
unique gauge transformation $\zeta_F(\phi)$ that takes $\hat\phi_F$ to
$\phi$:
\be
\phi = \hat\phi_F^{\zeta_F(\phi)}.
\ee
For $\phi$ near $\hat\phi_F$, $\zeta_F(\phi)$ will be near the identity and
can be parametrized by $\epsilon^B_F(\phi)$, the same coordinates used in
the previous problem. For such $\phi$ we have
\be
F^A(\phi) = \delta_BF^A(\hat\phi_F)\epsilon^B_F(\phi),
\ee
and we can write the factor $\Delta^F_{\rm FP}(\phi)\delta(F^A(\phi))$
appearing in the gauge-fixed path integral (22) in terms of
$\epsilon^B_F(\phi)$:
\bea
\Delta^F_{\rm FP}(\phi)\delta\left(F^A(\phi)\right) &=& \Delta^F_{\rm
FP}(\hat\phi_F)\delta\left(F^A(\phi)\right)\nonumber \\
&=&
\det\left(\delta_BF^A(\hat\phi_F)\right)\delta\left(\delta_BF^A(\hat\phi_F)\epsilon^B_F(\phi)\right)\nonumber
\\
&=& \delta\left(\epsilon^B_F(\phi)\right).
\eea

Defining $\zeta_G(\phi)$ in the same way, we have,
\be
\zeta_G\left(\phi^{\zeta_G^{-1}\zeta_F(\phi)}\right) =
\zeta_F\left(\phi\right).
\ee
Defining
\be
\phi' \equiv \phi^{\zeta_G^{-1}\zeta_F(\phi)},
\ee
it follows from (29) that
\be
\Delta^F_{\rm FP}(\phi)\delta\left(F^A(\phi)\right) = \Delta^G_{\rm
FP}(\phi')\delta\left(G^A(\phi')\right).
\ee
It is now straightforward to prove that the gauge-fixed path integral is
independent of the choice of gauge:
\bea
\int[d\phi]\,e^{-S(\phi)}\Delta^F_{\rm
FP}(\phi)\delta\left(F^A(\phi)\right)f(\phi) &=&
\int[d\phi']e^{-S(\phi')}\Delta^G_{\rm
FP}(\phi')\delta\left(G^A(\phi')\right)f(\phi')\nonumber \\
&=& \int[d\phi]\,e^{-S(\phi)}\Delta^G_{\rm
FP}(\phi)\delta\left(G^A(\phi)\right)f(\phi).\nonumber \\
\eea
In the first line we simultaneously used (32) and the gauge invariance of
the measure $[d\phi]e^{-S(\phi)}$ and the insertion $f(\phi)$; in the
second line we renamed the variable of integration from $\phi'$ to $\phi$.

\subsection{Problem 3.7}

Let us begin by expressing (3.4.19) in momentum space, to know what we're
aiming for. The Ricci scalar, to lowest order in the metric perturbation
$h_{ab}=g_{ab}-\delta_{ab}$, is
\be
R \approx (\partial_a\partial_b - \delta_{ab}\partial^2)h_{ab}.
\ee
In momentum space, the Green's function defined by (3.4.20) is
\be
\tilde{G}(p) \approx -\frac{1}{p^2}
\ee
(again to lowest order in $h_{ab}$), so the exponent of (3.4.19) is
\be
-\frac{a_1}{8\pi}
\int\frac{d^2\!p}{(2\pi)^2}\tilde{h}_{ab}(p)\tilde{h}_{cd}(-p)
\left(
\frac{p_ap_bp_cp_d}{p^2}
-2\delta_{ab}p_cp_d+
\delta_{ab}\delta_{cd}p^2
\right).
\ee

To first order in $h_{ab}$, the Polyakov action (3.2.3a) is
\be
S_{X} = \frac{1}{2}\int d^2\!\sigma
\left(\partial_aX\partial_aX + 
(\frac{1}{2}h\delta_{ab}-h_{ab})\partial_aX\partial_bX\right),
\ee
where $h\equiv h_{aa}$ (we have set $2\pi\alpha'$ to 1). We will use
dimensional regularization, which breaks conformal invariance because
the graviton trace couples to $X$ when $d\neq2$. The traceless part of
$h_{ab}$ in $d$ dimensions is $h'_{ab}=h_{ab}-h/d$. This leaves a coupling
between $h$ and $\partial_aX\partial_aX$ with coefficient
$1/2-1/d$. The momentum-space vertex for $h'_{ab}$ is
\be
\tilde{h}'_{ab}(p)k_a(k_b+p_b),
\ee
while that for $h$ is
\be
-\frac{d-2}{2d}\tilde{h}(p)k\cdot(k+p).
\ee
There are three one-loop diagrams with two external gravitons, depending on
whether the gravitons are traceless or trace.

We begin by dispensing with the $hh$ diagram. In dimensional
regularization, divergences in loop integrals show up as poles in the $d$
plane. Arising as they do in the form of a gamma function, these are always
simple poles. But the diagram is multiplied by two factors of $d-2$ from
the two $h$ vertices, so it vanishes when we take $d$ to 2.

The $hh'_{ab}$ diagram is multiplied by only one factor of $d-2$, so part
of it (the divergent part that would normally be subtracted off) might
survive. It is equal to
\be
-\frac{d-2}{4d}\int\frac{d^dp}{(2\pi)^d}\tilde{h}'_{ab}(p)\tilde{h}(-p)
\int\frac{d^dk}{(2\pi)^d}\frac{k_a(k_b+p_b)k\cdot(k+p)}{k^2(k+p)^2}.
\ee
The $k$ integral can be evaluated by the usual tricks:
\bea
\lefteqn{\int_0^1dx\int\frac{d^dk}{(2\pi)^d}
\frac{k_a(k_b+p_b)k\cdot(k+p)}{(k^2+2xp\cdot k+xp^2)^2}}\qquad\quad \\
&& = \int_0^1dx\int\frac{d^dq}{(2\pi)^d}
\frac{(q_a-xp_a)(q_b+(1-x)p_b)(q-xp)\cdot(q+(1-x)p)}
{\left(q^2+x(1-x)p^2\right)^2}. \nonumber
\eea
Discarding terms that vanish due to the tracelessness of $h'_{ab}$ or that
are finite in the limit $d\to2$ yields
\be
p_ap_b\int_0^1dx\,(\frac{1}{2}-3x-3x^2)
\int\frac{d^dq}{(2\pi)^d}\frac{q^2}{(q^2+x(1-x)p^2)^2}.
\ee
The divergent part of the $q$ integral is independent of $x$, and the $x$
integral vanishes, so this diagram vanishes as well.

We are left with just the $h'_{ab}h'_{cd}$ diagram, which (including a
symmetry factor of 4 for the identical vertices and identical propagators)
equals
\be
\frac{1}{4}\int\frac{d^dp}{(2\pi)^d}\tilde{h}'_{ab}(p)\tilde{h}'_{cd}(-p)
\int\frac{d^dk}{(2\pi)^d}\frac{k_a(k_b+p_b)k_c(k_d+p_d)}{k^2(k+p)^2}.
\ee
The usual tricks, plus the symmetry and tracelessness of $h'_{ab}$, allow
us to write the $k$ integral in the following way:
\be
\int_0^1dx\int\frac{d^dq}{(2\pi)^d}\frac
{\frac{2}{d(d+2)}\delta_{ac}\delta_{bd}q^4
+\frac{1}{d}(1-2x)^2\delta_{ac}p_bp_dq^2
+x^2(1-x)^2p_ap_bp_cp_d}
{(q^2+x(1-x)p^2)^2}.
\ee
The $q^4$ and $q^2$ terms in the numerator give rise to divergent
integrals. Integrating these terms over $q$ yields
\bea
\lefteqn{\frac{1}{8\pi}\int_0^1dx\,
\Gamma(1-\frac{d}{2})\left(\frac{x(1-x)p^2}{4\pi}\right)^{d/2-1}}
\qquad\qquad\qquad\qquad\qquad\qquad\qquad \\
&& \times\left[-\frac{2}{d}x(1-x)\delta_{ac}\delta_{bd}p^2
+(1-2x)^2\delta_{ac}p_bp_d\right]. \nonumber
\eea
The divergent part of this is
\be
\frac{\delta_{ac}\delta_{bd}p^2-2\delta_{ac}p_bp_d}{24\pi(d-2)}.
\ee
However, it is a fact that the symmetric part of the product of two
symmetric, traceless, $2\times2$ matrices is proportional to the identity
matrix, so the two terms in the numerator are actually equal after
multiplying by $\tilde{h}'_{ab}(p)\tilde{h}'_{cd}(-p)$---we see that
dimensional regularization has already discarded the divergence for us. The
finite part of (45) is (using this trick a second time)
\be
\frac{\delta_{ac}\delta_{bd}p^2}{8\pi}\int_0^1dx\,
\left[
\left(-\gamma-\ln\left(\frac{x(1-x)p^2}{4\pi}\right)\right)
(\frac{1}{2}-3x+3x^2)
-x(1-x)
\right].
\ee
Amazingly, this also vanishes upon performing the $x$ integral. It remains
only to perform the integral for the last term in the numerator of (44),
which is convergent at $d=2$:
\be
\int_0^1dx\int
\frac{d^2q}{(2\pi)^2}\frac{x^2(1-x)^2p_ap_bp_cp_d}{(q^2+x(1-x)p^2)^2}
= \frac{p_ap_bp_cp_d}{4\pi p^2}\int_0^1dx\,x(1-x) 
= \frac{p_ap_bp_cp_d}{24\pi p^2}.
\ee
Plugging this back into (43), we find for the 2-graviton contribution to
the vacuum amplitude,
\be
\frac{1}{96\pi}\int\frac{d^2\!p}{(2\pi)^2}
\tilde{h}_{ab}(p)\tilde{h}_{cd}(-p)
\left(
\frac{p_ap_bp_cp_d}{p^2}
-\delta_{ab}p_cp_d
+\frac{1}{4}\delta_{ab}\delta_{cd}p^2
\right).
\ee

This result does not agree with (36), and is furthermore quite peculiar. 
It is Weyl invariant (since
the trace $h$ decoupled), but not diff invariant. It therefore
appears that, instead of a Weyl anomaly, we have
discovered a gravitational anomaly. However, just because dimensional
regularization has (rather amazingly) thrown away the divergent parts of
the loop integrals for us, does not mean that renormalization becomes
unnecessary. We must still choose
renormalization conditions, and introduce counterterms to satisfy them. In
this case, we will impose diff invariance, which is
more important than Weyl invariance---without it, it would be impossible to
couple this CFT consistently to gravity. Locality in real space demands
that the counterterms be of the same form as the last two terms in the
parentheses in (49). We are therefore free to adjust the
coefficients of these two terms in order to achieve diff invariance. Since
(36) is manifestly diff invariant, it is clearly the desired expression,
with $a_1$ taking the value $-1/12$. (It is worth pointing out that there
is no local counterterm quadratic in $h_{ab}$ that one could add that is
diff invariant by itself, and that would therefore have to be fixed by some
additional renormalization condition. This is because diff-invariant
quantities are constructed out of the Ricci scalar, and $\int d^2\sigma
R^2$ has the wrong dimension.)

\subsection{Problem 3.9}

Fix coordinates such that the boundary lies at $\sigma_2=0$. Following the
prescription of problem 2.10 for normal ordering operators in the presence
of a boundary, we include in the contraction the image term:
\be
\Delta_{\rm b}(\sigma,\sigma') =
\Delta(\sigma,\sigma')+\Delta(\sigma,\sigma'^*),
\ee
where $\sigma^*_1=\sigma_1$, $\sigma^*_2=-\sigma_2$. If $\sigma$ and
$\sigma'$ both lie on the boundary, then the contraction is effectively
doubled:
\be
\Delta_{\rm b}(\sigma_1,\sigma'_1) =
2\Delta\left((\sigma_1,\sigma_2=0),(\sigma_1',\sigma_2'=0)\right).
\ee
If $\mathcal{F}$ is a boundary operator, then
the $\sigma_2$ and $\sigma_2'$ integrations in the definition (3.6.5) of
$[\mathcal{F}]_{\rm r}$ can be done trivially:
\be
[\mathcal{F}]_{\rm r} =
\exp\left(\frac{1}{2}\int
d\sigma_1d\sigma_1'\,\Delta_{\rm b}(\sigma_1,\sigma_1')\frac{\delta}{\delta
X^\nu(\sigma_1,\sigma_2=0)}\frac{\delta}{\delta
X_\nu(\sigma_1',\sigma_2'=0)}\right)\mathcal{F}.
\ee
Equation (3.6.7) becomes
\be
\delta_{\rm W}[\mathcal{F}]_{\rm r} = [\delta_{\rm W}\mathcal{F}]_{\rm
r}+\frac{1}{2}\int d\sigma_1d\sigma_1'\,\delta_{\rm
W}\Delta_{\rm b}(\sigma_1,\sigma_1')\frac{\delta}{\delta
X^\nu(\sigma_1)}\frac{\delta}{\delta X_\nu(\sigma_1')}[\mathcal{F}]_{\rm r}.
\ee

The tachyon vertex operator (3.6.25) is
\be
V_0 = g_{\rm o}\int_{\sigma_2=0} d\sigma_1\,g_{11}^{1/2}(\sigma_1)[e^{ik\cdot
X(\sigma_1)}]_{\rm r},
\ee
and its Weyl variation (53) is
\bea
\delta_{\rm W}V_0 &=& g_{\rm o}\int
d\sigma_1\,g_{11}^{1/2}(\sigma_1)\left(\delta\omega(\sigma_1)+\delta_{\rm
W}\right)[e^{ik\cdot X(\sigma_1)}]_{\rm r}\nonumber \\
&=& g_{\rm o}\int
d\sigma_1\,g_{11}^{1/2}(\sigma_1)\left(\delta\omega(\sigma_1)-\frac{k^2}{2}\delta_{\rm
W}\Delta_{\rm b}(\sigma_1,\sigma_1)\right)[e^{ik\cdot X(\sigma_1)}]_{\rm
r}\nonumber \\
&=&  (1-\alpha'k^2)g_{\rm o}\int
d\sigma_1\,g_{11}^{1/2}(\sigma_1)\delta\omega(\sigma_1)[e^{ik\cdot
X(\sigma_1)}]_{\rm r},
\eea
where we have used (3.6.11) in the last equality:
\be
\delta_{\rm W}\Delta_{\rm b}(\sigma_1,\sigma_1) = 2\delta_{\rm
W}\Delta(\sigma_1,\sigma_1') = 2\alpha'\delta\omega(\sigma_1).
\ee
Weyl invariance thus requires
\be
k^2 = \frac{1}{\alpha'}.
\ee

The photon vertex operator (3.6.26) is
\be
V_1 = -i\frac{g_{\rm o}}{\sqrt{2\alpha'}}e_\mu\int_{\sigma_2=0}
d\sigma_1\,[\partial_1X^\mu(\sigma_1)e^{ik\cdot X(\sigma_1)}]_{\rm r}.
\ee
The spacetime gauge equivalence,
\be
V_1(k,e) = V_1(k,e+\lambda k),
\ee
is clear from the fact that $k_\mu\partial_1X^\mu e^{ik\cdot X}$ is a total
derivative. The expression (58) has no explicit metric dependence, so the
variation of $V_1$ comes entirely from the variation of the renormalization
contraction:
\bea
\lefteqn{\delta_{\rm W}[\partial_1X^\mu(\sigma_1)e^{ik\cdot
X(\sigma_1)}]_{\rm r}}\qquad\qquad\nonumber \\
&=& \frac{1}{2}\int d\sigma_1'd\sigma_1''\,\delta_{\rm
W}\Delta_{\rm b}(\sigma_1',\sigma_1'')\frac{\delta}{\delta
X^\nu(\sigma_1')}\frac{\delta}{\delta
X_\nu(\sigma_1'')}[\partial_1X^\mu(\sigma_1)e^{ik\cdot X(\sigma_1)}]_{\rm
r}\nonumber \\
&=& ik^\mu\left.\partial_1\delta_{\rm
W}\Delta_{\rm b}(\sigma_1,\sigma_1'')\right|_{\sigma_1''=\sigma_1}[e^{ik\cdot
X(\sigma_1)}]_{\rm r}\nonumber \\
&& \qquad\qquad\qquad\qquad\qquad{}-\frac{k^2}{2}\delta_{\rm
W}\Delta_{\rm b}(\sigma_1,\sigma_1)[\partial_1X^\mu(\sigma_1)e^{ik\cdot
X(\sigma_1)}]_{\rm r}\nonumber \\
&=& i\alpha'k^\mu\partial_1\delta\omega(\sigma_1)[e^{ik\cdot
X(\sigma_1)}]_{\rm
r}-\alpha'k^2\delta\omega(\sigma_1)[\partial_1X^\mu(\sigma_1)e^{ik\cdot
X(\sigma_1)}]_{\rm r},
\eea
where in the last equality we have used (56) and (3.6.15a):
\be
\left.\partial_1\delta_{\rm W}\Delta_{\rm
b}(\sigma_1,\sigma_1')\right|_{\sigma_1'=\sigma_1} =
2\left.\partial_1\delta_{\rm
W}\Delta(\sigma_1,\sigma_1')\right|_{\sigma_1'=\sigma_1} =
\alpha'\partial_1\delta\omega(\sigma_1).
\ee
Integration by parts yields
\be
\delta_{\rm W}V_1 = -i\sqrt{\frac{\alpha'}{2}}g_{\rm o}(e\cdot
kk_\mu-k^2e_\mu)\int
d\sigma_1\,\delta\omega(\sigma_1)[\partial_1X^\mu(\sigma_1)e^{ik\cdot
X(\sigma_1)}]_{\rm r}.
\ee
For this quantity to vanish for arbitrary $\delta\omega(\sigma_1)$ requires
the vector $e\cdot kk-k^2e$ to vanish. This will happen if $e$ and $k$ are
collinear, but by (59) $V_1$ vanishes in this case. The other possibility
is
\be
k^2 = 0,\qquad e\cdot k = 0.
\ee

\subsection{Problem 3.11}

Since we are interested in the $H^2$ term, let us assume $G_{\mu\nu}$ to be
constant, $\Phi$ to vanish, and $B_{\mu\nu}$ to be linear in $X$, implying
that
\be
H_{\omega\mu\nu} = 3\partial_{[\omega}B_{\mu\nu]}
\ee
is constant. With these simplifications, the sigma model action becomes
\bea
S_\sigma &=& \frac{1}{4\pi\alpha'}\int
d^2\!\sigma\,g^{1/2}\left(G_{\mu\nu}g^{ab}\partial_aX^\mu\partial_bX^\nu+i\partial_\omega
B_{\mu\nu}\epsilon^{ab}X^\omega\partial_aX^\mu\partial_bX^\nu\right)\nonumber
\\
&=& \frac{1}{4\pi\alpha'}\int
d^2\!\sigma\,g^{1/2}\left(G_{\mu\nu}g^{ab}\partial_aX^\mu\partial_bX^\nu+\frac{i}{3}H_{\omega\mu\nu}\epsilon^{ab}X^\omega\partial_aX^\mu\partial_bX^\nu\right).\nonumber
\\
\eea
In the second line we have used the fact that
$\epsilon^{ab}X^\omega\partial_aX^\mu\partial_bX^\nu$ is totally
antisymmetric in $\omega,\mu,\nu$ (up to integration by parts) to
antisymmetrize $\partial_\omega B_{\mu\nu}$.

Working in conformal gauge on
the worldsheet and transforming to complex coordinates, 
\bea
g^{1/2}g^{ab}\partial_aX^\mu\partial_bX^\nu &=& 4\partial X^{(\mu}\bar\partial
X^{\nu)}, \\
g^{1/2}\epsilon^{ab}\partial_aX^\mu\partial_bX^\nu &=& -4i\partial
X^{[\mu}\bar\partial X^{\nu]}, \\
d^2\!\sigma &=& \frac{1}{2}d^2\!z,
\eea
the action becomes
\bea
S_\sigma &=& S_{\rm f}+S_{\rm i}, \\
S_{\rm f} &=& \frac{1}{2\pi\alpha'}G_{\mu\nu}\int d^2\!z\,\partial
X^\mu\bar\partial X^\nu, \\
S_{\rm i} &=& \frac{1}{6\pi\alpha'}H_{\omega\mu\nu}\int
d^2\!z\,X^\omega\partial X^\mu\bar\partial X^\nu,
\eea
where we have split it into the action for a free CFT and an interaction
term. The path integral is now
\bea
\langle\dots\rangle_\sigma &=& \langle e^{-S_{\rm i}}\dots\rangle_{\rm
f}\nonumber \\
&=& \langle\dots\rangle_{\rm f}-\langle S_{\rm i}\dots\rangle_{\rm
f}+\frac{1}{2}\langle S_{\rm i}^2\dots\rangle_{\rm f}+\cdots,
\eea
where $\langle\,\rangle_{\rm f}$ is the path integral calculated using only
the free action (70). The Weyl variation of the first term gives rise to
the $D-26$ Weyl anomaly calculated in section 3.4, while that of the second
gives rise to the term in $\beta^B_{\mu\nu}$ that is linear in $H$
(3.7.13b). It is the Weyl variation of the third term, quadratic in $H$,
that we are interested in, and in particular the part proportional to
\be
\int d^2\!z\,\langle:\partial X^\mu\bar\partial X^\nu:\dots\rangle_{\rm f},
\ee
whose coefficient gives the $H^2$ term in $\beta^G_{\mu\nu}$. This third
term is
\bea
\lefteqn{\frac{1}{2}\langle S_{\rm i}^2\dots\rangle_{\rm f} =
\frac{1}{2(6\pi\alpha')^2}H_{\omega\mu\nu}H_{\omega'\mu'\nu'}} \\
&& \times\int d^2\!zd^2\!z'\,\langle:X^\omega(z,\bar z)\partial X^\mu(z)\bar\partial
X^\nu(\bar z)::X^{\omega'}(z',\bar
z')\partial'X^{\mu'}(z')\bar\partial'X^{\nu'}(\bar z'):\dots\rangle_{\rm
f},\nonumber
\eea
where we have normal-ordered the interaction vertices. The Weyl variation
of this integral will come from the singular part of the OPE when $z$ and
$z'$ approach each other. Terms in the OPE containing exactly two $X$
fields (which will yield (73) after the $z'$ integration is performed) are
obtained by performing two cross-contractions. There are 18 different pairs of
cross-contractions one can apply to the integrand of (74), but, since they
can all be obtained from each other by integration by parts and permuting
the indices $\omega,\mu,\nu$, they all give the same result. The
contraction derived from the free action (70) is
\be
X^\mu(z,\bar z)X^\nu(z',\bar z') = :X^\mu(z,\bar z)X^\nu(z',\bar
z'):-\frac{\alpha'}{2}G^{\mu\nu}\ln|z-z'|^2,
\ee
so, picking a representative pair of cross-contractions, the part of (74)
we are interested in is
\bea
\lefteqn{\frac{18}{2(6\pi\alpha')^2}H_{\omega\mu\nu}H_{\omega'\mu'\nu'}}\nonumber
\\
&& \times\int
d^2\!zd^2\!z'\,\left(-\frac{\alpha'}{2}\right)G^{\omega\mu'}\partial'\ln|z-z'|^2\left(-\frac{\alpha'}{2}\right)G^{\nu\omega'}\bar\partial\ln|z-z'|^2\nonumber
\\
&& \qquad\qquad\qquad\qquad\qquad\qquad\qquad\qquad\qquad\qquad\quad\times\langle:\partial
X^\mu(z)\bar\partial'X^{\nu'}(\bar z'):\dots\rangle_{\rm f}\nonumber \\
&=& -\frac{1}{16\pi^2}H_{\mu\lambda\omega}H_\nu{}^{\lambda\omega}\int
d^2\!zd^2\!z'\,\frac{1}{|z'-z|^2}\langle:\partial
X^\mu(z)\bar\partial'X^\nu(\bar z'):\dots\rangle_{\rm f}.
\eea
The Weyl variation of this term comes from cutting off the logarithmically
divergent integral of $|z'-z|^{-2}$ near $z'=z$, so we can drop the less
singular terms coming from the Taylor expansion of $\bar\partial'X^\nu(\bar
z')$:
\be
-\frac{1}{16\pi^2}H_{\mu\lambda\omega}H_\nu{}^{\lambda\omega}\int
d^2\!z\,\langle:\partial X^\mu(z)\bar\partial X^\nu(\bar
z):\dots\rangle_{\rm f}\int d^2\!z'\,\frac{1}{|z'-z|^2}.
\ee
The diff-invariant distance between $z'$ and $z$ is (for short distances)
$e^{\omega(z)}|z'-z|$, so a diff-invariant cutoff would be at
$|z'-z|=\epsilon e^{-\omega(z)}$. The Weyl-dependent part of the second
integral of (77) is then
\be
\int d^2\!z'\,\frac{1}{|z'-z|^2} \sim -2\pi\ln(\epsilon e^{-\omega(z)}) = -2\pi\ln\epsilon+2\pi\omega(z),
\ee
and the Weyl variation of (76) is
\be
-\frac{1}{8\pi}H_{\mu\lambda\omega}H_\nu{}^{\lambda\omega}\int
d^2\!z\,\delta\omega(z)\langle:\partial X^\mu(z)\bar\partial X^\nu(\bar
z):\dots\rangle_{\rm f}.
\ee
Using (66) and (68), and the fact that the difference between
$\langle\,\rangle_\sigma$ and $\langle\,\rangle_{\rm f}$ involves higher
powers of $H$ (see (72)) which we can neglect, we can write this as
\be
-\frac{1}{16\pi}H_{\mu\lambda\omega}H_\nu{}^{\lambda\omega}\int
d^2\!\sigma\,g^{1/2}\delta\omega g^{ab}\langle:\partial_aX^\mu\partial_bX^\nu:\dots\rangle_\sigma.
\ee
This is of the form of (3.4.6), with
\be
T'^a{}_a =
\frac{1}{8}H_{\mu\lambda\omega}H_\nu{}^{\lambda\omega}g^{ab}\partial_aX^\mu\partial_bX^\nu
\ee
being the contribution of this term to the stress tensor. According to
(3.7.12), $T'^a{}_a$ in turn contributes the following term to
$\beta^G_{\mu\nu}$:
\be
-\frac{\alpha'}{4}H_{\mu\lambda\omega}H_\nu{}^{\lambda\omega}.
\ee

\subsection{Problem 3.13}

If the dilaton $\Phi$ is constant and $D=d+3$, then the equations of motion
(3.7.15) become, to leading order in $\alpha'$,
\bea
R_{\mu\nu}-\frac{1}{4}H_{\mu\lambda\omega}H_\nu{}^{\lambda\omega}
&=& 0, \\
\nabla^\omega H_{\omega\mu\nu} &=& 0, \\
\frac{d-23}{\alpha'}-\frac{1}{4}H_{\mu\nu\lambda}H^{\mu\nu\lambda} &=& 0.
\eea
Letting $i,j,k$ be indices on the 3-sphere and $\alpha,\beta,\gamma$ be
indices on the flat $d$-dimensional spacetime, we apply the ansatz
\be
H_{ijk} = h\epsilon_{ijk},
\ee
where $h$ is a constant and $\epsilon$ is the volume form on the
sphere, with all other components vanishing. (Note that this form for $H$
cannot be obtained as the exterior derivative of a non-singular gauge field
$B$; $B$ must have a Dirac-type singularity somewhere on the sphere.)
Equation (84) is then immediately satisfied, because the volume form is
always covariantly constant on a manifold, so $\nabla^{i}H_{ijk}=0$, and
all other components vanish trivially. Since
$\epsilon_{ijk}\epsilon^{ijk}=6$, equation (85) fixes $h$ in terms of $d$:
\be
h^2 = \frac{2(d-23)}{3\alpha'},
\ee
implying that there are solutions only for $d>23$. The Ricci tensor on a
3-sphere of radius $r$ is given by
\be
R_{ij} = \frac{2}{r^2}G_{ij}.
\ee
Similarly,
\be
\epsilon_{ikl}\epsilon_j{}^{kl} = 2G_{ij}.
\ee
Most components of equation (83) vanish trivially, but those for which both
indices are on the sphere fix $r$ in terms of $h$:
\be
r^2 = \frac{4}{h^2} = \frac{6\alpha'}{d-23}.
\ee

\setcounter{equation}{0}

\newpage
\section{Chapter 4}

\subsection{Problem 4.1}

To begin, let us recall the spectrum of the open string at level $N=2$ in
light-cone quantization. In representations of SO($D-2$), we had a
symmetric rank 2 tensor, 
\be
f_{ij}\alpha^i_{-1}\alpha^j_{-1}|0;k\rangle,
\ee
and a vector,
\be
e_i\alpha^i_{-2}|0;k\rangle.
\ee
Together, they make up the traceless symmetric rank 2 tensor representation
of SO($D-1$), whose dimension is $D(D-1)/2-1$. This is what we expect to
find.

In the OCQ, the general state at level 2 is
\be
|f,e;k\rangle =
\left(f_{\mu\nu}\alpha_{-1}^\mu\alpha_{-1}^\nu+e_\mu\alpha_{-2}^\mu\right)|0;k\rangle,
\ee
a total of $D(D+1)/2+D$ states. Its norm is
\bea
\langle e,f;k|e,f;k'\rangle &=&
\langle0;k|\left(f^*_{\rho\sigma}\alpha_1^\rho\alpha_1^\sigma+e^*_\rho\alpha^\rho_2\right)\left(f_{\mu\nu}\alpha_{-1}^\mu\alpha_{-1}^\nu+e_\mu\alpha_{-2}^\mu\right)|0;k'\rangle\nonumber \\ 
&=& 2\left(f^*_{\mu\nu}f^{\mu\nu}+e^*_\mu e^\mu\right)\langle0;k|0;k'\rangle.
\eea
The terms in the mode expansion of the Virasoro generator relevant here
are as follows:
\bea
L_0 &=& \alpha'p^2+\alpha_{-1}\cdot\alpha_1+\alpha_{-2}\cdot\alpha_2+\cdots
\\
L_1 &=& \sqrt{2\alpha'}p\cdot\alpha_1+\alpha_{-1}\cdot\alpha_2+\cdots \\
L_2 &=&
\sqrt{2\alpha'}p\cdot\alpha_2+\frac{1}{2}\alpha_1\cdot\alpha_1+\cdots \\
L_{-1} &=& \sqrt{2\alpha'}p\cdot\alpha_{-1}+\alpha_{-2}\cdot\alpha_1+\cdots
\\
L_{-2} &=&
\sqrt{2\alpha'}p\cdot\alpha_{-2}+\frac{1}{2}\alpha_{-1}\cdot\alpha_{-1}+\cdots.
\eea
As in the cases of the tachyon and photon, the $L_0$ condition yields the
mass-shell condition:
\bea
0 &=& (L_0-1)|f,e;k\rangle\nonumber \\
&=& (\alpha'k^2+1)|f,e;k\rangle,
\eea
or $m^2=1/\alpha'$, the same as in the light-cone quantization. Since the
particle is massive, we can go to its rest frame for simplicity:
$k_0=1/\sqrt{\alpha'}$, $k_i=0$. The $L_1$ condition fixes $e$ in terms of
$f$, removing $D$ degrees of freedom:
\bea
0 &=& L_1|f,e;k\rangle\nonumber \\
&=&
2\left(\sqrt{2\alpha'}f_{\mu\nu}k^\nu+e_\mu\right)\alpha_{-1}^\mu|0;k\rangle,
\eea
implying
\be
e_\mu = \sqrt{2}f_{0\mu}.
\ee
The $L_2$ condition adds one more constraint:
\bea
0 &=& L_2|f,e;k\rangle\nonumber \\
&=& \left(2\sqrt{2\alpha'}k_\mu e^\mu+f_\mu^\mu\right)|0;k\rangle.
\eea
Using (12), this implies
\be
f_{ii} = 5f_{00},
\ee
where $f_{ii}$ is the trace on the spacelike part of $f$.

There are $D+1$ independent spurious states at this level:
\bea
|g,\gamma;k\rangle &=&
 \left(L_{-1}g_\mu\alpha^\mu_{-1}+L_{-2}\gamma\right)|0;k\rangle
 \\
&=&
 \left(\sqrt{2\alpha'}g_{(\mu}k_{\nu)}+\frac{\gamma}{2}\eta_{\mu\nu}\right)\alpha_{-1}^\mu\alpha_{-1}^\nu|0;k\rangle+\left(g_\mu+\sqrt{2\alpha'}\gamma
 k_\mu\right)\alpha^\mu_{-2}|0;k\rangle.\nonumber
\eea
These states are physical and therefore null for $g_0=\gamma=0$. Removing
these $D-1$ states from the spectrum leaves $D(D-1)/2$ states, the extra
one with respect to the light-cone quantization being the SO($D-1$) scalar,
\be
f_{ij} = f\delta_{ij},\qquad f_{00} = \frac{D-1}{5}f,\qquad e_0 =
\frac{\sqrt{2}(D-1)}{5}f,
\ee
with all other components zero. (States with vanishing $f_{00}$ must be
traceless by (14), and this is the unique state satisfying (12) and (14)
that is orthogonal to all of these.) The norm of this state is proportional
to
\be
f_{\mu\nu}^*f^{\mu\nu}+e_\mu^*e^\mu = \frac{(D-1)(26-D)f^2}{25},
\ee
positive for $D<26$ and negative for $D>26$. In the case $D=26$, this state
is spurious, corresponding to (15) with $\gamma=2f$,
$g_0=3\sqrt{2}f$. Removing it from the spectrum leaves us with the states
$f_{ij}$, $f_{ii}=0$, $e=0$---precisely the traceless symmetric rank 2
tensor of SO(25) we found in the light-cone quantization.

\setcounter{equation}{0}

\newpage
\section{Chapter 5}

\subsection{Problem 5.1}

\paragraph{(a)}
Our starting point is the following formal expression for the path
integral:
\be
Z(X_0,X_1) = \int_{\renewcommand{\arraystretch}{0.7}\begin{array}{c}\scriptstyle X(0)=X_0\\\scriptstyle X(1)=X_1\end{array}}\frac{[dXde]}{V_{\rm diff}}\exp\left(-S_{\rm
m}[X,e]\right),
\ee
where the action for the ``matter'' fields $X^\mu$ is
\be
S_{\rm m}[X,e] = \frac{1}{2}\int_0^1d\tau\,e\left(e^{-1}\partial X^\mu
e^{-1}\partial X_\mu+m^2\right)
\ee
(where $\partial\equiv d/d\tau).$ We have fixed the coordinate range for
$\tau$ to be [0,1]. Coordinate diffeomorphisms $\zeta:[0,1]\to[0,1]$, under
which the $X^\mu$ are scalars,
\be
X^{\mu\zeta}(\tau^\zeta) = X^\mu(\tau),
\ee
and the einbein $e$ is a ``co-vector,''
\be
e^\zeta(\tau^\zeta) = e(\tau)\frac{d\tau}{d\tau^\zeta},
\ee
leave the action (2) invariant. $V_{\rm diff}$ is the volume of this group
of diffeomorphisms. The $e$ integral in (1) runs over positive functions on
[0,1], and the integral
\be
l \equiv \int_0^1d\tau\,e
\ee
is diffeomorphism invariant and therefore a modulus; the moduli space is
$(0,\infty)$.

In order to make sense of the functional integrals in (1) we will need to
define an inner product on the space of functions on [0,1], which will
induce measures on the relevant function spaces. This inner product will
depend on the einbein $e$ in a way that is uniquely determined by the
following two constraints: (1) the inner product must be diffeomorphism
invariant; (2) it must depend on $e(\tau)$ only locally, in other words, it
must be of the form
\be
(f,g)_e = \int_0^1d\tau\,h(e(\tau))f(\tau)g(\tau),
\ee
for some function $h$.  As we will see, these conditions will be necessary
to allow us to regularize the infinite products that will arise in carrying
out the functional integrals in (1), and then to renormalize them by
introducing a counter-term action, in a way that respects the symmetries of
the action (2). For $f$ and $g$ scalars, the inner product satisfying these
two conditions is
\be
(f,g)_e \equiv \int_0^1d\tau\,efg.
\ee
We can express the matter action (2) using this inner product:
\be
S_{\rm m}[X,e] = \frac{1}{2}(e^{-1}\partial X^\mu,e^{-1}\partial
X_\mu)_e+\frac{lm^2}{2}.
\ee

We now wish to express the path integral (1) in a slightly less formal
way by choosing a fiducial einbein $e_l$ for each point $l$ in the
moduli space, and replacing the integral over einbeins by an integral over
the moduli space times a Faddeev-Popov determinant $\Delta_{\rm
FP}[e_l]$. Defining $\Delta_{\rm FP}$ by
\be
1 = \Delta_{\rm FP}[e]\int_0^\infty
dl\int[d\zeta]\,\delta[e-e_l^\zeta],
\ee
we indeed have, by the usual sequence of formal manipulations,
\be
Z(X_0,X_1) = \int_0^\infty
dl\int_{\renewcommand{\arraystretch}{0.7}\begin{array}{c}\scriptstyle
X(0)=X_0\\\scriptstyle X(1)=X_1\end{array}}[dX]\,\Delta_{\rm
FP}[e_l]\exp\left(-S_{\rm m}[X,e_l]\right).
\ee
To calculate the Faddeev-Popov determinant (9) at the point $e=e_l$, we
expand $e$ about $e_l$ for small diffeomorphisms $\zeta$ and small changes
in the modulus:
\be
e_l-e_{l+\delta l}^\zeta = \partial\gamma-\frac{de_l}{dl}\delta l,
\ee
where $\gamma$ is a scalar function parametrizing small diffeomorphisms:
$\tau^\zeta=\tau+e^{-1}\gamma$; to respect the fixed coordinate range,
$\gamma$ must vanish at 0 and 1.  Since the change (11) is, like $e$, a
co-vector, we will for simplicity multiply it by $e_l^{-1}$ in order to
have a scalar, and then bring into play our inner product (7) in order to
express the delta functional in (9) as an integral over scalar functions
$\beta$:
\be
\Delta_{\rm FP}^{-1}[e_l] = \int d\delta l[d\gamma d\beta]\,\exp\left(2\pi
i(\beta,e_l^{-1}\partial\gamma-e_l^{-1}\frac{de_l}{dl}\delta l)_{e_l}\right)
\ee
The integral is inverted by replacing the bosonic variables $\delta l$,
$\gamma$, and $\beta$ by Grassman variables $\xi$, $c$, and $b$:
\bea
\Delta_{\rm FP}[e_l] &=& \int
d\xi[dcdb]\,\exp\left(\frac{1}{4\pi}(b,e_l^{-1}\partial c-e_l^{-1}\frac{de_l}{dl}\xi)_{e_l}\right)\nonumber
\\
&=&
\int[dcdb]\,\frac{1}{4\pi}(b,e_l^{-1}\frac{de_l}{dl})_{e_l}\exp\left(\frac{1}{4\pi}(b,e_l^{-1}\partial
c)_{e_l}\right).
\eea
We can now write the path integral (10) in a more explicit form:
\bea
\lefteqn{Z(X_0,X_1) = \int_0^\infty
dl\int_{\renewcommand{\arraystretch}{0.7}\begin{array}{c}\scriptstyle
X(0)=X_0\\\scriptstyle
X(1)=X_1\end{array}}[dX]\int_{c(0)=c(1)=0}[dcdb]\,\frac{1}{4\pi}(b,e_l^{-1}\frac{de_l}{dl})_{e_l}}
\\
&& \qquad\qquad\qquad\qquad\qquad\qquad\qquad\qquad\qquad\quad\times\exp\left(-S_{\rm
g}[b,c,e_l]-S_{\rm m}[X,e_l]\right),\nonumber
\eea
where
\be
S_{\rm g}[b,c,e_l] = -\frac{1}{4\pi}(b,e_l^{-1}\partial c)_{e_l}.
\ee

\paragraph{(b)}
At this point it becomes convenient to work in a specific gauge, the
simplest being
\be
e_l(\tau) = l.
\ee
Then the inner product (7) becomes simply
\be
(f,g)_l = l\int_0^1d\tau\,fg.
\ee

In order to evaluate the Faddeev-Popov determinant (13), let us decompose
$b$ and $c$ into normalized eigenfunctions of the operator
\be
\Delta = -(e_l^{-1}\partial)^2 = -l^{-2}\partial^2:
\ee
\bea
b(\tau) &=& \frac{b_0}{\sqrt{l}}+\sqrt{\frac{2}{l}}\sum_{j=1}^\infty
b_j\cos(\pi j\tau), \\
c(\tau) &=& \sqrt{\frac{2}{l}}\sum_{j=1}^\infty c_j\sin(\pi j\tau),
\eea
with eigenvalues
\be
\nu_j = \frac{\pi^2j^2}{l^2}.
\ee
The ghost action (15) becomes
\be
S_{\rm g}(b_j,c_j,l) = -\frac{1}{4l}\sum_{j=1}^\infty jb_jc_j.
\ee
The zero mode $b_0$ does not enter into the action, but it is singled out
by the insertion appearing in front of the exponential in (13):
\be
\frac{1}{4\pi}(b,e_l^{-1}\frac{de_l}{dl})_{e_l} = \frac{b_0}{4\pi\sqrt{l}}.
\ee
The Faddeev-Popov determinant is, finally,
\bea
\Delta_{\rm FP}(l) &=& \int\prod_{j=0}^\infty db_j\prod_{j=1}^\infty
dc_j\,\frac{b_0}{4\pi\sqrt{l}}\exp\left(\frac{1}{4l}\sum_{j=1}^\infty
jb_jc_j\right)\nonumber \\
&=& \frac{1}{4\pi\sqrt{l}}\prod_{j=1}^\infty\frac{j}{4l}\nonumber \\
&=&
\frac{1}{4\pi\sqrt{l}}{\det}'\left(\frac{\Delta}{16\pi^2}\right)^{1/2},
\eea
the prime on the determinant denoting omission of the zero eigenvalue.

\paragraph{(c)}
Let us decompose $X^\mu(\tau)$ into a part which obeys the classical
equations of motion,
\be
X^\mu_{\rm cl}(\tau) = X_0+(X_1-X_0)\tau,
\ee
plus quantum fluctuations; the fluctuations vanish at 0 and 1, and can
therefore be decomposed into the same normalized eigenfunctions of $\Delta$
as $c$ was (20):
\be
X^\mu(\tau) = X^\mu_{\rm cl}(\tau)+\sqrt{\frac{2}{l}}\sum_{j=1}^\infty
x_j^\mu\sin(\pi j\tau).
\ee
The matter action (8) becomes
\be
S_{\rm m}(X_0,X_1,x_j) =
\frac{(X_1-X_0)^2}{2l}+\frac{\pi^2}{l^2}\sum_{j=1}^\infty
j^2x_j^2+\frac{lm^2}{2},
\ee
and the matter part of the path integral (10)
\bea
\lefteqn{\int_{\renewcommand{\arraystretch}{0.7}\begin{array}{c}\scriptstyle
X(0)=X_0\\\scriptstyle X(1)=X_1\end{array}}[dX]\,\exp\left(-S_{\rm
m}[X,e_l]\right)}\qquad\qquad\nonumber \\
&=&
\exp\left(-\frac{(X_1-X_0)^2}{2l}-\frac{lm^2}{2}\right)\int\prod_{\mu=1}^D\prod_{j=1}^\infty
dx^\mu_j\,\exp\left(-\frac{\pi^2}{l^2}\sum_{j=1}^\infty
j^2x_j^2\right)\nonumber \\
&=& \exp\left(-\frac{(X_1-X_0)^2}{2l}-\frac{lm^2}{2}\right){\det}'\left(\frac{\Delta}{\pi}\right)^{-D/2},
\eea
where we have conveniently chosen to work in a Euclidean spacetime in order
to make all of the Gaussian integrals convergent.

\paragraph{(d)}
Putting together the results (10), (24), and (28), and dropping the
irrelevant constant factors multiplying the operator $\Delta$ in the
infinite-dimensional determinants, we have:
\be
Z(X_0,X_1) = \int_0^\infty
dl\,\frac{1}{4\pi\sqrt{l}}\exp\left(-\frac{(X_1-X_0)^2}{2l}-\frac{lm^2}{2}\right)\left({\det}'\Delta\right)^{(1-D)/2}.
\ee
We will regularize the determinant of $\Delta$ in the same way as it is
done in Appendix A.1, by dividing by the determinant of the operator
$\Delta+\Omega^2$:
\bea
\frac{{\det}'\Delta}{{\det}'(\Delta+\Omega^2)} &=&
\prod_{j=1}^\infty\frac{\pi^2j^2}{\pi^2j^2+\Omega^2l^2}\nonumber \\
&=& \frac{\Omega l}{\sinh\Omega l}\nonumber \\
&\sim& 2\Omega l\exp\left(-\Omega l\right),
\eea
where the last line is the asymptotic expansion for large $\Omega$. The
path integral (29) becomes
\bea
\lefteqn{Z(X_0,X_1)}\quad \\
&=& \frac{1}{4\pi(2\Omega)^{(D-1)/2}}\int_0^\infty dl\,l^{-D/2}\exp\left(-\frac{(X_1-X_0)^2}{2l}-\frac{l(m^2-(D-1)\Omega)}{2}\right).\nonumber
\eea
The inverse divergence due to the factor of $\Omega^{(1-D)/2}$ in front of
the integral can be dealt with by a field renormalization, but since we
will not concern ourselves with the overall normalization of the path
integral we will simply drop all of the factors that appear in front. The
divergence coming from the $\Omega$ term in the exponent can be cancelled
by a (diffeomorphism invariant) counterterm in the action,
\be
S_{\rm ct} = \int_0^1d\tau\,eA = lA
\ee
The mass $m$ is renormalized by what is left over after the cancellation of
infinities,
\be
m^2_{\rm phys} = m^2-(D-1)\Omega-2A,
\ee
but for simplicity we will assume that a renormalization condition has been
chosen that sets $m_{\rm phys}=m$.

We can now proceed to the integration over moduli space:
\be
Z(X_0,X_1) = \int_0^\infty
dl\,l^{-D/2}\exp\left(-\frac{(X_1-X_0)^2}{2l}-\frac{lm^2}{2}\right).
\ee
The integral is most easily done after passing to momentum space:
\bea
\tilde{Z}(k) &\equiv& \int d^D\!X\,\exp\left(ik\cdot X\right)Z(0,X)\nonumber \\
&=& \int_0^\infty dl\,l^{-D/2}\exp\left(-\frac{lm^2}{2}\right)\int
d^D\!X\,\exp\left(ik\cdot X-\frac{X^2}{2l}\right)\nonumber \\
&=& \left(\frac{\pi}{2}\right)^{D/2}\int_0^\infty
dl\,\exp\left(-\frac{l(k^2+m^2)}{2}\right)\nonumber \\
&=& \left(\frac{\pi}{2}\right)^{D/2}\frac{2}{k^2+m^2};
\eea
neglecting the constant factors, this is precisely the momentum space
scalar propagator.

\renewcommand{\no}{{}^{{}_\star}_{{}^\star}}
\setcounter{equation}{0}
\newpage
\section{Chapter 6}

\subsection{Problem 6.1}

In terms of $u=1/z$, (6.2.31) is
\bea
\delta^d(\sum k_i)\prod_{i<j}\left|\frac{1}{u_i}-\frac{1}{u_j}\right|
^{\alpha'k_i\cdot k_j}
&=& \delta^d(\sum k_i)
\prod_{i<j}\left(|u_{ij}|^{\alpha'k_i\cdot k_j}
|u_iu_j|^{-\alpha'k_i\cdot k_j}\right) \nonumber \\
&=& \delta^d(\sum k_i)
\prod_{i<j}|u_{ij}|^{\alpha'k_i\cdot k_j}\prod_i|u_i|^{\alpha'k_i^2}.
\eea
Since this is an expectation value of closed-string tachyon vertex
operators, $\alpha'k_i^2=4$ and the expectation value is smooth at
$u_i=0$.

\subsection{Problem 6.3}

For any $n\geq2$ numbers $z_i$, we have
\be
\sum_{i=1}^n (-1)^i \prod_{j<k \atop j,k\neq i} z_{jk} = 
\left|
\begin{array}{cccccc}
1      & 1      & z_1    & z_1^2  & \cdots & z_1^{n-2} \\
1      & 1      & z_2    & z_2^2  & \cdots & z_2^{n-2} \\
\vdots & \vdots & \vdots & \vdots & \ddots & \vdots    \\
1      & 1      & z_n    & z_n^2  & \cdots & z_n^{n-2}
\end{array}
 \right|
= 0.
\end{equation}
The minor of the matrix with respect to the first entry in the $i$th row is
a Vandermonde matrix for the other $z_j$, so its determinant provides the
$i$th term in the sum. Specializing to the
case $n=5$, relabeling $z_5$ by $z_1'$, and dividing by
$z_{11'}z_{21'}z_{31'}z_{41'}$ yields
\be
-\frac{z_{23}z_{24}z_{34}}{z_{11'}}+\frac{z_{13}z_{14}z_{34}}{z_{21'}}-\frac{z_{12}z_{14}z_{24}}{z_{31'}}+\frac{z_{12}z_{13}z_{23}}{z_{41'}}
= \frac{z_{12}z_{13}z_{14}z_{23}z_{24}z_{34}}{z_{11'}z_{21'}z_{31'}z_{41'}},
\end{equation}
which is what we are required to prove.

\subsection{Problem 6.5}

\paragraph{(a)}
We have
\be
I(s,t) =
\frac{\Gamma(-1-\alpha's)\Gamma(-1-\alpha't)}{\Gamma(-2-\alpha's-\alpha't)},
\end{equation}
so the pole at $\alpha's=J-1$ arises from the first gamma function in the
numerator. The residue of $\Gamma(z)$ at $z$ a non-positive integer is
$(-1)^z/\Gamma(1-z)$, so the residue of $I(s,t)$ is
\be
\frac{(-1)^J\Gamma(-1-\alpha't)}{\Gamma(J+1)\Gamma(-1-J-\alpha't)} =
\frac{1}{J!}(2+\alpha't)(3+\alpha't)\cdots(J+1+\alpha't),
\end{equation}
a polynomial of degree $J$ in $t$. Using
\be
s+t+u = -\frac{4}{\alpha'},
\end{equation}
once $s$ is fixed at $(J-1)/\alpha'$, $t$ can be expressed in terms of
$t-u$:
\be
t = \frac{t-u}{2}-\frac{J+3}{2\alpha'},
\end{equation}
so (5) is also a polynomial of degree $J$ in $t-u$.

\paragraph{(b)}
The momentum of the intermediate state in the $s$ channel is
$k_1+k_2=-(k_3+k_4)$, so in its rest frame we have
\be
k_1^i = -k_2^i,\qquad k_3^i=-k_4^i,\qquad k_1^0 = k_2^0 = -k_3^0 = -k_4^0 =
\frac{\sqrt{s}}{2}.
\end{equation}
Specializing to the case where all the external particles are tachyons
($k^2=1/\alpha'$) and the intermediate state is at level 2 ($s=1/\alpha'$),
we further have
\be
k_1^ik_1^i = k_2^ik_2^i = k_3^ik_3^i = k_4^ik_4^i = \frac{5}{4\alpha'}.
\end{equation}
It also determines $t$ in terms of $k_1^ik_3^i$:
\bea
t &=& -(k_1+k_3)^2\nonumber \\
&=& -\frac{5}{2\alpha'}-2k_1^ik_3^i.
\eea
Using (5), the residue of the pole in $I(s,t)$ at $\alpha's=1$ is
\be
\frac{1}{2}(2+\alpha't)(3+\alpha't) = -\frac{1}{8}+2\alpha'^2(k_1^ik_3^i)^2.
\end{equation}

The operator that projects matrices onto multiples of the identity matrix
in $D-1$ dimensional space is
\be
P^0_{ij,kl} = \delta_{ij}\frac{1}{D-1}\delta_{kl},
\end{equation}
while the one that projects them onto traceless symmetric matrices is
\be
P^2_{ij,kl} =
\frac{1}{2}\left(\delta_{ik}\delta_{jl}+\delta_{il}\delta_{jk}\right)-P^0_{ij,kl}.
\end{equation}
Inserting the linear combination $\beta_0P^0+\beta_2P^2$ between the matrices
$k_1^ik_1^j$ and $k_3^kk_3^l$ yields
\be
(\beta_0-\beta_2)\frac{k_1^ik_1^ik_3^jk_3^j}{D-1}+\beta_2(k_1^ik_3^i)^2 =
(\beta_0-\beta_2)\frac{25}{16\alpha'^2(D-1)}+\beta_2(k_1^ik_3^i)^2.
\end{equation}
Comparison with (11) reveals
\be
\beta_0 = \frac{2\alpha'^2(26-D)}{25},\qquad\beta_2 = 2\alpha',
\end{equation}
so that, as promised, $\beta_0$ is positive, zero, or negative depending on
whether $D$ is less than, equal to, or greater than 26.

What does all this have to do with the open string spectrum at level 2? The
amplitude $I$ has a pole in $s$ wherever $s$ equals the mass-squared of an
open string state, allowing the intermediate state in the $s$ channel to go
on shell. The residue of this pole can be written schematically as
\be
\langle f|S\left(\sum_o\frac{|o\rangle\langle o|}{\langle
o|o\rangle}\right)S|i\rangle,
\end{equation}
where the sum is taken over open string states at level
$J=\alpha's+1$ with momentum equal to that of the initial and final states;
we have not assumed that the intermediate states are normalized, to allow
for the possibility that some of them might have negative norm. More
specifically, 
\be
|i\rangle = |0;k_1\rangle|0;k_2\rangle,\qquad\langle f| =
\langle0;-k_3|\langle0;-k_4|.
\end{equation}
The open string spectrum at level 2 was worked out as a function
of $D$ in problem 4.1. For any $D$ it includes $D(D-1)/2-1$ positive-norm
states transforming in the spin 2 representation of the little group
SO($D-1$). Working in the rest frame of such a state, the
$S$-matrix elements involved in (16) are fixed by SO($D-1$) invariance and
(anti-)linearity in the polarization matrix $a$:
\be
\langle a|S|0;k_1\rangle|0;k_2\rangle \propto a^*_{ij}k_1^i
k_2^j,\qquad\langle 0;-k_3|\langle 0;-k_4|S|a\rangle \propto a_{kl}k_3^k
k_4^l.
\end{equation}
Summing over an orthonormal basis in the space of symmetric traceless
matrices yields the contribution of these states to (16):
\be
\sum_aa^*_{ij}k_1^ik_2^ja_{kl}k_3^kk_4^l = k_1^ik_1^jP^2_{ij,kl}k_3^kk_3^l.
\end{equation}
This explains the positive value of $\beta_2$ found in (15).

For $D\neq26$, there is another state $|b\rangle$ in the spectrum whose
norm is positive for $D<26$ and negative for $D>26$. Since this state is an
SO($D-1$) scalar, the $S$-matrix elements connecting it to the initial and
final states are given by:
\be
\langle b|S|0;k_1\rangle|0;k_2\rangle \propto \delta_{ij}k_1^ik_2^j,\qquad
\langle 0;-k_3|\langle 0;-k_4|S|b\rangle \propto \delta_{kl}k_3^kk_4^l.
\end{equation}
Its contribution to (16) is therefore clearly a positive multiple of
\be
k_1^ik_1^jP^0_{ij,kl}k_3^kk_3^l
\end{equation}
if $D<26$, and a negative multiple if $D>26$.

\subsection{Problem 6.7}

\paragraph{(a)}
The $X$ path integral (6.5.11) follows immediately from (6.2.36), where 
\be
v^\mu(y_1) =
-2i\alpha'\left(\frac{k_2^\mu}{y_{12}}-\frac{k_3^\mu}{y_{13}}\right)
\end{equation}
(we leave out the contraction between $\dot{X}^\mu(y_1)$ and $e^{ik_1\cdot
X(y_1)}$ because their product is already renormalized in the path
integral). Momentum conservation, $k_1+k_2+k_3=0$, and the mass shell
conditions imply
\bea
0 = k_1^2 = (k_2+k_3)^2 = \frac{2}{\alpha'}+2k_2\cdot k_3, \\
\frac{1}{\alpha'} = k_3^2 = (k_1+k_2)^2 = \frac{1}{\alpha'}+2k_1\cdot k_2, \\
\frac{1}{\alpha'} = k_2^2 = (k_1+k_3)^2 = \frac{1}{\alpha'}+2k_1\cdot k_3,
\eea
so that $2\alpha'k_2\cdot k_3=-2$, while $2\alpha'k_1\cdot
k_2=2\alpha'k_1\cdot k_3=0$. Equation (6.5.11) therefore simplifies to
\bea
\lefteqn{\left\langle\no\dot{X}^\mu e^{ik_1\cdot X}(y_1)\no\no e^{ik_2\cdot
X}(y_2)\no\no e^{ik_3\cdot
X}(y_3)\no\right\rangle_{D_2}}\qquad\qquad\qquad\qquad\qquad\quad \\
&& =
2\alpha'C^X_{D_2}(2\pi)^{26}\delta^{26}(k_1+k_2+k_3)\frac{1}{y_{23}^2}\left(\frac{k_2^\mu}{y_{12}}+\frac{k_3^\mu}{y_{13}}\right).\nonumber
\eea
The ghost path integral is given by (6.3.2):
\be
\left\langle c(y_1)c(y_2)c(y_3)\right\rangle_{D_2} = C^{\rm
g}_{D_2}y_{12}y_{13}y_{23}.
\end{equation}
Putting these together with (6.5.10) and using (6.4.14),
\be
\alpha'g_{\rm o}^2e^{-\lambda}C^X_{D_2}C^{\rm g}_{D_2} = 1,
\end{equation}
yields
\bea
\lefteqn{S_{D_2}(k_1,a_1,e_1;k_2,a_2;k_3,a_3)} \\
&& = -2ig_{\rm
o}'(2\pi)^{26}\delta^{26}(k_1+k_2+k_3)\frac{y_{13}e\cdot
k_2+y_{12}e\cdot k_3}{y_{23}}{\rm
Tr}(\lambda^{a_1}\lambda^{a_2}\lambda^{a_3})+(2\leftrightarrow3).\nonumber 
\eea
Momentum conservation and the physical state condition $e_1\cdot k_1=0$
imply
\be
e_1\cdot k_2 = -e_1\cdot k_3 = \frac{1}{2}e_1\cdot k_{23},
\end{equation}
so
\bea
\lefteqn{S_{D_2}(k_1,a_1,e_1;k_2,a_2;k_3,a_3)}\qquad\qquad\qquad\qquad\quad \\
&& = -ig_{\rm 0}'(2\pi)^{26}\delta^{26}(k_1+k_2+k_3)e_1\cdot k_{23}{\rm
Tr}(\lambda^{a_1}[\lambda^{a_2},\lambda^{a_3}]),\nonumber
\eea
in agreement with (6.5.12).

\paragraph{(b)}
Using equations (6.4.17), (6.4.20), and (6.5.6), we see that the
four-tachyon amplitude near $s=0$ is given by
\bea
\lefteqn{S_{D_2}(k_1,a_1;k_2,a_2;k_3,a_3;k_4,a_4)}\nonumber \\
&&
= \frac{ig_{\rm o}^2}{\alpha'}(2\pi)^{26}\delta^{26}(\sum_ik_i)\nonumber \\
&& \quad\times{\rm
Tr}(\lambda^{a_1}\lambda^{a_2}\lambda^{a_4}\lambda^{a_3}+\lambda^{a_1}\lambda^{a_3}\lambda^{a_4}\lambda^{a_2}-\lambda^{a_1}\lambda^{a_2}\lambda^{a_3}\lambda^{a_4}-\lambda^{a_1}\lambda^{a_4}\lambda^{a_3}\lambda^{a_2})\frac{u-t}{2s}\nonumber
\\
&& = -\frac{ig_{\rm o}^2}{2\alpha'}(2\pi)^{26}\delta^{26}(\sum_ik_i){\rm
Tr}([\lambda^{a_1},\lambda^{a_2}][\lambda^{a_3},\lambda^{a_4}])\frac{u-t}{s}.
\eea

We can calculate the same quantity using unitarity. By analogy with
equation (6.4.13), the 4-tachyon amplitude near the pole at $s=0$ has the
form 
\bea
\lefteqn{S_{D_2}(k_1,a_1;k_2,a_2;k_3,a_3;k_4,a_4)}\nonumber \\
&&
=
i\int\frac{d^{26}\!k}{(2\pi)^{26}}\,\sum_{a,e}\frac{S_{D_2}(-k,a,e;k_1,a_1;k_2,a_2)S_{D_2}(k,a,e;k_3,a_3;k_4,a_4)}{-k^2+i\epsilon}\nonumber \\
&& =
i\int\frac{d^{26}\!k}{(2\pi)^{26}}\,\sum_{a,e}\frac{1}{-k^2+i\epsilon}(-i)g_{\rm 0}'(2\pi)^{26}\delta^{26}(k_1+k_2-k)e\cdot k_{12}{\rm
Tr}(\lambda^a[\lambda^{a_1},\lambda^{a_2}])\nonumber \\
&& \qquad\qquad\qquad\qquad\qquad\times(-i)g_{\rm 0}'(2\pi)^{26}\delta^{26}(k+k_3+k_4)e\cdot k_{34}{\rm
Tr}(\lambda^a[\lambda^{a_3},\lambda^{a_4}])\nonumber \\
&& = -ig_{\rm o}'^2(2\pi)^{26}\delta^{26}(\sum_ik_i)\sum_a{\rm
Tr}(\lambda^a[\lambda^{a_1},\lambda^{a_2}]){\rm
Tr}(\lambda^a[\lambda^{a_3},\lambda^{a_4}])\frac{\sum_ee\cdot k_{12}e\cdot
k_{34}}{s+i\epsilon}\nonumber\\
&& = -ig_{\rm o}'^2(2\pi)^{26}\delta^{26}(\sum_ik_i){\rm Tr}([\lambda^{a_1},\lambda^{a_2}][\lambda^{a_3},\lambda^{a_4}])\frac{u-t}{s+i\epsilon}.
\eea
In the second equality we have substituted equation (31) (or (6.5.12)).
The polarization vector $e$ is summed over an orthonormal basis of
(spacelike) vectors obeying the physical state condition $e\cdot k=0$,
which after the integration over $k$ in the third equality becomes
$e\cdot(k_1+k_2)=e\cdot(k_3+k_4)=0$. If we choose one of the vectors in
this basis to be $e'=k_{12}/|k_{12}|$, then none of the others will
contribute to the sum in the second to last line, which becomes,
\be
\sum_ee\cdot k_{12}e\cdot k_{34} = k_{12}\cdot k_{34} = u-t.
\end{equation}
In the last equality of (33) we have also applied equation
(6.5.9). Comparing (32) and (33), we see that
\be
g_{\rm o}' = \frac{g_{\rm o}}{\sqrt{2\alpha'}},
\end{equation}
in agreement with (6.5.14).

This result confirms the normalization of the photon vertex operator as
written in equation (3.6.26). The state-operator mapping gives the same
normalization: in problem 2.9, we saw that the photon vertex
operator was
\be
e_\mu\alpha^\mu_{-1}|0;0\rangle \cong i\sqrt{\frac{2}{\alpha'}}e_\mu\no\partial
X^\mu e^{ik\cdot X}\no = i\sqrt{\frac{2}{\alpha'}}e_\mu\no\bar\partial X^\mu
e^{ik\cdot X}\no.
\end{equation}
Since the boundary is along the $\sigma^1$-axis, the derivative can be
written using (2.1.3):
\be
\dot{X} = \partial_1X = (\partial+\bar\partial)X = 2\partial X.
\end{equation}
Hence the vertex operator is
\be
\frac{i}{\sqrt{2\alpha'}}e_\mu\no\dot{X}^\mu e^{ik\cdot X}\no,
\end{equation}
which, after multiplying by the factor $-g_{\rm o}$ and integrating over
the position on the boundary, agrees with (3.6.26).

\subsection{Problem 6.9}

\paragraph{(a)}
There are six cyclic orderings of the four vertex operators on the boundary
of the disk, illustrated in figure 6.2. Consider first the ordering
$(3,4,1,2)$ shown in figure 6.2(a). If we fix the positions of the vertex
operators for gauge bosons 1, 2, and 3, with
\be
-\infty<y_3<y_1<y_2<\infty,
\end{equation}
then we must integrate the position of the fourth gauge boson vertex
operator from $y_3$ to $y_1$. The contribution this ordering makes to the
amplitude is
\bea
\lefteqn{e^{-\lambda}g_{\rm o}^4(2\alpha')^{-2}
{\rm Tr}(\lambda^{a_3}\lambda^{a_4}\lambda^{a_1}\lambda^{a_2})
e^1_{\mu_1}e^2_{\mu_2}e^3_{\mu_3}e^4_{\mu_4}} \nonumber \\
&& \qquad\times\int_{y_3}^{y_1}dy_4\, \left\langle
\no c^1\dot{X}^{\mu_3}e^{ik_3\cdot X}(y_3) \no
\no \dot{X}^{\mu_4}e^{ik_4\cdot X}(y_4) \no \right.\nonumber \\
&& \qquad\qquad\qquad\qquad \times
\left.\no c^1\dot{X}^{\mu_1}e^{ik_1\cdot X}(y_1) \no
\no c^1\dot{X}^{\mu_2}e^{ik_2\cdot X}(y_2) \no
\right\rangle \\
&=& e^{-\lambda}g_{\rm o}^4(2\alpha')^{-2}
{\rm Tr}(\lambda^{a_3}\lambda^{a_4}\lambda^{a_1}\lambda^{a_2})
e^1_{\mu_1}e^2_{\mu_2}e^3_{\mu_3}e^4_{\mu_4}
iC^X_{D_2}C^{\rm
g}_{D_2}(2\pi)^{26}\delta^{26}(\sum_ik_i) \nonumber \\
&& \qquad \times|y_{12}|^{2\alpha'k_1\cdot k_2+1}
|y_{13}|^{2\alpha'k_1\cdot k_3+1}
|y_{23}|^{2\alpha'k_2\cdot k_3+1} \nonumber \\
&& \qquad \times\int_{y_3}^{y_1}dy_4\,
|y_{14}|^{2\alpha'k_1\cdot k_4}
|y_{24}|^{2\alpha'k_2\cdot k_4}
|y_{34}|^{2\alpha'k_3\cdot k_4} \nonumber \\
&& \qquad\qquad\qquad \times \left\langle
[v^{\mu_3}(y_3)+q^{\mu_3}(y_3)]
[v^{\mu_4}(y_4)+q^{\mu_4}(y_4)] \right.\nonumber \\
&& \qquad\qquad\qquad\qquad \times \left.
[v^{\mu_1}(y_1)+q^{\mu_1}(y_1)]
[v^{\mu_2}(y_2)+q^{\mu_2}(y_2)]
\right\rangle.
\nonumber
\eea
The $v^\mu$ that appear in the path integral in the last two lines are
linear in the momenta; for instance
\be
v^\mu(y_3) 
= -2i\alpha'(k_1^\mu y_{31}^{-1}+k_2^\mu y_{32}^{-1}+k_4^\mu y_{34}^{-1}).
\end{equation}
They therefore contribute terms in which the polarization vectors $e_i$ are
dotted with the momenta. Since we are looking only for terms in which the
$e_i$ appear in the particular combination $e_1\cdot e_2\,e_3\cdot e_4$, we
can neglect the $v^\mu$. The terms we are looking for arise from the
contraction of the $q^\mu$ with each other. Specifically, the singular part
of the OPE of $q^\mu(y)$ with $q^\nu(y')$ is
\be
-2\alpha'(y-y')^{-2}\eta^{\mu\nu},
\end{equation}
so the combination $e_1\cdot e_2\,e_3\cdot e_4$ arises from the
contractions of $q^{\mu_1}(y_1)$ with $q^{\mu_2}(y_2)$ and
$q^{\mu_3}(y_3)$ with $q^{\mu_4}(y_4)$:
\bea
\lefteqn{ig_{\rm o}^2\alpha'^{-1}
{\rm Tr}(\lambda^{a_3}\lambda^{a_4}\lambda^{a_1}\lambda^{a_2})
e_1\cdot e_2\,e_3\cdot e_4
(2\pi)^{26}\delta^{26}(\sum_ik_i)} \nonumber \\
&& \qquad\qquad\times |y_{12}|^{2\alpha'k_1\cdot k_2-1}
|y_{13}|^{2\alpha'k_1\cdot k_3+1}
|y_{23}|^{2\alpha'k_2\cdot k_3+1} \nonumber \\
&& \qquad\qquad \times \int_{y_3}^{y_1}dy_4\,
|y_{14}|^{2\alpha'k_1\cdot k_4}
|y_{24}|^{2\alpha'k_2\cdot k_4}
|y_{34}|^{2\alpha'k_3\cdot k_4-2}.
\eea
We can choose $y_1$, $y_2$, and $y_3$ as we like, so long as we obey (39),
and the above expression simplifies if we take the limit $y_2\to\infty$
while keeping $y_1$ and $y_3$ fixed. Then $|y_{12}|\sim|y_{23}|\sim y_2$
and (since $y_3<y_4<y_1$) $|y_{24}|\sim y_2$ as well. Making these
substitutions above, $y_2$ appears with a total power of
\bea
2\alpha'k_1\cdot k_2-1+2\alpha'k_2\cdot k_3+1+2\alpha'k_2\cdot k_4 &=&
2\alpha'(k_1+k_3+k_4)\cdot k_2 \nonumber \\
&=& -2\alpha'k_2^2 \nonumber \\
&=& 0.
\eea
We can simplify further by setting $y_3=0$ and $y_1=1$. Since $s=-2k_3\cdot
k_4$ and $u=-2k_1\cdot k_4$, the integral above reduces to:
\be
\int_0^1dy_4\,(1-y_4)^{-\alpha'u}y_4^{-\alpha's-2} =
B(-\alpha'u+1,-\alpha's-1).
\end{equation}

If we now consider a different cyclic ordering of the vertex operators, we
can still fix $y_1$, $y_2$, and $y_3$ while integrating over $y_4$.
Equation (43) will remain the same, with two exceptions: the order of the
$\lambda^a$ matrices appearing in the trace, and the limits of integration
on $y_4$, will change to reflect the new order. The limits of integration
will be whatever positions immediately precede and succede $y_4$, while the
position that is opposite $y_4$ will be taken to infinity. It can
easily be seen that the trick that allowed us to take $y_2$ to infinity
(equation (44)) works equally well for $y_1$ or $y_3$. The lower and upper
limits of integration can be fixed at 0 and 1 respectively as before, and the
resulting integral over $y_4$ will once again give a beta
function. However, since different factors in the integrand of (43) survive
for different orderings, the beta function will have different arguments in
each case. Putting together the results from the six cyclic orderings,
we find that the part of the four gauge boson amplitude proportional to
$e_1\cdot e_2\,e_3\cdot e_4$ is
\bea
\lefteqn{\frac{ig_{\rm o}^2}{\alpha'}e_1\cdot e_2\,e_3\cdot e_4
(2\pi)^{26}\delta^{26}(\sum_ik_i)} \\
&& \qquad\times\left[{\rm Tr}
(\lambda^{a_1}\lambda^{a_2}\lambda^{a_4}\lambda^{a_3}
+\lambda^{a_1}\lambda^{a_3}\lambda^{a_4}\lambda^{a_2})
B(-\alpha't+1,-\alpha's-1) \right. \nonumber \\
&& \qquad\quad +{\rm Tr}
(\lambda^{a_1}\lambda^{a_3}\lambda^{a_2}\lambda^{a_4}
+\lambda^{a_1}\lambda^{a_4}\lambda^{a_2}\lambda^{a_3})
B(-\alpha't+1,-\alpha'u+1) \nonumber \\
&& \qquad\quad \left.+{\rm Tr}
(\lambda^{a_1}\lambda^{a_2}\lambda^{a_3}\lambda^{a_4}
+\lambda^{a_1}\lambda^{a_4}\lambda^{a_3}\lambda^{a_2})
B(-\alpha'u+1,-\alpha's-1) \right]. \nonumber
\eea

\paragraph{(b)}
There are four tree-level diagrams that contribute to four-boson scattering
in Yang-Mills theory: the $s$-channel, the $t$-channel, the $u$-channel,
and the four-point vertex. The four-point vertex diagram (which is
independent of momenta) includes the following term proportional to
$e_1\cdot e_2\,e_3\cdot e_4$:
\be
-ig_{\rm o}'^2 e_1\cdot e_2\,e_3\cdot e_4
(2\pi)^{26}\delta^{26}(\sum_ik_i)
\sum_e\left(f^{a_1a_3e}f^{a_2a_4e}+f^{a_1a_4e}f^{a_2a_3e}\right)
\end{equation}
(see Peskin and Schroeder, equation A.12). The Yang-Mills coupling is
\be
g_{\rm o}' = (2\alpha')^{-1/2}g_{\rm o}
\end{equation}
(6.5.14), and the $f^{abc}$ are the gauge group structure constants:
\be
f^{abc} = {\rm Tr}\left([\lambda^a,\lambda^b]\lambda^c\right).
\end{equation}
We can therefore re-write (47) in the following form:
\bea
\lefteqn{-\frac{ig_{\rm o}^2}{\alpha'}e_1\cdot e_2\,e_3\cdot e_4
(2\pi)^{26}\delta^{26}(\sum_ik_i)
{\rm Tr}\bigg(
\lambda^{a_1}\lambda^{a_3}\lambda^{a_2}\lambda^{a_4}
+ \lambda^{a_1}\lambda^{a_4}\lambda^{a_2}\lambda^{a_3} }
\\
&& \qquad\quad\left.
-\frac{1}{2}(
\lambda^{a_1}\lambda^{a_2}\lambda^{a_4}\lambda^{a_3}
+\lambda^{a_1}\lambda^{a_3}\lambda^{a_4}\lambda^{a_2}
+\lambda^{a_1}\lambda^{a_2}\lambda^{a_3}\lambda^{a_4}
+\lambda^{a_1}\lambda^{a_4}\lambda^{a_3}\lambda^{a_2}
)
\right).
\nonumber
\eea
Of the three diagrams that contain three-point vertices, only the
$s$-channel diagram contains a term proportional to 
$e_1\cdot e_2\,e_3\cdot e_4$. It is
\bea
\lefteqn{-ig_{\rm o}'^2e_1\cdot e_2\,e_3\cdot e_4
(2\pi)^{26}\delta^{26}(\sum_ik_i)
\frac{u-t}{s}
\sum_ef^{a_1a_2e}f^{a_3a_4e}} \nonumber \\
&& \quad = -\frac{ig_{\rm o}^2}{\alpha'}e_1\cdot e_2\,e_3\cdot e_4
(2\pi)^{26}\delta^{26}(\sum_ik_i)
\frac{t-u}{2s} \\
&& \qquad\quad\times {\rm Tr}\left(
\lambda^{a_1}\lambda^{a_2}\lambda^{a_3}\lambda^{a_4}
+\lambda^{a_1}\lambda^{a_4}\lambda^{a_3}\lambda^{a_2}
-\lambda^{a_1}\lambda^{a_2}\lambda^{a_4}\lambda^{a_3}
-\lambda^{a_1}\lambda^{a_3}\lambda^{a_4}\lambda^{a_2}
\right). \nonumber
\eea
Combining (50) and (51) and using
\be
s+t+u=0,
\end{equation}
we obtain, for the part of the four-boson amplitude proportional to
$e_1\cdot e_2\,e_3\cdot e_4$, at tree level,
\bea
\lefteqn{-\frac{ig_{\rm o}^2}{\alpha'}e_1\cdot e_2\,e_3\cdot e_4
(2\pi)^{26}\delta^{26}(\sum_ik_i)} \\
&& \qquad\times\left[{\rm Tr}
(\lambda^{a_1}\lambda^{a_2}\lambda^{a_4}\lambda^{a_3}
+\lambda^{a_1}\lambda^{a_3}\lambda^{a_4}\lambda^{a_2})
\left(-1-\frac{t}{s}\right)
\right. \nonumber \\
&& \qquad\quad +{\rm Tr}
(\lambda^{a_1}\lambda^{a_3}\lambda^{a_2}\lambda^{a_4}
+\lambda^{a_1}\lambda^{a_4}\lambda^{a_2}\lambda^{a_3})
\nonumber \\
&& \qquad\quad +{\rm Tr}
(\lambda^{a_1}\lambda^{a_2}\lambda^{a_3}\lambda^{a_4}
+\lambda^{a_1}\lambda^{a_4}\lambda^{a_3}\lambda^{a_2})
\left(-1-\frac{u}{s}\right)
\bigg]. \nonumber
\eea
This is intentionally written in a form suggestively similar to equation
(46). It is clear that (46) reduces to (53) (up to an overall sign) if we
take the limit $\alpha'\to0$ with $s$, $t$, and $u$ fixed, since in that
limit
\bea
B(-\alpha't+1,-\alpha's-1) &\approx& -1-\frac{t}{s}, \nonumber \\
B(-\alpha't+1,-\alpha'u+1) &\approx& 1, \\
B(-\alpha'u+1,-\alpha's-1) &\approx& -1-\frac{u}{s}. \nonumber
\eea
Thus this single string theory diagram reproduces, at momenta small
compared to the string scale, the sum of the field theory Feynman
diagrams.

\subsection{Problem 6.11}

\paragraph{(a)}

The $X$ path integral is given by (6.2.19):
\bea
\lefteqn{\left\langle
:\partial X^\mu\bar\partial X^\nu e^{ik_1\cdot X}(z_1,\bar z_1):
:e^{ik_2\cdot X}(z_2,\bar z_2):
:e^{ik_3\cdot X}(z_3,\bar z_3):
\right\rangle_{S_2}}\qquad\qquad\nonumber \\
&& = -iC^X_{S_2}\frac{\alpha'^2}{4}(2\pi)^{26}\delta^{26}(k_1+k_2+k_3) \\
&& \qquad\times|z_{12}|^{\alpha'k_1\cdot k_2}
|z_{13}|^{\alpha'k_1\cdot k_3}
|z_{23}|^{\alpha'k_2\cdot k_3}
\left(\frac{k_2^\mu}{z_{12}}+\frac{k_3^\mu}{z_{13}}\right)
\left(\frac{k_2^\nu}{\bar z_{12}}+\frac{k_3^\nu}{\bar z_{13}}\right). \nonumber
\eea
The ghost path integral is given by (6.3.4):
\be
\left\langle
:c(z_1)\tilde c(\bar z_1)::c(z_2)\tilde c(\bar z_2)::c(z_3)\tilde c(\bar z_3):
\right\rangle_{S_2}
= C^{\rm g}_{S_2}|z_{12}|^2|z_{13}|^2|z_{23}|^2.
\end{equation}
The momentum-conserving delta function and the mass shell conditions
$k_1^2=0$, $k_2^2=k_3^2=4/\alpha'$ imply
\be
k_1\cdot k_2=k_1\cdot k_3=0,\qquad k_2\cdot k_3 = -4/\alpha'.
\end{equation}
Using (57), the transversality of the polarization tensor,
\be
e_{1\mu\nu}k_1^\mu = 0,
\end{equation}
and the result (6.6.8),
\be
C_{S_2} \equiv e^{-2\lambda}C^X_{S_2}C^{\rm g}_{S_2} =
\frac{8\pi}{\alpha'g_{\rm c}^2},
\end{equation}
we can put together the full amplitude for two closed-string tachyons and
one massless closed string on the sphere:
\bea
\lefteqn{S_{S_2}(k_1,e_1;k_2;k_3)} \nonumber \\
&& = g_{\rm c}^2g_{\rm c}'e^{-2\lambda}e_{1\mu\nu} \nonumber \\
&& \quad\qquad\times\left\langle
:\tilde cc\partial X^\mu\bar\partial X^\nu e^{ik_1\cdot X}(z_1,\bar z_1):
:\tilde cce^{ik_2\cdot X}(z_2,\bar z_3):
:\tilde cce^{ik_2\cdot X}(z_2,\bar z_3):
\right\rangle_{S_2} \nonumber  \\
&& = -iC_{S_2}g_{\rm c}^2g_{\rm c}'\frac{\alpha'^2}{4}
(2\pi)^{26}\delta^{26}(k_1+k_2+k_3) \nonumber \\
&& \qquad\qquad\qquad\qquad\qquad\qquad\times
e_{1\mu\nu}\frac{|z_{12}|^2|z_{13}|^2}{|z_{23}|^2}
\left(\frac{k_2^\mu}{z_{12}}+\frac{k_3^\mu}{z_{13}}\right)
\left(\frac{k_2^\nu}{\bar z_{12}}+\frac{k_3^\nu}{\bar z_{13}}\right)
\nonumber \\
&& = -i2\pi\alpha'g_{\rm c}'(2\pi)^{26}\delta^{26}(k_1+k_2+k_3) \nonumber \\
&& \qquad\qquad\qquad\qquad\qquad\qquad\times
e_{1\mu\nu}\frac{|z_{12}|^2|z_{13}|^2}{|z_{23}|^2}
\left(\frac{k_{23}^\mu}{2z_{12}}-\frac{k_{23}^\mu}{2z_{13}}\right)
\left(\frac{k_{23}^\nu}{2\bar z_{12}}-\frac{k_{23}^\nu}{2\bar
z_{13}}\right) \nonumber \\
&& = -i\frac{\pi\alpha'}{2}g_{\rm c}'
(2\pi)^{26}\delta^{26}(k_1+k_2+k_3)e_{1\mu\nu}k_{23}^\mu k_{23}^\nu.
\eea

\paragraph{(b)}

Let us calculate the amplitude for massless closed string exchange between
closed string tachyons (this is a tree-level field theory calculation but
for the vertices we will use the amplitude calculated above in string
theory). We will restrict ourselves to the $s$-channel diagram, because we
are interested in comparing the result with the pole at $s=0$ in the
Virasoro-Shapiro amplitude. Here the propagator for the massless
intermediate string provides the pole at $s=0$:
\be
-i(2\pi)^{26}\delta^{26}(\sum k_i)
\frac{\pi^2\alpha'^2g_{\rm c}'^2}{4s}
\sum_ee_{\mu\nu}k_{12}^\mu k_{12}^\nu e_{\mu'\nu'}k_{34}^{\mu'}k_{34}^{\nu'}.
\end{equation}
Here $e$ is summed over an orthonormal basis of symmetric polarization
tensors obeying the condition $e_{\mu\nu}(k_1^\mu+k_2^\mu)=0$. We could
choose as one element of that basis the tensor
\be
e_{\mu\nu} = \frac{k_{12\mu}k_{12\nu}}{k_{12}^2},
\end{equation}
which obeys the transversality condition by virtue of the fact that
$k_1^2=k_2^2$. With this choice, none of the other elements of the basis
would contribute to the sum, which reduces to
\be
\left(k_{12}\cdot k_{34}\right)^2 = (u-t)^2.
\end{equation}
The amplitude (61) is thus
\be
-i(2\pi)^{26}\delta^{26}(\sum k_i)\frac{\pi^2\alpha'^2g_{\rm c}'^2}{4}
\frac{(u-t)^2}{s}.
\end{equation}

Now, the Virasoro-Shapiro amplitude is
\be
i(2\pi)^{26}\delta^{26}(\sum k_i)\frac{16\pi^2g_{\rm c}^2}{\alpha'}
\frac{\Gamma(a)\Gamma(b)\Gamma(c)}{\Gamma(a+b)\Gamma(a+c)\Gamma(b+c)},
\end{equation}
where
\bea
a = -1-\frac{\alpha's}{4}, \qquad b &=& -1-\frac{\alpha't}{4}, 
\qquad c = -1-\frac{\alpha'u}{4}.
\eea
The pole at $s=0$ arises from the factor of $\Gamma(a)$, which is, to
lowest order in $s$,
\be
\Gamma(a) \approx \frac{4}{\alpha's}.
\end{equation}
To lowest order in $s$ the other gamma functions simplify to
\bea
\frac{\Gamma(b)\Gamma(c)}{\Gamma(a+b)\Gamma(a+c)\Gamma(b+c)}
&\approx& \frac{\Gamma(b)\Gamma(c)}{\Gamma(b-1)\Gamma(c-1)\Gamma(2)}
\nonumber \\
&=& (b-1)(c-1) \nonumber \\
&=& -\frac{\alpha'^2}{64}(u-t)^2.
\eea
Thus the part of the amplitude we are interested in is
\be
-i(2\pi)^{26}\delta^{26}(\sum k_i)\pi^2g_{\rm c}^2
\frac{(u-t)^2}{s}.
\end{equation}
Comparison with (64) shows that
\be
g_{\rm c} = \frac{\alpha'g_{\rm c}'}{2}.
\end{equation}

\paragraph{(c)}

In Einstein frame the tachyon kinetic term decouples from the dilaton, as
the tachyon action (6.6.16) becomes
\be
S_T = -\frac{1}{2}\int d^{26}\!x\,(-\tilde G)^{1/2}
\left(\tilde G^{\mu\nu}\partial_\mu T\partial_\nu T
-\frac{4}{\alpha'}e^{\tilde\Phi/6}T^2\right).
\end{equation}
If we write a metric perturbation in the following form,
\be
\tilde G_{\mu\nu} = \eta_{\mu\nu}+2\kappa e_{\mu\nu}f,
\end{equation}
where $e_{\mu\nu}e^{\mu\nu}=1$, then the kinetic term for $f$ will be
canonically normalized. To lowest order in $f$ and $T$, the interaction
Lagrangian is
\be
\mathcal{L}_{\rm int} = 
\kappa e^{\mu\nu}f\partial_\mu T\partial_\nu T 
+ \kappa e^\mu_\mu f
\left(\frac{2}{\alpha'}T^2-\frac{1}{2}\partial^\mu T\partial_\mu T\right)
\end{equation}
(from now on all indices are raised and lowered with $\eta_{\mu\nu}$). The
second term, proportional to the trace of $e$, makes a vanishing
contribution to the amplitude on-shell:
\be
-i\kappa e^\mu_\mu\left(\frac{4}{\alpha'}+k_2\cdot k_3\right)
(2\pi)^{26}\delta^{26}(k_1+k_2+k_3) = 0.
\end{equation}
(The trace of the massless closed
string polarization tensor $e$ used in the string calculations of parts (a)
and (b) above represents the dilaton, not the trace of the (Einstein frame)
graviton, which can always be gauged away.) The amplitude from the first
term of (73) is
\be
2i\kappa e_{\mu\nu}k_2^\mu k_3^\nu(2\pi)^{26}\delta^{26}(k_1+k_2+k_3)
= -i\frac{\kappa}{2}e_{\mu\nu}k_{23}^\mu k_{23}^\nu
(2\pi)^{26}\delta^{26}(k_1+k_2+k_3),
\end{equation}
where we have used the transversality of the graviton polarization
$e_{\mu\nu}k_1^\mu=0$. Comparison with the amplitude (6.6.14) shows that
\be
\kappa = \pi\alpha'g_{\rm c}'.
\end{equation}

\subsection{Problem 6.12}

We can use the three CKVs of the upper half-plane to fix the position $z$ of
the closed-string vertex operator and the position $y_1$ of one of the
upper-string vertex operators. We integrate over the position $y_2$ of the
unfixed open-string vertex operator:
\bea
\lefteqn{S_{D_2}(k_1,k_2,k_3)}\qquad\qquad \nonumber\\
&& =
g_{\rm c}g_{\rm o}^2e^{-\lambda}
\int dy_2\,\left\langle
:c\tilde ce^{ik_1\cdot X}(z,\bar z):
\no c^1e^{ik_2\cdot X}(y_1)\no
\no e^{ik_3\cdot X}(y_2)\no
\right\rangle_{D_1} \nonumber \\
&& = g_{\rm c}g_{\rm o}^2e^{-\lambda}
C_{D_2}^{\rm g}|z-y_1||\bar z-y_1||z-\bar z|
iC_{D_2}^{\rm X}(2\pi)^{26}\delta^{26}(k_1+k_2+k_3) \nonumber \\
&& \qquad\times
|z-\bar z|^{\alpha'k_1^2/2}|z-y_1|^{2\alpha'k_1\cdot k_2}
\int dy_2\,|z-y_2|^{2\alpha'k_1\cdot k_3}|y_1-y_2|^{2\alpha'k_2\cdot k_3}
\nonumber \\
&& =
iC_{D_2}g_{\rm c}g_{\rm o}^2
(2\pi)^{26}\delta^{26}(k_1+k_2+k_3) \nonumber \\
&& \qquad\times
|z-y_1|^{-2}|z-\bar z|^3\int dy_2\,|z-y_2|^{-4}|y_1-y_2|^2.
\eea
The very last line is equal to $4\pi$, independent of $z$ and $y_1$ (as it
should be), as can be calculated by contour integration in the complex
plane. Taking into account (6.4.14), the result is
\be
\frac{4\pi ig_{\rm c}}{\alpha'}(2\pi)^{26}\delta^{26}(k_1+k_2+k_3).
\end{equation}

\setcounter{equation}{0}
\newpage

\section{Chapter 7}

\subsection{Problem 7.1}

Equation (6.2.13) applied to the case of the torus tells us
\begin{multline}
\left\langle
\prod_{i=1}^n:e^{ik_i\cdot X(w_i,\bar w_i)}:
\right\rangle_{T2} = \\
iC^X_{T^2}(\tau)(2\pi)^d\delta^d(\sum k_i)
\exp\left(-\sum_{i<j}k_i\cdot k_j G'(w_i,w_j)
-\frac{1}{2}\sum_i k_i^2G'_\te{r}(w_i,w_i)\right).
\end{multline}
$G'$ is the Green's function (7.2.3),
\begin{equation}
G'(w_i,w_j) = 
-\frac{\alpha'}{2}
\ln\left|\vartheta_1\left(\frac{w_{ij}}{2\pi}|\tau\right)\right|^2
+\alpha'\frac{(\I w_{ij})^2}{4\pi\tau_2}
+k(\tau),
\end{equation}
while $G'_\te{r}$ is the renormalized Green's function, defined by
(6.2.15),
\begin{equation}
G'_\te{r}(w_i,w_j) = G'(w_i,w_j)+\frac{\alpha'}{2}\ln |w_{ij}|^2,
\end{equation}
designed to be finite in the limit $w_j\to w_i$:
\begin{align}
\lim_{w_j\to w_i}G'_\te{r}(w_i,w_j) 
&= -\frac{\alpha'}{2}
\ln\left|\lim_{w_{ij}\to 0}
\frac{\vartheta_1\left(\frac{w_{ij}}{2\pi}|\tau\right)}{w_{ij}}
\right|^2
+k(\tau)
\notag \\
&=
-\frac{\alpha'}{2}
\ln\left|\frac{\partial_\nu\vartheta_1(\nu=0|\tau)}{2\pi}\right|^2
+k(\tau).
\end{align}
The argument of the exponential in (1) is thus
\begin{align}
-\sum_{i<j}k_i&\cdot k_jG'(w_i,w_j)
-\frac{1}{2}\sum_i k_i^2G'_\te{r}(w_i,w_i) \notag\\
=& \sum_{i<j}\alpha'k_i\cdot k_j\left(
\ln\left|\vartheta_1\left(\frac{w_{ij}}{2\pi}|\tau\right)\right|
-\frac{(\I w_{ij})^2}{4\pi\tau_2}
\right)
+\frac{\alpha'}{2}
\ln\left|\frac{\partial_\nu\vartheta_1(0|\tau)}{2\pi}\right|
\sum_ik_i^2
\notag \\
&\qquad-\frac{1}{2}k(\tau)\sum_{i,j}k_i\cdot k_j
\notag\\
=& \sum_{i<j}\alpha'k_i\cdot k_j\left(
\ln\left|\frac{2\pi}{\partial_\nu\vartheta_1(0|\tau)}
\vartheta_1\left(\frac{w_{ij}}{2\pi}|\tau\right)\right|
-\frac{(\I w_{ij})^2}{4\pi\tau_2}
\right),
\end{align}
where in the second equality we have used the overall momentum-conserving
delta function, which implies $\sum_{i,j}k_i\cdot k_j=0$. Plugging this
into (1) yields (7.2.4).

Under the modular transformation $\tau\to\tau'=-1/\tau$ the coordinate $w$
is mapped to $w'=w/\tau$. The weights of the vertex operator
$:\exp(ik_i\cdot X(w_i,\bar w_i)):$ are (2.4.17)
\begin{equation}
h = \tilde h = \frac{\alpha'k_i^2}{4}
\end{equation}
so, according to (2.4.13), the product of vertex operators on the LHS of
(1) transforms to
\begin{equation}
\left\langle\prod_{i=1}^n:e^{ik_i\cdot X(w'_i,\bar w'_i)}:\right\rangle_{T^2}
= |\tau|^{\sum_i\alpha'k_i^2/2}
\left\langle\prod_{i=1}^n:e^{ik_i\cdot X(w_i,\bar w_i)}:\right\rangle_{T^2}.
\end{equation}
On the RHS of (1), the vacuum amplitude 
\begin{equation}
C^X_{T^2} = (4\pi\alpha'\tau_2)^{-d/2}|\eta(\tau)|^{-2d}
\end{equation}
is invariant, since
\begin{equation}
\tau_2' = \frac{\tau_2}{|\tau|^2},
\end{equation}
and (7.2.4b)
\begin{equation}
\eta(\tau') = (-i\tau)^{1/2}\eta(\tau).
\end{equation}
According to (7.2.40d),
\begin{equation}
\vartheta_1\left(\frac{w_{ij}'}{2\pi}|\tau'\right) = 
-i(-i\tau)^{1/2}\exp\left(\frac{iw_{ij}^2}{4\pi\tau}\right)
\vartheta_1\left(\frac{w_{ij}}{2\pi}|\tau\right)
\end{equation}
and
\begin{equation}
\partial_\nu\vartheta_1(0|\tau') = 
(-i\tau)^{3/2}\partial_\nu\vartheta_1(0|\tau),
\end{equation}
so
\begin{equation}
\ln\left|\frac{2\pi}{\partial_\nu\vartheta_1(0|\tau')}
\vartheta_1\left(\frac{w'_{ij}}{2\pi}|\tau'\right)\right| =
\ln\left|\frac{2\pi}{\tau\partial_\nu\vartheta_1(0|\tau)}
\vartheta_1\left(\frac{w_{ij}}{2\pi}|\tau\right)\right|
-\I\left(\frac{w_{ij}^2}{4\pi\tau}\right).
\end{equation}
The second term on the right is equal to
\begin{equation}
\I\left(\frac{w_{ij}^2}{4\pi\tau}\right) =
\frac{1}{4\pi|\tau|^2}
\left(
2\tau_1\I w_{ij}\R w_{ij}-
\tau_2(\R w_{ij})^2+\tau_2(\I w_{ij})^2
\right),
\end{equation}
and it cancels the change in the last term on the RHS of (5):
\begin{equation}
\frac{(\I w'_{ij})^2}{4\pi\tau_2'}
= \frac{1}{4\pi\tau_2|\tau|^2}
\left(-2\tau_1\tau_2\I w_{ij}\R w_{ij}
+\tau^2_2(\R w_{ij})^2+\tau_1^2(\I w_{ij})^2\right).
\end{equation}
The only change, then, is the new factor of $|\tau|^{-1}$ in the logarithm
on the RHS of (13), which gets taken to the power
\begin{equation}
\sum_{i<j}\alpha'k_i\cdot k_j = -\frac{1}{2}\sum_i\alpha'k_i^2;
\end{equation}
we finally arrive, as expected, at the transformation law (7).

\subsection{Problem 7.3}

As usual it is convenient to solve Poisson's equation (6.2.8) in momentum
space. Because the torus is compact, the momentum space is a lattice,
generated by the complex numbers
$k_a = 1-i\tau_1/\tau_2$ and $k_b = i/\tau_2$.
The Laplacian in momentum space is
\begin{equation}
-|k|^2 = 
-|n_ak_a+n_bk_b|^2
= -\frac{|n_b-n_a\tau|^2}{\tau_2^2}
\equiv -\omega_{n_an_b}^2.
\end{equation}
In terms of the normalized Fourier components
\begin{equation}
\mathsf{X}_{n_an_b}(w) =
\frac{1}{2\pi\tau_2^{1/2}}e^{i(n_ak_a+n_bk_b)\cdot w}
\end{equation}
(where the dot product means, as usual, $A\cdot B\equiv\R AB^*$),
the Green's function in real space is (6.2.7)
\begin{equation}
G'(w,w') =
\sum_{(n_a,n_b)\neq(0,0)}\frac{2\pi\alpha'}{\omega^2_{n_an_b}}
\mathsf{X}_{n_an_b}(w)^*\mathsf{X}_{n_an_b}(w').
\end{equation}
Rather than show that (19) is equal to (7.2.3) (or (2)) directly, we will
show that (7.2.3) has the correct Fourier coefficients, that is,
\begin{align}
\int_{T^2}d^2\!w\,\mathsf{X}_{00}(w)G'(w,w') &= 0, \\
\int_{T^2}d^2\!w\,\mathsf{X}_{n_an_b}(w)G'(w,w') &=
\frac{2\pi\alpha'}{\omega^2_{n_an_b}}\mathsf{X}_{n_an_b}(w'),
\qquad (n_a,n_b)\neq(0,0).
\end{align}
The $w$-independent part of the Green's function is left as the unspecified
constant $k(\tau)$ in (7.2.3), which is adjusted (as a function of $\tau$)
to satisfy equation (20). To prove (21), we first divide both sides by
$\mathsf{X}_{n_an_b}(w')$, and use the fact that $G'$ depends only on the
difference $w-w'$ to shift the variable of integration:
\begin{equation}
\int_{T^2}d^2\!w\,e^{i(n_ak_a+n_bk_b)\cdot w}G'(w,0)
= \frac{2\pi\alpha'}{\omega^2_{n_an_b}}.
\end{equation}
To evaluate the LHS, let us use coordinates $x$ and $y$ on the torus
defined by $w=2\pi(x+y\tau)$. The Jacobian for this change of coordinates
is $(2\pi)^2\tau_2$, so we have
\begin{equation}
(2\pi)^2\tau_2\int_0^1dy\int_0^1dx\,e^{2\pi i(n_ax+n_by)}
\left[-\frac{\alpha'}{2}\ln\left|\vartheta_1(x+y\tau|\tau)\right|^2
+\alpha'\pi y^2\tau_2\right].
\end{equation}
Using the infinite-product representation of the theta function (7.2.38d),
we can write,
\begin{multline}
\ln\vartheta_1(x+y\tau|\tau) = 
K(\tau)-i\pi(x+y\tau)
\\+\sum_{m=0}^\infty\ln\left(1-e^{2\pi i(x+(y+m)\tau)}\right)
+\sum_{m=1}^\infty\ln\left(1-e^{-2\pi i(x+(y-m)\tau)}\right),
\end{multline}
where we've collected the terms that are independent of $x$ and $y$
into the function $K(\tau)$, which drops out of the integral. Expression
(23) thus becomes
\begin{multline}
-2\pi^2\alpha'\tau_2\int_0^1dy\int_0^1dx\,
e^{2\pi i(n_ax+n_by)} 
\Bigg[2\pi\tau_2y(1-y) \\
\left.+ \sum_{m=0}^\infty\ln
\left|1-e^{2\pi i(x+(y+m)\tau)}\right|^2
+ \sum_{m=1}^\infty\ln\left|1-e^{-2\pi i(x+(y-m)\tau)}\right|^2
\right].
\end{multline}
The integration of the first term in the brackets is straightforward:
\begin{equation}
\int_0^1dx\int_0^1dy\,
e^{2\pi i(n_ax+n_by)}2\pi\tau_2y(1-y)
= 
\begin{cases}-\frac{\tau_2}{\pi n_b^2}, & n_a=0 \\
0, & n_a\neq0
\end{cases}.
\end{equation}
To integrate the logarithms, we first convert the infinite sums into infinite
regions of integration in $y$ (using the periodicity of the first factor
under $y\to y+1$):
\begin{multline}
\int_0^1dy\,
e^{2\pi i(n_ax+n_by)} \\
\shoveright{
\times\left[
\sum_{m=0}^\infty\ln
\left|1-e^{2\pi i(x+(y+m)\tau)}\right|^2
+ \sum_{m=1}^\infty\ln\left|1-e^{-2\pi i(x+(y-m)\tau)}\right|^2
\right]
} \\
\shoveleft{\quad= \int_0^\infty dy\,e^{2\pi i(n_ax+n_by)}
\ln\left|1-e^{2\pi i(x+y\tau)}\right|^2} \\
+ \int_{-\infty}^0 dy\,e^{2\pi i(n_ax+n_by)}
\ln\left|1-e^{-2\pi i(x+y\tau)}\right|^2.
\end{multline}
The $x$ integral can now be performed by separating the logarithms into
their holomorphic and anti-holomorphic pieces and Taylor expanding. For
example,
\begin{align}
\int_0^1dx\,e^{2\pi i(n_ax+n_by)}
\ln\left(1-e^{2\pi i(x+y\tau)}\right)
&= -\sum_{n=1}^\infty\frac{1}{n}
\int_0^1dx\,
e^{2\pi i((n_a+n)x+(n_b+n\tau)y)} \notag \\
&= \begin{cases}
\frac{1}{n_a}e^{2\pi i(n_b-n_a\tau)y}, & n_a<0 \\
0, & n_a\geq0
\end{cases}.
\end{align}
The $y$ integral is now straightforward:
\begin{equation}
\int_0^\infty dy\,\frac{1}{n_a}e^{2\pi i(n_b-n_a\tau)y}
= -\frac{1}{2\pi in_a(n_b-n_a\tau)}.
\end{equation}
Similarly,
\begin{align}
\int_0^\infty dy\int_0^1dx\,e^{2\pi i(n_ax+n_by)}
\ln\left(1-e^{-2\pi i(x+y\bar\tau)}\right)
&= \begin{cases}\frac{1}{2\pi in_a(n_b-n_a\bar\tau)}, & n_a>0 \\
0, & n_a\leq0
\end{cases}, \\
\int_{-\infty}^0 dy\int_0^1dx\,e^{2\pi i(n_ax+n_by)}
\ln\left(1-e^{-2\pi i(x+y\tau)}\right)
&= \begin{cases}-\frac{1}{2\pi in_a(n_b-n_a\tau)}, & \!n_a>0 \\
0, & \!n_a\leq0
\end{cases}, \\
\int_{-\infty}^0 dy\int_0^1dx\,e^{2\pi i(n_ax+n_by)}
\ln\left(1-e^{2\pi i(x+y\bar\tau)}\right)
&= \begin{cases}\frac{1}{2\pi in_a(n_b-n_a\bar\tau)}, & n_a<0 \\
0, & n_a\geq0
\end{cases}.
\end{align}
Expressions (26) and (29)--(32) can be added up to give a single expression
valid for any sign of $n_a$:
\begin{equation}
-\frac{\tau_2}{\pi|n_b-n_a\tau|^2};
\end{equation}
multiplying by the prefactor $-2\pi^2\alpha'\tau_2$ in front of the
integral in (25) indeed yields precisely the RHS of (22), which is what we
were trying to prove.

\subsection{Problem 7.5}

In each case we hold $\nu$ fixed while taking $\tau$ to its appropriate
limit.

\paragraph{(a)}

When $\I\tau\to\infty$, $q\equiv\exp(2\pi i\tau)\to0$, and it is clear
from either the infinite sum expressions (7.2.37) or the infinite product
expressions (7.2.38) that in this limit
\begin{align}
\vartheta_{00}(\nu,\tau) &\to 1, \\
\vartheta_{10}(\nu,\tau) &\to 1, \\
\vartheta_{01}(\nu,\tau) &\to 0, \\
\vartheta_{11}(\nu,\tau) &\to 0.
\end{align} 
Note that all of these limits are independent of $\nu$.

\paragraph{(b)}

Inverting the modular transformation (7.2.40a), we have
\begin{align}
\vartheta_{00}(\nu,\tau) &= 
(-i\tau)^{-1/2}e^{-\pi i\nu^2/\tau}\vartheta_{00}(\nu/\tau,-1/\tau) 
\notag \\
&= (-i\tau)^{-1/2}\sum_{n=-\infty}^\infty e^{-\pi i(\nu-n)^2/\tau}.
\end{align}
When we take $\tau$ to 0 along the imaginary axis, each term in the series
will go either to 0 (if $\R(\nu-n)^2>0$) or to infinity (if
$\R(\nu-n)^2\leq0$). Since different terms in the series cannot
cancel for arbitrary $\tau$, the theta function can go to 0 only if every
term in the series does so:
\begin{equation}
\forall n\in Z,\quad\R(\nu-n)^2>0;
\end{equation}
otherwise it will diverge. For $|\R\nu|\leq1/2$, condition (39) is
equivalent to
\begin{equation}
|\I\nu|<|\R\nu|;
\end{equation}
in general, for $|\R\nu-n|\leq1/2$, the theta function goes to 0 if
\begin{equation}
|\I\nu|<|\R\nu-n|.
\end{equation}

Since $\vartheta_{01}(\nu,\tau)=\vartheta_{00}(\nu+1/2,\tau)$, the region
in which it goes to 0 in the limit $\tau\to0$ is simply shifted by $1/2$
compared to the case treated above.

For $\vartheta_{10}$, the story is the same as for $\vartheta_{00}$, since
\begin{align}
\vartheta_{10}(\nu,\tau) &= 
(-i\tau)^{-1/2}e^{-\pi i\nu^2/\tau}\vartheta_{01}(\nu/\tau,-1/\tau) 
\notag \\
&= (-i\tau)^{-1/2}\sum_{n=-\infty}^\infty(-1)^ne^{-\pi i(\nu-n)^2/\tau};
\end{align}
the sum will again go to 0 where (39) is obeyed, and infinity elsewhere.

Finally, $\vartheta_{11}$ goes to 0 in the same region as $\vartheta_{01}$,
since the sum is the same as (42) except over the half-odd-integers,
\begin{align}
\vartheta_{11}(\nu,\tau) &= 
i(-i\tau)^{-1/2}e^{-\pi i\nu^2/\tau}\vartheta_{11}(\nu/\tau,-1/\tau) 
\notag \\
&= (-i\tau)^{-1/2}\sum_{n=-\infty}^\infty(-1)^ne^{-\pi i(\nu-n+1/2)^2/\tau},
\end{align}
so the region (39) is shifted by $1/2$.

\paragraph{(c)}

According to (7.2.39) and (7.2.40), under the modular transformations
\begin{align}
\tau' &= \tau+1, \\
\tau' &= -\frac{1}{\tau},
\end{align}
the theta functions are exchanged with each other and multiplied by
factors that are finite as long as $\nu$ and $\tau$ are finite.
Also, under (45) $\nu$ is transformed to
\begin{equation}
\nu' = \frac{\nu}{\tau}.
\end{equation}
We are considering limits where $\tau$ approaches some non-zero real value
$\tau_0$ along a path parallel to the imaginary axis, in other words, we set
$\tau=\tau_0+i\epsilon$ and take $\epsilon\to0^+$. The
property of approaching the real axis along a path parallel to the
imaginary axis is preserved by the modular transformations (to first order
in $\epsilon$):
\begin{align}
\tau' &= \tau_0+1+i\epsilon, \\
\tau' &=
-\frac{1}{\tau_0}+i\frac{\epsilon}{\tau_0^2}+\mathcal{O}(\epsilon^2),
\end{align}
under (44) and (45) respectively. By a sequence of transformations (44) and
(45) one can reach any rational limit point $\tau_0$ starting with
$\tau_0=0$, the case considered in part (b) above. During these
transformations (which always begin with (44)), the region (39), in which
$\vartheta_{00}$ goes to 0, will repeatedly be shifted by $1/2$ and rescaled
by $\tau_0$ (under (46)). (Note that the limiting value, being either 0 or
infinity, is insensitive to the finite prefactors involved in the
transformations (7.2.39) and (7.2.40).) It is easy to see that this
cumulative sequence of rescalings will telescope into a single rescaling by
a factor $q$, where $p/q$ is the final value of $\tau_0$ in reduced form.

As for the case when $\tau_0$ is irrational, I can only conjecture that the
theta functions diverge (almost) everywhere on the $\nu$ plane in that
limit.

\subsection{Problem 7.7}

The expectation value for fixed open string tachyon vertex operators on the
boundary of the cylinder is very similar to the corresponding formula (7.2.4)
for closed string tachyon vertex operators on the torus. The major
difference comes from the fact that the Green's function is
doubled. The method of images gives the Green's function for the cylinder
in terms of that for the torus (7.2.3):
\begin{equation}
G'_{C_2}(w,w') = 
G'_{T^2}(w,w') + G'_{T^2}(w,-\bar w').
\end{equation}
However, since the boundary of $C_2$ is given by those points that are
invariant under the involution $w\to-\bar w$, the two terms on the RHS
above are equal if either $w$ or $w'$ is on the boundary. The renormalized
self-contractions are also doubled, so we have
\begin{multline}
\left\langle\prod_{i=1}^n
\no e^{ik_i\cdot X(w_i,\bar w_i)}\no
\right\rangle_{C_2}
= \\
iC^X_{C_2}(t)(2\pi)^{26}\delta^{26}(\sum_i k_i)
\prod_{i<j}\left|
\eta(it)^{-3}\vartheta_1\left(\frac{w_{ij}}{2\pi}|it\right)
\exp\left[-\frac{(\I w_{ij})^2}{4\pi t}\right]
\right|^{2\alpha'k_i\cdot k_j}.
\end{multline}

The boundary of the cylinder breaks into two connected components, and the
vertex operator positions $w_i$ must be integrated over both
components. If there are Chan-Paton factors, then the integrand will
include two traces, one
for each component of the boundary, and the order of the factors in each
trace will be the order of the operators on the corresponding component. We
will denote the product of these two traces $T(w_1,\dots,w_n)$, and of
course it will also depend implicitly on the Chan-Paton factors themselves
$\lambda^{a_i}$. Borrowing from the cylinder vacuum amplitude given in
(7.4.1), we can write the $n$-tachyon amplitude as
\begin{multline}
S_{C_2}(k_1,a_1;\dots;k_n,a_n) = \\
\shoveleft{i(2\pi)^{26}\delta^{26}(\sum_ik_i)g_\te{o}^n
\int_0^\infty\frac{dt}{2t}(8\pi^2\alpha't)^{-13}\eta(it)^{-24}} 
\prod_{i=1}^n\left(\int_{\partial C_2}dw_i\right) \\
\times T(w_1,\dots,w_n)\prod_{i<j}\left|
\eta(it)^{-3}\vartheta_1\left(\frac{w_{ij}}{2\pi}|it\right)
\exp\left[-\frac{(\I w_{ij})^2}{4\pi t}\right]
\right|^{2\alpha'k_i\cdot k_j}.
\end{multline}

\subsection{Problem 7.8}

In this problem we consider the part of the amplitude (51) in which the
first $m\geq2$ vertex operators are on one of the cylinder's boundaries,
and the other $n-m\geq2$ are on the other one. For concreteness let us put
the first set on the boundary at $\R w=0$ and the second set on the
boundary at $\R w=\pi$, and then double the amplitude (51). Since we will be
focusing on the region of the moduli space where $t$ is very small, it is
convenient to scale the vertex operator positions with $t$, so
\begin{equation}
w_i = \begin{cases}
2\pi ix_it,&i=1,\dots,m \\ \pi+2\pi ix_it,&i=m+1,\dots,n
\end{cases}.
\end{equation}
The $x_i$ run from 0 to 1 independent of $t$, allowing us to change of the
order of integration in (51). Using (16) and the mass
shell condition $\alpha'k_i^2=1$ we can write the part of the
amplitude we're interested in as follows:
\begin{multline}
S'(k_1,a_1;\dots;k_n,a_n) =
i(2\pi)^{26}\delta^{26}(\sum_ik_i)
g_\te{o}^n2^{-13}(2\pi)^{n-26}\alpha'{}^{-13} \\
\shoveleft{\times
\prod_i\left(\int_0^1 dx_i\right)T_1(x_1,\dots,x_m)T_2(x_{m+1},\dots,x_n)} \\
\times\int_0^\infty dt\,t^{n-14}\eta(it)^{3n-24}
\prod_{i<j}\left|
\vartheta_{1,2}(ix_{ij}t|it)\exp(-\pi x_{ij}^2t)
\right|^{2\alpha'k_i\cdot k_{j}}.
\end{multline}
In the last product the type of theta function to use depends on whether
the vertex operators $i$ and $j$ are on the same boundary (in which case
$\vartheta_1$) or on opposite boundaries (in which case
$\vartheta_2(ix_{ij}t|it)=-\vartheta_1(ix_{ij}t-1/2|it)$).
Concentrating now on the last line of (53), let us apply the modular
transformations (7.2.40b), (7.2.40d), and (7.2.44b), and change the
variable of integration to $u=1/t$:
\begin{equation}
\int_0^\infty du\,\eta(iu)^{3n-24}
\prod_{i<j}\left|\vartheta_{1,4}(x_{ij}|iu)\right|^{2\alpha'k_i\cdot k_j}.
\end{equation}

For large $u$, each of the factors in the integrand of (54) can be written as a
fractional power of $q\equiv e^{-2\pi u}$ times a power series in $q$
(with coefficients that may depend on the $x_{ij}$; see (7.2.37), (7.2.38),
and (7.2.43)):
\begin{align}
\eta(iu) &= q^{1/24}(1-q+\dots); \\
\vartheta_1(x|iu) &= 
2q^{1/8}\sin(\pi x)\left(1-(1+2\cos(2\pi x))q+\dots\right); \\
\vartheta_4(x|iu) &= \sum_{k=-\infty}^\infty(-1)^kq^{k^2/2}e^{2\pi ikx} 
= 1-2\cos(2\pi x)q^{1/2}+\dots.
\end{align}
For $\eta$ and $\vartheta_1$ this power series involves only integer powers
of $q$, whereas for $\vartheta_4$ it mixes
integer and half-integer powers. Substituting (55)-(57) into (54) yields
\begin{multline}
\prod_{\substack{i<j \\ i\simeq j}}|2\sin\pi x_{ij}|^{2\alpha'k_i\cdot k_j}
\int_0^\infty du\,q^{n/8-1+
\alpha'
\sum_{\substack{\scriptscriptstyle i<j \\ \scriptscriptstyle i\simeq j}}
k_i\cdot k_j/4}
(1+\dots)
 \\
= \prod_{\substack{i<j \\ i\simeq j}}|2\sin\pi x_{ij}|^{2\alpha'k_i\cdot k_j}
\int_0^\infty du\,q^{-1-\alpha's/4}(1+\dots)
\end{multline}
The symbol $i\simeq j$ means that the sum or product is only over pairs of
vertex operators on the same boundary of the cylinder. We obtain the second
line from the first by using
\begin{align}
\sum_{\substack{i<j \\ i\simeq j}}k_i\cdot k_j
&= \sum_{i<j}k_i\cdot k_j - \sum_{i=1}^m\sum_{j=m+1}^nk_i\cdot k_j \notag \\
&= -\frac{1}{2}\sum_{i=1}^nk_i^2 - 
\left(\sum_{i=1}^mk_i\right)\cdot\left(\sum_{j=m+1}^nk_j\right) \notag \\
&= -\frac{n}{2\alpha'}-s,
\end{align}
where $s=-(\sum_{i=1}^nk_i)^2$ is the mass squared of the
intermediate state propagating along the long cylinder. (The two $n$'s
we have cancelled against each other in (58) both come from
$\alpha'\sum_ik_i^2$, so in fact we could have done without the mass shell
condition in the derivation.) Each power of $q$ appearing in the power
series $(1+\dots)$ in (58) will produce, upon performing the $u$
integration, a pole in $s$:
\begin{equation}
\int_0^\infty du\,q^{-1-\alpha's/4+k}
= \frac{2}{\pi(4k-4-\alpha's)}.
\end{equation}
Since every (non-negative) integer power $k$ appears in the expansion of
(54), we have the entire sequence of closed string masses at
$s=4(k-1)/\alpha'$.

What about the half-integer powers of $q$ that appear in the expansion of
$\vartheta_4$, (57)? The poles from these terms in fact vanish,
but only after integrating over the vertex operator positions. To see this,
consider the effect of uniformly translating all the vertex operator
positions on just one of the boundaries by an amount $y$: 
$x_i\to x_i+y$, $i=1,\dots,m$. This translation changes the relative
position $x_{ij}$ only if the vertex operators $i$ and $j$ are on opposite
boundaries; it thus leaves the Chan-Paton factors $T_1$ and $T_2$
in (53) invariant, as well as all the factors in the integrand of (54)
except those involving $\vartheta_4$; those become
\begin{equation}
\prod_{i=1}^m\prod_{j=m+1}^n
|\vartheta_4(x_{ij}+y|iu)|^{2\alpha'k_i\cdot k_j}.
\end{equation}
Expanding this out using (57), each term will be of the form 
\begin{equation}
cq^{\sum_lk_l^2/2}e^{2\pi iy\sum_lk_l},
\end{equation}
where the $k_l$ are integers and the coefficient $c$ depends on the $x_i$
and the momenta $k_i$. Since $y$ is effectively integrated over when one
integrates over the vertex operator positions, only the terms for which 
$\sum_lk_l=0$ will survive. This condition implies that
$\sum_lk_l^2$ must be even, so only integer powers of $q$ produce poles
in the amplitude.

\subsection{Problem 7.9}

We wish to consider the result of the previous problem in the simplest
case, when $n=4$, $m=2$, and there are no Chan-Paton indices. We are
interested in particular in the first pole, at $s=-4/\alpha'$, where the
intermediate closed string goes on shell as a tachyon. Neglecting the
``$\dots$'' and approximating the exponents $2\alpha'k_1\cdot
k_2=2\alpha'k_3\cdot k_4=-\alpha's-2$ by 2, (58) becomes
\begin{equation}
-\sin^2(\pi x_{12})\sin^2(\pi x_{34})\frac{32}{\pi(\alpha's+4)}.
\end{equation}
After integrating over the positions $x_i$, the amplitude (53) is
\begin{equation}
S'(k_1,\dots,k_4) =
-i(2\pi)^{26}\delta^{26}(\sum_ik_i)g_\te{o}^42^{-9}(2\pi)^{-23}\alpha'{}^{-14}
\frac{1}{s+4/\alpha'}.
\end{equation}

In Problem 6.12, we calculated the three-point vertex for two open-string
tachyons to go to a closed-string tachyon:
\begin{equation}
\frac{4\pi ig_\te{c}}{\alpha'}(2\pi)^{26}\delta^{26}(k_1+k_2+k_3).
\end{equation}
We can reproduce the $s$-channel pole of (64) by simply writing down the
Feynman diagram using this vertex and the closed-string tachyon propagator,
$i/(s+4/\alpha')$:
\begin{equation}
-i(2\pi)^{26}\delta^{26}(\sum_ik_i)g_\te{c}^24(2\pi)^2\alpha'{}^{-2}
\frac{1}{s+4/\alpha'}.
\end{equation}
We can now compare the residues of the poles in (64) and (66) to obtain the
following relation between $g_\te{c}$ and $g_\te{o}$ (note that, since the
vertex (65) is valid only on-shell, it is only appropriate to compare the
residues of the poles at $s=-4/\alpha'$ in (64) and (66), not the detailed
dependence on $s$ away from the pole):
\begin{equation}
\frac{g_\te{o}^2}{g_\te{c}} = 2^{11/2}(2\pi)^{25/2}\alpha'{}^6.
\end{equation}
This is in agreement with (8.7.28).

\subsection{Problem 7.10}

When the gauge group is a product of $U(n_i)$ factors, the generators are
block diagonal. A tree-level diagram is proportional to
$\Tr(\lambda^{a_1}\cdots\lambda^{a_l})$, and vanishes unless all the
Chan-Paton factors are in the same block. Unitarity requires
\begin{equation}
\Tr(\lambda^{a_1}\cdots\lambda^{a_i}\lambda^{a_{i+1}}\cdots\lambda^{a_l})
= \Tr(\lambda^{a_1}\cdots\lambda^{a_i}\lambda^e)
\Tr(\lambda^e\lambda^{a_{i+1}}\cdots\lambda^{a_l}),
\end{equation}
which is an identity for any product of $U(n_i)$s. (The LHS vanishes if
all the $\lambda^a$ are not in the same block, and equals the usual $U(n)$
value if the are. The RHS has the same property, because the $\lambda^e$
must be in the same block both with the first group of $\lambda^a$s and with
the second group for a non-zero result.)

Now consider the cylinder with two vertex operators on each boundary, with
Chan-Paton factors $\lambda^a$ and $\lambda^b$ on one boundary and
$\lambda^c$ and $\lambda^d$ on the other. This amplitude is proportional to
$\Tr(\lambda^a\lambda^b)\Tr(\lambda^c\lambda^d)$, i.e. we only need the two
Chan-Paton factors in each pair to be in the same block. However, if we
make a unitary cut in the open
string channel, then the cylinder becomes a disk with an open string
propagator connecting two points on the boundary. If on the intermediate
state we sum only over block-diagonal generators, then this amplitude will
vanish unless $\lambda^a,\lambda^b,\lambda^c,\lambda^d$ are all in the same
block. For this to be consistent there can only be one block, i.e. the
gauge group must be simple.

\subsection{Problem 7.11}

We will be using the expression (7.3.9) for the point-particle vacuum
amplitude to obtain a generalized version of the cylinder vacuum amplitude
(7.4.1). Since (7.3.9) is an integral over the circle modulus $l$, whereas
(7.4.1) is an integral over the cylinder modulus $t$, we need to know the
relationship between these quantities. The modulus $l$ is defined with
respect to the point-particle action (5.1.1), which (after choosing the
analog of unit gauge for the einbein $e$) is
\begin{equation}
\frac{1}{2}\int_0^l d\tau
\left((\partial_\tau x)^2+m^2\right).
\end{equation}
The Polyakov action in unit gauge (2.1.1), on a cylinder with modulus $t$,
is
\begin{equation}
\frac{1}{4\pi\alpha'}\int_0^\pi dw_1\int_0^{2\pi t}dw_2
\left((\partial_1X)^2+(\partial_2X)^2\right).
\end{equation}
Decomposing $X(w_1,w_2)$ into its center-of-mass motion $x(w_2)$ and its
internal state $y(w_1,w_2)$ (with $\int dw_1\,y=0$), the Polyakov action
becomes
\begin{equation}
\frac{1}{4\alpha'}\int_0^{2\pi t}dw_2
\left((\partial_2x)^2+
\frac{1}{\pi}\int_0^\pi dw_1\left((\partial_1y)^2+(\partial_2y)^2\right)
\right).
\end{equation}
We can equate (69) and (71) by making the identifications
$\tau=2\alpha'w_2$, $l=4\pi\alpha't$, and
\begin{equation}
m^2 = 
\frac{1}{4\pi\alpha'{}^2}
\int_0^\pi dw_1\left((\partial_1y)^2+(\partial_2y)^2\right).
\end{equation}

Using the relation $l=4\pi\alpha't$ to translate between the circle and the
cylinder moduli, we can now sum the circle vacuum amplitude (7.3.9) over
the open string spectrum, obtaining the second line of (7.4.1):
\begin{equation}
Z_{C_2} = iV_D\int_0^\infty\frac{dt}{2t}(8\pi^2\alpha't)^{-D/2}
\sum_{i\in\mathcal{H}^\perp_\te{o}}e^{-2\pi t\alpha'm^2_i}.
\end{equation}
Taking a spectrum with $D'$ net sets of oscillators and a ground state at
$m^2=-1/\alpha'$, the sum is evaluated in the usual way (with $q=e^{-2\pi
t}$):
\begin{align}
\sum_{i\in\mathcal{H}^\perp_\te{o}}q^{\alpha'm^2_i} &=
\left(\prod_{i=1}^{D'}\prod_{n=1}^\infty\sum_{N_{in}=0}^\infty\right)
q^{-1+\sum_{n=1}^\infty nN_{in}} \notag \\
&= q^{-1}\prod_{i=1}^{D'}\prod_{n=1}^\infty
\left(\sum_{N_{in}=0}^\infty q^{nN_{in}}\right) \notag \\
&= q^{-1}\prod_{i=1}^{D'}\prod_{n=1}^\infty
\frac{1}{1-q^n} \notag \\
&= q^{D'/24-1}\eta(it)^{-D'}.
\end{align}
As in Problem 7.8 above, we are interested in studying the propagation of
closed string modes along long, thin cylinders, which corresponds to the
region $t\ll1$. So let us change the variable of integration to $u=1/t$,
\begin{align}
Z_{C_2} &= \frac{iV_D}{2(8\pi^2\alpha')^{D/2}}\int_0^\infty dt\,
t^{-D/2-1}e^{-2\pi t(D'/24-1)}\eta(it)^{-D'} \notag \\
&= \frac{iV_D}{2(8\pi^2\alpha')^{D/2}}\int_0^\infty du\,
u^{(D-D'-2)/2}e^{-2\pi(D'/24-1)/u}\eta(iu)^{-D'},
\end{align}
and add a factor of $e^{-\pi u\alpha'k^2/2}$ to the integrand, where $k$ is
the momentum flowing along the cylinder:
\begin{equation}
\int_0^\infty du\,
u^{(D-D'-2)/2}e^{-2\pi(D'/24-1)/u-\pi u\alpha'k^2/2}\eta(iu)^{-D'}.
\end{equation}
Now we can expand the eta-function for large $u$ using the product
representation (7.2.43). Each term will be of the form
\begin{equation}
\int_0^\infty du\,
u^{(D-D'-2)/2}e^{-2\pi(D'/24-1)/u-2\pi u(\alpha'k^2/4+m-D'24)},
\end{equation}
where $m$ is a non-negative integer.

In the case $D=26$, $D'=24$, (77) reveals the expected series of
closed-string poles, as found in Problem 7.8. If $D'\neq24$, the integral
yields a modified Bessel function,
\begin{equation}
\int_0^\infty du\,u^{c-1}e^{-a/u-b/u} = 
2b^{-c/2}a^{c/2}K_c\left(2\sqrt{ab}\right),
\end{equation}
which has a branch cut along the negative real axis, hence the constraint
$D'=24$. When $D'=24$, (77) simplifies to
\begin{equation}
\int_0^\infty du\,u^{(D-26)/2}e^{-2\pi u(\alpha'k^2/4+m-1)}
= \frac{\Gamma\left(\frac{D-24}{2}\right)}
{\left(2\pi(\alpha'k^2/4+m-1)\right)^{(D-24)/2}}.
\end{equation}
For $D$ odd there is a branch cut. For $D$ even but less than 26, the gamma
function is infinite; even if one employs a ``minimal subtraction'' scheme
to remove the infinity, the remainder has a logarithmic branch cut in
$k^2$. For even $D\geq26$ there is indeed a pole, but this pole is simple
(as one expects for a particle propagator) only for $D=26$.

\subsection{Problem 7.13}

We follow the same steps in calculating the vacuum Klein bottle amplitude
as Polchinski does in calculating the vacuum torus amplitude, starting with
finding the scalar partition function:
\begin{align}
Z_X(t) &= \langle1\rangle_{X,K_2} \notag \\
&= \Tr(\Omega e^{-2\pi tH}) \notag \\
&= q^{-13/6}\Tr(\Omega q^{L_0+\tilde{L}_0}).
\end{align}
The operator $\Omega$ implements the orientation-reversing boundary
condition in the Euclidean time ($\sigma^2$) direction of the Klein
bottle. Since $\Omega$ switches left-movers and right-movers, it is
convenient to work with states and operators that have definite properties
under orientation reversal. We therefore define the raising and lowering
operators $\alpha^\pm_m = (\alpha_m\pm\tilde{\alpha}_m)/\sqrt{2}$, so that
$\alpha_m^+$ commutes with $\Omega$, while $\alpha^-_m$ anti-commutes with
it (we have suppressed the spacetime index $\mu$). These are
normalized to have the usual commutation relations, and can be used to
build up the spectrum of the closed string in the usual way. The ground
states $|0;k\rangle$ of the closed string are invariant under orientation
reversal. The $\Omega q^{L_0+\tilde{L}_0}$ eigenvalue of a ground state is
$q^{\alpha'k^2/2}$; each raising operator $\alpha^+_{-m}$
multiplies that eigenvalue by $q^m$, while each raising operator
$\alpha^-_{-m}$ multiplies it by $-q^m$. Summing over all combinations of
such operators, the partition function is
\begin{align}
Z_X(t) &= q^{-13/6}V_{26}\int\frac{d^{26}k}{(2\pi)^{26}}
q^{\alpha'k^2/2}\prod_{m=1}^\infty(1-q^m)^{-1}(1+q^m)^{-1} \notag \\
&= iV_{26}(4\pi^2\alpha't)^{-13}\eta(2it)^{-26}.
\end{align}

For $bc$ path integrals on the Klein bottle, it is again convenient to
introduce raising and lowering operators with definite properties under
orientation reversal: $b^\pm_m = (b_m\pm\tilde{b}_m)/\sqrt{2}$, $c^\pm_m =
(c_m\pm\tilde{c}_m)/\sqrt{2}$. The particular path integral we will be
interested in is 
\begin{equation}
\left\langle c^\pm_mb_0^+\right\rangle_{K_2} = 
q^{13/6}\Tr\left((-1)^F\Omega c^\pm_mb^+_0q^{L_0+\tilde{L}_0}\right).
\end{equation}
Build up the $bc$ spectrum by starting with a ground state
$|\downarrow\downarrow\rangle$ that is annihilated by $b^\pm_m$ for
$m\geq0$ and $c^\pm_m$ for $m>0$, and acting on it with the raising
operators $b^\pm_m$ ($m<0$) and $c^\pm_m$ ($m\leq0$). The operator
$(-1)^F\Omega q^{L_0+\tilde{L}_0}$ is diagonal in this basis, whereas the
operator $c^\pm_mb_0^+$ takes basis states to other basis states (if it
does not annihilate them). Therefore the trace (81) vanishes. The only
exception is the case of the operator $c^+_0b_0^+$, which is diagonal in
this basis; it projects onto the subspace of states that are built up from
$c_0^+|\downarrow\downarrow\rangle$. This state has eigenvalue $-q^{-2}$ 
under $(-1)^F\Omega c^+_0b^+_0q^{L_0+\tilde{L}_0}$. Acting with $c^-_0$
does not change this eigenvalue; acting with $b^+_{-m}$ or $c^+_{-m}$
($m>0$) multiplies it by $-q^m$, and with $b^-_{-m}$ or $c^-_{-m}$ by
$q^m$. We thus have
\begin{equation}
\left\langle c^+_0b^+_0\right\rangle_{K_2} =
2q^{1/6}\prod_{m=1}^\infty(1-q^m)^2(1+q^m)^2 =
2\eta(2it)^2.
\end{equation}
(We have taken the absolute value of the result.)

The Klein bottle has only one CKV, which translates in the $\sigma^2$
direction. Let us temporarily include an arbitrary number of vertex
operators in the path integral, and fix the $\sigma^2$ coordinate of the
first one $\mathcal{V}_1$. According to (5.3.9), the amplitude is
\begin{equation}
S = \int_0^\infty\frac{dt}{4}
\left\langle
\int d\sigma^1_1\,c^2\mathcal{V}_1(\sigma^1_1,\sigma^2_1)
\frac{1}{4\pi}(b,\partial_t\hat{g})
\prod_{i=2}^n\int\frac{d^2w_i}{2}\mathcal{V}_i(w_i,\bar w_i)
\right\rangle_{K_2}.
\end{equation}
The overall factor of $1/4$ is from the discrete symmetries of the Klein
bottle, with $1/2$ from $w\to\bar{w}$ and $1/2$ from $w\to-w$.

To evaluate the $b$ insertion, let us temporarily fix the coordinate region
at that for $t=t_0$ and let the metric vary with $t$:
\begin{equation}
\hat{g}(t) = \begin{pmatrix} 1 & 0 \\ 0 & t^2/t_0^2 \end{pmatrix}.
\end{equation}
Then
\begin{equation}
\partial_t\hat{g}(t_0) = \begin{pmatrix} 0 & 0 \\ 0 & 2/t_0 \end{pmatrix}
\end{equation}
and
\begin{align}
\frac{1}{4\pi}(b,\partial_t\hat{g}) &=
\frac{1}{4\pi} \int d\sigma^1d\sigma^2 \frac{2}{t}b_{22} \notag \\
&= -\int d\sigma^1 (b_{ww}+b_{\bar{w}\bar{w}}) \notag \\
&= 2\pi(b_0+\tilde{b}_0) \notag \\
&= 2\sqrt{2}\pi b_0^+.
\end{align}
If we expand the $c$ insertion in the path integral in terms of the $c_m$
and $\tilde{c}_m$,
\begin{equation}
c^2(\sigma^1,\sigma^2) =
\frac{1}{2}\left(
\frac{c(z)}{z}+\frac{\tilde{c}(\bar{z})}{\bar{z}}
\right) =
\frac{1}{2}\sum_m\left(\frac{c_m}{z^m}+\frac{\tilde{c}_m}{\bar{z}^m}\right),
\end{equation}
then, as we saw above, only the $m=0$ term, which is $c_0^+/\sqrt{2}$, will
contribute to the ghost path integral. This allows us to factor the $c$
ghost out of the integal over the first vertex operator position in (83),
and put all the vertex operators on the same footing:
\begin{equation}
S = \int_0^\infty\frac{dt}{4t}
\left\langle
c_0^+b_0^+
\prod_{i=1}^n\int\frac{d^2w_i}{2}\mathcal{V}_i(w_i,\bar w_i)
\right\rangle_{K_2}.
\end{equation}
We can now extrapolate to the case where there are no vertex operators
simply by setting $n=0$ above. Using (80) and (82), this gives
\begin{equation}
Z_{K_2} =
iV_{26}\int_0^\infty\frac{dt}{2t}(4\pi^2\alpha't)^{-13}\eta(2it)^{-24}.
\end{equation}
This is off from Polchinski's result (7.4.15) by a factor of 2.

\subsection{Problem 7.15}

\paragraph{(a)}

If the $\sigma^2$ coordinate is periodically identified (with period
$2\pi$), then a cross-cap at $\sigma^1=0$ implies the identification
\begin{equation}
(\sigma^1,\sigma^2) \cong (-\sigma^1,\sigma^2+\pi).
\end{equation}
This means the following boundary conditions on the scalar and ghost
fields:
\begin{align}
\partial_1X^\mu(0,\sigma^2) = -\partial_1X^\mu(0,\sigma^2+\pi), \quad&\quad
\partial_2X^\mu(0,\sigma^2) = \partial_2X^\mu(0,\sigma^2+\pi), \\
c^1(0,\sigma^2) = -c^1(0,\sigma^2+\pi), \quad&\quad
c^2(0,\sigma^2) = c^2(0,\sigma^2+\pi), \\
b_{12}(0,\sigma^2) = -b_{12}(0,\sigma^2+\pi), \quad&\quad
b_{11}(0,\sigma^2) = b_{11}(0,\sigma^2+\pi).
\end{align}
These imply the following conditions on the modes at $\sigma^1=0$:
\begin{equation}
\alpha^\mu_n+(-1)^n\tilde{\alpha}^\mu_{-n} =
c_n + (-1)^n\tilde{c}_{-n} =
b_n - (-1)^n \tilde{b}_{-n} = 0
\end{equation}
(for all $n$). The state corresponding to the cross-cap is then
\begin{equation}
|C\rangle \propto 
\exp\left[-\sum_{n=1}^\infty(-1)^n
\left(\frac{1}{n}\alpha_{-n}\cdot\tilde{\alpha}_{-n}
+b_{-n}\tilde{c}_{-n}+\tilde{b}_{-n}c_{-n}\right)\right]
(c_0+\tilde{c}_0)|0;0;\downarrow\downarrow\rangle.
\end{equation}

\paragraph{(b)}

The Klein bottle vacuum amplitude is
\begin{equation}
\int_0^\infty ds\,\langle C|c_0b_0e^{-s(L_0+\tilde{L}_0)}|C\rangle.
\end{equation}
Since the raising and lowering operators for different oscillators commute
with each other (or, in the case of the ghost oscillators, commute in
pairs), we can factorize the integrand into a separate amplitude for each
oscillator:
\begin{multline}
e^{2s}\langle\downarrow\downarrow|
(b_0+\tilde{b}_0)c_0b_0(c_0+\tilde{c}_0)|\downarrow\downarrow\rangle
\langle0|e^{-s\alpha'p^2/2}|0\rangle \\
\shoveleft\times\prod_{n=1}^\infty\Bigg(
\langle0|e^{-(-1)^nc_n\tilde{b}_n}e^{-sn(b_{-n}c_n+\tilde{c}_{-n}\tilde{b}_n)}
e^{-(-1)^nb_{-n}\tilde{c}_{-n}}|0\rangle \\
\times\langle0|e^{-(-1)^n\tilde{c}_nb_n}e^{-sn(\tilde{b}_{-n}\tilde{c}_n+c_{-n}b_n)}
e^{-(-1)^n\tilde{b}_{-n}c_{-n}}|0\rangle \\
\times\left.\prod_{\mu=0}^{25}
\langle0|e^{-(-1)^n\tilde{\alpha}_n^\mu\alpha_{n\mu}/n}
e^{
-s(\alpha_{-n}^\mu\alpha_{n\mu}+\tilde{\alpha}_{-n}^\mu\tilde{\alpha}_{n\mu})}
e^{-(-1)^n\alpha_{-n}^\mu\tilde{\alpha}_{-n\mu}/n}
|0\rangle\right)
\end{multline}
(no summation over $\mu$ in the last line). We have used the expressions
(4.3.17) for the adjoints of the raising and lowering operators. The
zero-mode amplitudes are independent of $s$, so we won't bother with
them. The exponentials of the ghost raising operators truncate after the
second term:
\begin{align}
e^{-(-1)^nb_{-n}\tilde{c}_{-n}}|0\rangle &= 
|0\rangle-(-1)^nb_{-n}\tilde{c}_{-n}|0\rangle, \\
\langle0|e^{-(-1)^nc_n\tilde{b}_n} &=
\langle0|-(-1)^n\langle0|c_n\tilde{b}_n.
\end{align}
Both terms in (99) are eigenstates of
$b_{-n}c_n+\tilde{c}_{-n}\tilde{b}_n$, with eigenvalues of 0 and 2
respectively. The first ghost amplitude is thus:
\begin{equation}
\left(\langle0|-(-1)^n\langle0|c_n\tilde{b}_n\right)
\Big(|0\rangle-e^{-2sn}(-1)^nb_{-n}\tilde{c}_{-n}|0\rangle\Big) =
1-e^{-2sn}.
\end{equation}
The second ghost amplitude gives the same result. To evaluate the scalar
amplitudes, we must expand out the $|C\rangle$ exponential:
\begin{equation}
e^{-(-1)^n\alpha_{-n}^\mu\tilde{\alpha}_{n\mu}/n}|0\rangle =
\sum_{m=0}^\infty
\frac{1}{m!n^m}(-1)^{(n+1)m}(\alpha_{-n}^\mu\tilde{\alpha}_{n\mu})^m
|0\rangle.
\end{equation}
Each term in the series is an eigenstate of
$\alpha_{-n}^\mu\alpha_{n\mu}+\tilde{\alpha}_{-n}^\mu\tilde{\alpha}_{n\mu}$,
with eigenvalue $2nm$, and each term has unit norm, so the amplitude is
\begin{equation}
\sum_{m=0}^\infty e^{-2snm} = \frac{1}{1-e^{-2sn}}.
\end{equation}
Finally, we find that the total amplitude (98) is proportional to
\begin{equation}
e^{2s}\prod_{n=1}^\infty(1-e^{-2sn})^{-24} =
\eta(is/\pi)^{-24}.
\end{equation}
This is the $s$-dependent part of the Klein bottle vacuum amplitude, and
agrees with the integrand of (7.4.19).

The vacuum amplitude for the M\"obius strip is
\begin{equation}
\int_0^\infty ds\langle B|c_0b_0e^{-s(L_0+\tilde{L}_0)}|C\rangle.
\end{equation}
The only difference from the above analysis is the absence of the factor
$(-1)^n$ multiplying the bras. Thus the ghost amplitude (101) becomes
instead
\begin{equation}
\left(\langle0|-\langle0|c_n\tilde{b}_n\right)
\Big(|0\rangle-e^{-2sn}(-1)^nb_{-n}\tilde{c}_{-n}|0\rangle\Big) =
1-(-1)^ne^{-2sn},
\end{equation}
while the scalar amplitude (103) becomes
\begin{equation}
\sum_{m=0}^\infty(-1)^{nm}e^{-2snm} = \frac{1}{1-(-1)^ne^{-2sn}}.
\end{equation}
The total amplitude is then
\begin{align}
e^{2s}\prod_{n=1}^\infty\left(1-(-1)^ne^{-2sn}\right)^{-24} &=
e^{2s}\prod_{n=1}^\infty\left(1+e^{-4s(n-1/2)}\right)^{-24}
\prod_{n=1}^\infty\left(1-e^{-4sn}\right)^{-24} \notag \\
&= \vartheta_{00}(0,2is/\pi)^{-12}\eta(2is/\pi)^{-12},
\end{align}
in agreement with the integrand of (7.4.23).

\setcounter{equation}{0}
\newpage

\section{Chapter 8}

\subsection{Problem 8.1}

\paragraph{(a)}

The spatial world-sheet coordinate $\sigma^1$ should be chosen in the range
$-\pi<\sigma^1<\pi$ for (8.2.21) to work. In other words, define
$\sigma^1 = -\I\ln z$ (with the branch cut for the logarithm on the
negative real axis). The only non-zero commutators involved are
\begin{equation}
[x_L,p_L] = i, \qquad [\alpha_m,\alpha_n] = m\delta_{m,-n}.
\end{equation}
Hence 
\begin{align}
[X_L(z_1),X_L(z_2)] \notag \\
=& -i\frac{\alpha'}{2}\ln z_2\,[x_L,p_L]
-i\frac{\alpha'}{2}\ln z_1\,[p_L,x_L]
-\frac{\alpha'}{2}\sum_{m,n\neq 0}\frac{[\alpha_m,\alpha_n]}{mnz_1^mz_2^n}
\notag \\
=& \frac{\alpha'}{2}
\left(
\ln z_2 - \ln z_1 - \sum_{n\neq0}\frac{1}{n}\left(\frac{z_1}{z_2}\right)^n
\right) \notag \\
=& \frac{\alpha'}{2}
\left(
\ln z_2 - \ln z_1 
+ \ln\left(1-\frac{z_1}{z_2}\right)-\ln\left(1-\frac{z_2}{z_1}\right)
\right) \notag \\
=& \frac{\alpha'}{2}
\left(
\ln z_2 - \ln z_1 
+ \ln\left(\frac{1-\frac{z_1}{z_2}}{1-\frac{z_2}{z_1}}\right)
\right) \notag \\
=& \frac{\alpha'}{2}
\left(
\ln z_2 - \ln z_1 
+ \ln\left(-\frac{z_1}{z_2}\right)
\right) \notag \\
=& \frac{i\alpha'}{2}
\left(\sigma^1_1-\sigma^1_2+(\sigma_2^1-\sigma_1^1\pm\pi)\right).
\end{align}
Because of where we have chosen to put the branch cut for the logarithm,
the quantity in the inner parentheses must be between $-\pi$ and $\pi$. The
upper sign is therefore chosen if $\sigma_1^1>\sigma_2^1$, and the lower
otherwise. (Note that the fourth equality is legitimate because the
arguments of both
$1-\frac{z_1}{z_2}$ and $1-\frac{z_2}{z_1}$ are in the range
$(-\pi/2,\pi/2)$.)

\paragraph{(b)}

Inspection of the above derivation shows that
\begin{equation}
[X_R(z_1),X_R(z_2)] = -\frac{\pi i\alpha'}{2}\sign(\sigma_1^1-\sigma_2^1).
\end{equation}
The CBH formula tells us that, for two operators $A$ and $B$ whose
commutator is a scalar,
\begin{equation}
e^Ae^B = e^{[A,B]}e^Be^A.
\end{equation}
In passing $\mathcal{V}_{k_Lk_R}(z,\bar{z})$ through
$\mathcal{V}_{k'_Lk'_R}(z',\bar{z}')$, (4) will give signs from several
sources. The factors from the cocyles commuting past the operators
$e^{ik_Lx_L+ik_Rx_R}$ and $e^{ik'_Lx_L+ik'_Rx_R}$ are given in
(8.2.23). This is cancelled by the factor from commuting the normal ordered
exponentials past each other:
\begin{equation}
e^{-(k_Lk'_L-k_Rk'_R)\pi i\alpha'\sign(\sigma^1-{\sigma^1}')/2}
= (-1)^{nw'+n'w}.
\end{equation}

\subsection{Problem 8.3}

\paragraph{(a)}

In the sigma-model action, we can separate $X^{25}$ from the other scalars,
which we call $X^\mu$:
\begin{multline}
S_\sigma = 
\frac{1}{4\pi\alpha'}\int_Md^2\!\sigma\,g^{1/2}
\left(\left(g^{ab}G_{\mu\nu}+i\epsilon^{ab}B_{\mu\nu}\right)
\partial_aX^\mu\partial_bX^\nu \right. \\
\left.+2\left(g^{ab}G_{25\mu}+i\epsilon^{ab}B_{25\mu}\right)
\partial_aX^{25}\partial_bX^\mu
+g^{ab}G_{25,25}\partial_aX^{25}\partial_bX^{25}\right).
\end{multline}
(Since we won't calculate the shift in the dilaton, we're setting aside the
relevant term in the action.)

\paragraph{(b)}

We can gauge the $X^{25}$ translational symmetry by introducing a
worldsheet gauge field $A_a$:
\begin{multline}
S'_\sigma = 
\frac{1}{4\pi\alpha'}\int_Md^2\!\sigma\,g^{1/2}
\left(\left(g^{ab}G_{\mu\nu}+i\epsilon^{ab}B_{\mu\nu}\right)
\partial_aX^\mu\partial_bX^\nu \right. \\
{}+ 2\left(g^{ab}G_{25\mu}+i\epsilon^{ab}B_{25\mu}\right)
(\partial_aX^{25}+A_a)\partial_bX^\mu \\
\left.{} + g^{ab}G_{25,25}(\partial_aX^{25}+A_a)(\partial_bX^{25}+A_b)\right).
\end{multline}
This action is invariant under $X^{25}\to X^{25}+\lambda$, $A_a\to
A_a-\partial_a\lambda$. In fact, it's consistent to allow the gauge
parameter $\lambda(\sigma)$ to be periodic with the same periodicity as
$X^{25}$ (making the gauge group a compact U(1)). This periodicity will 
be necessary later, to allow us to unwind the string.

\paragraph{(c)}

We now add a Lagrange multiplier term to the action,
\begin{multline}
S''_\sigma = 
\frac{1}{4\pi\alpha'}\int_Md^2\!\sigma\,g^{1/2}
\left(\left(g^{ab}G_{\mu\nu}+i\epsilon^{ab}B_{\mu\nu}\right)
\partial_aX^\mu\partial_bX^\nu\right. \\
{}+ 2\left(g^{ab}G_{25\mu}+i\epsilon^{ab}B_{25\mu}\right)
(\partial_aX^{25}+A_a)\partial_bX^\mu \\
\left. {} + g^{ab}G_{25,25}(\partial_aX^{25}+A_a)(\partial_bX^{25}+A_b) 
+i\phi\epsilon^{ab}(\partial_aA_b-\partial_bA_a)\right).
\end{multline}
Integrating over $\phi$ forces $\epsilon^{ab}\partial_aA_b$ to vanish,
which on a topologically trivial worldsheet means that $A_a$ is
gauge-equivalent to 0, bringing us back to the action (6). Of course,
there's not much point in making $X^{25}$ periodic on a topologically
trivial worldsheet, and on a non-trivial worldsheet the gauge field may
have non-zero holonomies around closed loops. In order for these to be
multiples of $2\pi R$ (and therefore removable by a gauge transformation),
$\phi$ must also be periodic (with period $2\pi/R$). For details, see Rocek
and Verlinde (1992).

\paragraph{(d)}

Any $X^{25}$ configuration is gauge equivalent to $X^{25}=0$, this
condition leaving no additional gauge degrees of freedom. The action,
after performing an integration by parts (and ignoring the holonomy
issue) is
\begin{multline}
S'''_\sigma = 
\frac{1}{4\pi\alpha'}\int_Md^2\!\sigma\,g^{1/2}
\left(\left(g^{ab}G_{\mu\nu}+i\epsilon^{ab}B_{\mu\nu}\right)
\partial_aX^\mu\partial_bX^\nu + G_{25,25}g^{ab}A_aA_b \right. \\
\left. {}+ 2\left(G_{25\mu}g^{ab}\partial_bX^\mu
+ iB_{25\mu}\epsilon^{ab}\partial_bX^\mu
+ i\epsilon^{ab}\partial_b\phi\right)A_a \right).
\end{multline}
We can complete the square on $A_a$,
\begin{multline}
S'''_\sigma = 
\frac{1}{4\pi\alpha'}\int_Md^2\!\sigma\,g^{1/2}
\Big(\left(g^{ab}G_{\mu\nu}+i\epsilon^{ab}B_{\mu\nu}\right)
\partial_aX^\mu\partial_bX^\nu \\
\left. {} + G_{25,25}g^{ab}(A_a+G_{25,25}^{-1}B_a)(A_b+G_{25,25}^{-1}B_b)
-G_{25,25}^{-1}g^{ab}B_aB_b
\right),
\end{multline}
where
$B^a=G_{25\mu}g^{ab}\partial_bX^\mu+iB_{25\mu}\epsilon^{ab}\partial_bX^\mu+i\epsilon^{ab}\partial_b\phi$.
Integrating over $A_a$, and ignoring the result, and using the fact that in
two dimensions $g_{ac}\epsilon^{ab}\epsilon^{cd}=g_{bd}$, we have
\begin{multline}
S_\sigma = 
\frac{1}{4\pi\alpha'}\int_Md^2\!\sigma\,g^{1/2}
\left(\left(g^{ab}G'_{\mu\nu}+i\epsilon^{ab}B'_{\mu\nu}\right)
\partial_aX^\mu\partial_bX^\nu \right. \\
\left.+2\left(g^{ab}G'_{25\mu}+i\epsilon^{ab}B'_{25\mu}\right)
\partial_a\phi\partial_bX^\mu
+g^{ab}G'_{25,25}\partial_a\phi\partial_b\phi\right),
\end{multline}
where
\begin{align}
G'_{\mu\nu} &= 
G_{\mu\nu}
-G_{25,25}^{-1}G_{25\mu}G_{25\nu}+G_{25,25}^{-1}B_{25\mu}B_{25\nu}, \notag \\
B'_{\mu\nu} &= B_{\mu\nu}
-G_{25,25}^{-1}G_{25\mu}B_{25\nu}+G_{25,25}^{-1}B_{25\mu}G_{25\nu}, \notag \\
G'_{25\mu} &= G_{25,25}^{-1}B_{25\mu}, \notag \\
B'_{25\mu} &= G_{25,25}^{-1}G_{25\mu}, \notag \\
G'_{25,25} &= G_{25,25}^{-1}.
\end{align}
Two features are clearly what we expect to occur upon T-duality: the
inversion of $G_{25,25}$, and the exchange of $B_{25\mu}$ with $G_{25\mu}$,
reflecting the fact that winding states, which couple to the former, are
exchanged with compact momentum states, which couple to the latter.

\subsection{Problem 8.4}

The generalization to $k$ dimensions of the Poisson resummation formula
(8.2.10) is
\begin{equation}
\sum_{n\in Z^k}\exp\left(-\pi a^{mn}n_mn_n+2\pi ib^nn_n\right) =
(\det a_{mn})^{1/2}\sum_{m\in Z^k}
\exp\left(-\pi a_{mn}(m^m-b^m)(m^n-b^n)\right),
\end{equation}
where $a^{mn}$ is a symmetric matrix and $a_{mn}$ is its inverse. This can
be proven by induction using (8.2.10).
The Virasoro generators for the compactified $X$s are
\begin{gather}
L_0 =
\frac1{4\alpha'}v_L^2 + 
\sum_{n=1}^\infty\alpha_{-n}\cdot\alpha_n, \\
\tilde L_0 =
\frac1{4\alpha'}v_R^2 + 
\sum_{n=1}^\infty\tilde\alpha_{-n}\cdot\tilde\alpha_n,
\end{gather}
where products of vectors are taken with the metric $G_{mn}$. The partition
function is
\begin{equation}
(q\bar q)^{-1/24}\Tr\left(q^{L_0}\bar q^{\tilde L_0}\right) =
|\eta(\tau)|^{-2k}\sum_{n,w_1\in Z^k}\exp\left[
-\frac{\pi\tau_2}{\alpha'}(v^2+R^2w_1^2)+2\pi i\tau_1n\cdot w_1
\right].
\end{equation}
Using (13) we now get
\begin{equation}
V_kZ_X(\tau)^k\sum_{w_1,w_2\in Z^k}\exp\left[
-\frac{\pi R^2}{\alpha'\tau_2}|w_2-\tau w_1|^2-2\pi ib_{mn}w_1^nw_2^m
\right].
\end{equation}
This includes the expected phase factor (8.2.12)---but unfortunately with
the wrong sign! The volume factor $V_k=R^k(\det G_{mn})^{1/2}$ comes from
the integral over the zero-mode.

\subsection{Problem 8.5}

\paragraph{(a)}

With $l \equiv (n/r+mr/2,n/r-mr/2)$, we have
\begin{equation}
l\circ l' = nm'+n'm.
\end{equation}
Evenness of the lattice is obvious. The dual lattice is generated by the
vectors $(n,m)=(1,0)$ and $(0,1)$, which also generate the original
lattice; hence it is self-dual.

\paragraph{(b)}

With
\begin{equation}
l \equiv 
\frac{1}{\sqrt{2\alpha'}}(v+wR,v-wR),
\end{equation}
one can easily calculate
\begin{equation}
l \circ l' = n\cdot w'+n'\cdot w
\end{equation}
(in particular, $B_{mn}$ drops out). Again, at this point it is more or
less obvious that the lattice is even and self-dual.

\subsection{Problem 8.6}

The metric (8.4.37) can be written
\begin{equation}
G_{mn} = \frac{\alpha'\rho_2}{R^2}M_{mn}(\tau), \qquad
M(\tau) =\frac1{\tau_2}
\begin{pmatrix} 1 & \tau_1 \\ \tau_1 & |\tau|^2 \end{pmatrix}.
\end{equation}
Thus
\begin{equation}
G^{mn}\partial_\mu G_{np} =
\rho_2^{-1}\partial_\mu\rho_2\delta^m{}_p + (M^{-1}\partial_\mu M)^m{}_p.
\end{equation}
The decoupling between $\tau$ and $\rho$ is due to the fact that the
determinant of $M$ is constant (in fact it's 1), so that
$M^{-1}\partial_\mu M$ is traceless:
\begin{equation}
G^{mn}G^{pq}\partial_\mu G_{mp}\partial^\mu G_{nq} =
\frac{2\partial_\mu\rho_2\partial^\mu\rho_2}{\rho_2^2} +
\Tr(M^{-1}\partial_\mu M)^2.
\end{equation}
With a little algebra the second term can be shown to equal
$2\partial_\mu\tau\partial^\mu\bar\tau/\tau_2^2$. Meanwhile, the
antisymmetry of $B$ implies that
$G^{mn}G^{pq}\partial_\mu B_{mp}\partial^\mu B_{nq}$ essentially
calculates the determinant of the inverse metric $\det G^{mn} = 
(R^2/\alpha'\rho_2)^2$, multiplying it by
$2\partial_\mu B_{24,25}\partial^\mu B_{24,25} = 
2(\alpha'/R^2)^2\partial_\mu\rho_1\partial^\mu\rho_1$. Adding this to (23)
we arrive at (twice) (8.4.39).

If $\tau$ and $\rho$ are both imaginary, then $B=0$ and the torus is
rectangular with proper radii
\begin{gather}
R_{24} = \sqrt{G_{24,24}}R = \sqrt\frac{\alpha'\rho_2}{\tau_2}, \\
R_{25} = \sqrt{G_{25,25}}R = \sqrt{\alpha'\rho_2\tau_2}.
\end{gather}
Clearly switching $\rho$ and $\tau$ is a T-duality on $X^{24}$, while
$\rho\to-1/\rho$ is T-duality on both $X^{24}$ and $X^{25}$ combined with
$X^{24}\leftrightarrow X^{25}$.

\subsection{Problem 8.7}

\paragraph{(a)}

Since we have already done this problem for the case $p=25$ in problem
6.9(a), we can simply adapt the result from that problem (equation (46) in
the solutions to chapter 6) to general $p$. (In this case, the Chan-Paton
factors are trivial, and we must include contributions from all three
combinations of polarizations.) The open string coupling $g_{{\rm o},p}$
depends on $p$, and we can compute it either by T-duality or by comparing
to the low-energy action (8.7.2).

Due to the Dirichlet boundary conditions, there is no zero mode in the path
integral and therefore no momentum-conserving delta function in those
directions. Except for this fact, the three-tachyon and Veneziano
amplitudes calculated in section 6.4 go through unchanged, so we have
\begin{equation}
C_{D_2,p} = \frac{1}{\alpha'g_{{\rm o},p}^2}.
\end{equation}
The four-ripple amplitude is
\begin{multline}
S = \frac{2ig_{{\rm o},p}^2}{\alpha'}
(2\pi)^{p+1}\delta^{p+1}(\sum_ik_i) \\
\times\left(
e_1\cdot e_2 e_3\cdot e_4 F(t,u)+
e_1\cdot e_3 e_2\cdot e_4 F(s,u)+
e_1\cdot e_4 e_2\cdot e_3 F(s,t)
\right),
\end{multline}
where
\begin{multline}
F(x,y) \equiv \\ 
B(-\alpha'x+1,\alpha'x+\alpha'y-1) +
B(-\alpha'y+1,\alpha'x+\alpha'y-1) +
B(-\alpha'x+1,-\alpha'y+1).
\end{multline}

If we had instead obtained this amplitude by T-dualizing the answer to
problem 6.9(a), using the fact that $\kappa$, and
therefore $g_{{\rm o},25}^2$, transform according to (8.3.30), we would have
found the same result with $g_{{\rm o},p}^2$ replaced with 
$g_{{\rm o},25}^2/(2\pi \sqrt{\alpha'})^{25-p}$, so we find
\begin{equation}
g_{{\rm o},p} = \frac{g_{{\rm o},25}}{(2\pi \sqrt{\alpha'})^{(25-p)/2}}.
\end{equation}

\paragraph{(b)}

To examine the Regge limit, let us re-write the amplitude (27) in the
following way:
\begin{multline}
S = \frac{2ig_{{\rm o},p}^2}{\alpha'}
(2\pi)^{p+1}\delta^{p+1}(\sum_ik_i)
\left(
1-\cos\pi\alpha't+\tan\frac{\pi\alpha's}{2}\sin\pi\alpha't
\right) \\
\times\left(
e_1\cdot e_2 e_3\cdot e_4
\frac{\Gamma(\alpha's+\alpha't+1)\Gamma(-\alpha't+1)}{\Gamma(\alpha's+2)}
+ e_1\cdot e_3 e_2\cdot e_4 
\frac{\Gamma(\alpha's+\alpha't+1)\Gamma(-\alpha't-1)}{\Gamma(\alpha's)}
\right. \\
\left. {} + e_1\cdot e_4 e_2\cdot e_3
\frac{\Gamma(\alpha's+\alpha't-1)\Gamma(-\alpha't+1)}{\Gamma(\alpha's)}
\right).
\end{multline}
The factor with the sines and cosines gives a pole wherever $\alpha's$ is
an odd integer, while the last factor gives the overall behavior in the
limit $s\to\infty$. In that limit the coefficients of $e_1\cdot e_2e_3\cdot
e_4$ and $e_1\cdot e_4e_2\cdot e_3$ go like
$s^{\alpha't-1}\Gamma(-\alpha't+1)$, while the coefficient of $e_1\cdot
e_3e_2\cdot e_4$ goes like $s^{\alpha't+1}\Gamma(-\alpha't-1)$, and
therefore dominates (unless $e_1\cdot e_3$ or $e_2\cdot e_4$ vanishes). We
thus have Regge behavior.

For hard scattering, the amplitude (27) has the same exponential falloff
(6.4.19) as the Veneziano amplitude, since the only differences are shifts
of 2 in the arguments of some of the gamma functions, which will not affect
their asymptotic behavior.

\paragraph{(c)}

Expanding $F(x,y)$ for small $\alpha'$, fixing $x$ and $y$, we find (with
some assistance from {\it Mathematica}) that the leading term is quadratic:
\begin{equation}
F(x,y) = -\frac{\pi^2{\alpha'}^2}{2}xy + O({\alpha'}^3).
\end{equation}
Hence the low energy limit of the amplitude (27) is
\begin{equation}
S \approx
-i\pi^2\alpha'g_{{\rm o},p}^2
(2\pi)^{p+1}\delta^{p+1}(\sum_ik_i)
\left(
e_1\cdot e_2 e_3\cdot e_4 tu +
e_1\cdot e_3 e_2\cdot e_4 su +
e_1\cdot e_4 e_2\cdot e_3 st
\right).
\end{equation}

The D-brane is embedded in a flat spacetime, $G_{\mu\nu}=\eta_{\mu\nu}$,
with vanishing $B$ and $F$ fields and constant dilaton. We use a coordinate
system on the brane $\xi^a = X^a$, $a=0,\dots,p$, so the induced metric is
\begin{equation}
G_{ab} = \eta_{ab}+\partial_aX^m\partial_bX^m,
\end{equation}
where the fields $X^m$, $m=p+1,\dots,25$, are the fluctuations in the
transverse position of the brane, whose scattering amplitude we wish to
find. Expanding the action (8.7.2) to quartic order in the fluctuations, we
can use the formula
\begin{equation}
\det(I+A) = 1 + \Tr A + \frac{1}{2}(\Tr A)^2 - \frac{1}{2}\Tr A^2 + O(A^3),
\end{equation}
to find
\begin{equation}
\mathbf{S}_p = -\tau_p \int d^{p+1}\!\xi
\left(1+\frac{1}{2}\eta^{ab}\partial_aX^m\partial_bX^m
+\frac{1}{8}(\eta^{ab}\eta^{cd}-2\eta^{ac}\eta^{bd})
\partial_aX^m\partial_bX^m\partial_cX^n\partial_dX^n
\right).
\end{equation}
The fields $X^m$ are not canonically normalized, and the coupling constant
is in fact
\begin{equation}
\frac{1}{8\tau_p} = \frac{\pi^2\alpha'g_{{\rm o},p}^2}{4}
\end{equation}
where we have used (6.6.18), (8.7.26), and (8.7.28), and (24). All the ways
of contracting four $X^m$s with the interaction term yield
\begin{equation}
e_1\cdot e_2 e_3\cdot e_4
(8k_1\cdot k_2k_3\cdot k_4-8k_1\cdot k_3k_2\cdot k_4
-8k_1\cdot k_4k_2\cdot k_3)
= 4e_1\cdot e_2 e_3\cdot e_4tu
\end{equation}
plus similar terms for $e_1\cdot e_3 e_2\cdot e_4$ and 
$e_1\cdot e_4 e_2\cdot e_3$. Multiplying this by the coupling constant
(30), and a factor $-i(2\pi)^{p+1}\delta^{p+1}(\sum_ik_i)$, yields
precisely the amplitude (32), showing that the two ways of calculating
$g_{{\rm o},p}$ agree.

\subsection{Problem 8.9}

\paragraph{(a)}

There are two principal changes in the case of Dirichlet boundary
conditions from the derivation of the disk expectation value (6.2.33):
First, there is no zero mode, so there is no momentum-space delta
function (this corresponds to the fact that the D-brane breaks translation
invariance in the transverse directions and therefore does not conserve
momentum). Second, the image charge in the Green's function has the
opposite sign:
\begin{equation}
G'_{\rm D}(\sigma_1,\sigma_2) =
-\frac{\alpha'}{2}\ln|z_1-z_2|^2+\frac{\alpha'}{2}\ln|z_1-\bar z_2|^2.
\end{equation}
Denoting the parts of the momenta parallel and perpendicular to the D-brane
by $k$ and $q$ respectively, the expectation value becomes
\begin{multline}
\left\langle
\prod_{i=1}^n
:e^{i(k_i+q_i)\cdot X(z_i,\bar z_i)}:
\right\rangle_{D_2,p}
= \\
iC^X_{D_2,p}(2\pi)^{p+1}\delta^{p+1}(\sum_i k_i)
\prod_{i=1}^n|z_i-\bar z_i|^{\alpha'(k_i^2-q_i^2)/2}
\prod_{i<j}|z_i-z_j|^{\alpha'(k_i\cdot k_j+q_i\cdot q_j)}
|z_i-\bar z_j|^{\alpha'(k_i\cdot k_j-q_i\cdot q_j)}.
\end{multline}

\paragraph{(b)}

For expectation values including operators $\partial_aX^M$ in the interior,
one performs the usual contractions, but using the Green's function (38)
rather than (6.2.32) for the Dirichlet directions.

\subsection{Problem 8.11}

\paragraph{(a)}

The disk admits three real CKVs. Fixing the position of one of the
closed-string vertex operators eliminates two of these, leaving the
one which generates rotations about the fixed operator. We can
eliminate this last CKV by integrating the second vertex operator along a
line connecting the fixed vertex operator to the edge of the disk. The
simplest way to implement this on the upper half-plane is by fixing $z_2$
on the positive imaginary axis and integrating $z_1$ from 0 to $z_2$. The
amplitude is
\begin{equation}
S =
g_{\rm c}^2e^{-\lambda}\int_0^{z_2}dz_1
\left\langle
:c^1e^{i(k_1+q_1)\cdot X}(z_1,\bar z_1):
:c\tilde ce^{i(k_2+q_2)\cdot X}(z_2,\bar z_2):
\right\rangle_{D_2,p},
\end{equation}
where, as in problem 8.9, $k$ and $q$ represent the momenta parallel and
perpendicular to the D-brane respectively. The ghost path integral is
\begin{align}
\left\langle
\frac{1}{2}(c(z_1)+\tilde c(\bar z_1))c(z_2)\tilde c(\bar z_2)
\right\rangle_{D_2} & =
\frac{C^{\rm g}_{D_2}}{2}
\left((z_1-z_2)(z_1-\bar z_2)+(\bar z_1-z_2)(\bar z_1-\bar z_2)\right)
(z_2-\bar z_2) \notag \\
&= 2C^{\rm g}_{D_2}(z_1-z_2)(z_1+z_2)z_2.
\end{align}
To evaluate the $X$ path integral we use the result of problem 8.9(a):
\begin{align}
& \left\langle
:e^{i(k_1+q_1)\cdot X(z_1,\bar z_1)}::e^{i(k_2+q_2)\cdot X(z_2,\bar z_2)}:
\right\rangle_{D_2,p} \notag \\
& = iC_{D_2,p}^X(2\pi)^{p+1}\delta^{p+1}(k_1+k_2) \notag \\
& \qquad\qquad\qquad
\times |z_1-\bar z_1|^{\alpha'(k_1^2-q_1^2)/2}
|z_2-\bar z_2|^{\alpha'(k_2^2-q_2^2)/2}
|z_1-z_2|^{\alpha'(k_1\cdot k_2+q_1\cdot q_2)}
|z_1-\bar z_2|^{\alpha'(k_1\cdot k_2-q_1\cdot q_2)} \notag \\
& = iC_{D_2,p}^X(2\pi)^{p+1}\delta^{p+1}(k_1+k_2) \notag \\
& \qquad\qquad\qquad
\times 2^{2\alpha'k^2-4}|z_1|^{\alpha'k^2-2}|z_2|^{\alpha'k^2-2}
|z_1-z_2|^{-\alpha's/2-4}|z_1+z_2|^{\alpha's/2-2k^2+4}.
\end{align}
We have used the kinematic relations $k_1+k_2=0$ and
$(k_1+q_1)^2=(k_2+q_2)^2=4/\alpha'$, and defined the parameters
\begin{gather}
k^2 \equiv k_1^2 = k_2^2 = 4-q_1^2 = 4-q_2^2, \notag \\
s \equiv -(q_1+q_2)^2.
\end{gather}
So we have
\begin{align}
S &=
- g_{\rm c}^2C_{D_2,p}(2\pi)^{p+1}\delta^{p+1}(k_1+k_2) \notag \\
&\qquad\qquad\qquad\qquad
\times2^{2\alpha'k^2-3}|z_2|^{\alpha'k^2-1}
\int_0^{z_2}dz_1
|z_1|^{\alpha'k^2-2}|z_1-z_2|^{-\alpha's/2-3}|z_1+z_2|^{\alpha's/2-2k^2+5}
\notag \\
&= -ig_{\rm c}^2C_{D_2,p}(2\pi)^{p+1}\delta^{p+1}(k_1+k_2)
2^{2\alpha'k^2-3}
\int_0^1dx\,x^{\alpha'k^2-2}(1-x)^{-\alpha's/2-3}(1+x)^{\alpha's/2-2k^2+5}
\notag \\
&= -i\frac{\pi^{3/2}(2\pi\sqrt{\alpha'})^{11-p}g_{\rm c}}{32}
(2\pi)^{p+1}\delta^{p+1}(k_1+k_2)
B(\alpha'k^2-1,-\alpha's/4-1).
\end{align}
In the last line we have used (26), (29), and (8.7.28) to calculate
$g_{\rm c}^2C_{D_2,p}$.

\paragraph{(b)}

In the Regge limit we increase the scattering energy, $k^2\to-\infty$,
while holding fixed the momentum transfer $s$. As usual, the beta function
in the amplitude give us Regge behavior:
\begin{equation}
S \sim
(-k^2)^{\alpha's/4+1}\Gamma(-\alpha's/4-1).
\end{equation}
In the hard scattering limit we again take $k^2\to-\infty$, but this time
fixing $k^2/s$. As in the Veneziano amplitude, the beta function gives
exponential behavior in this limit.

\paragraph{(c)}

The beta function in the amplitude has poles at $\alpha'k^2=1,0,-1,\dots$,
representing on-shell intermediate open strings on the D-brane: the closed
string is absorbed and then later re-emitted by the D-brane. These poles
come from the region of the integral in (44) where $z_1$ approaches the
boundary.

Note that there are also poles at $s=0$ and $s=-4$ (the poles at positive
$s$ are kinematically forbidden), representing an on-shell intermediate
closed string: the tachyon ``decays'' into another tachyon and either a
massless or a tachyonic closed string, and the latter is then absorbed
(completely, without producing open strings) by the D-brane. These poles
come from the region of the integral in (44) where $z_1$ approaches $z_2$.

\setcounter{equation}{0}
\newpage

\section{Appendix A}

\subsection{Problem A.1}

\paragraph{(a)}
We proceed by the same method as in the example on pages 339-341. Our
orthonormal basis for the periodic functions on $[0,U]$ will be:
\bea
f_0(u) &=& \frac{1}{\sqrt{U}},\nonumber \\
f_j(u) &=& \sqrt{\frac{2}{U}}\cos\frac{2\pi ju}{U}, \\
g_j(u) &=& \sqrt{\frac{2}{U}}\sin\frac{2\pi ju}{U},\nonumber
\eea
where $j$ runs over the positive integers. These are eigenfunctions of
$\Delta=-\partial_u^2+\omega^2$ with eigenvalues
\be
\lambda_j = \left(\frac{2\pi j}{U}\right)^2+\omega^2.
\ee
Hence (neglecting the counter-term action)
\bea
{\rm Tr}\exp(-\hat HU) &=& \int[dq]_{\rm P}\exp(-S_{\rm E})\nonumber \\
&=& \left(\det{}_{\rm P}\frac{\Delta}{2\pi}\right)^{-1/2}\nonumber \\
&=&
\sqrt{\frac{2\pi}{\lambda_0}}\prod_{j=1}^{\infty}\frac{2\pi}{\lambda_j}\nonumber \\
&=& \frac{\sqrt{2\pi}}{\omega}\prod_{j=1}^{\infty}\frac{2\pi U^2}{4\pi^2j^2+\omega^2U^2}.
\eea
This infinite product vanishes, so we regulate it by dividing by the same
determinant with $\omega\rightarrow\Omega$:
\be
\frac{\Omega}{\omega}\prod_{j=1}^{\infty}\frac{1+\left(\frac{\Omega U}{2\pi
j}\right)^2}{1+\left(\frac{\omega U}{2\pi j}\right)^2} =
\frac{\sinh\frac{1}{2}\Omega U}{\sinh\frac{1}{2}\omega U}.
\ee
For large $\Omega$ this becomes
\be
\frac{e^{\Omega U/2}}{2\sinh\frac{1}{2}\omega U};
\ee
the divergence can easily be cancelled with a counter-term Lagrangian
$L_{\rm ct}=\Omega/2$, giving
\be
{\rm Tr}\exp(-\hat HU) = \frac{1}{2\sinh\frac{1}{2}\omega U}.
\ee

The eigenvalues of $\hat H$ are simply
\be
E_i = (i+\frac{1}{2})\omega,
\ee
for non-negative integer $i$, so using (A.1.32) gives
\bea
{\rm Tr}\exp(-\hat HU) &=& \sum_{i=0}^{\infty}\exp(-E_iU)\nonumber \\
&=& e^{-\omega U/2}\sum_{i=0}^{\infty}e^{-i\omega U}\nonumber \\
&=& \frac{1}{2\sinh\frac{1}{2}\omega U}.
\eea
To be honest, the overall normalization of (6) must be obtained by
comparison with this result.

\paragraph{(b)}

Our basis for the anti-periodic functions on $[0,U]$ consists
of the eigenfunctions of $\Delta$,
\bea
f_j(u) &=& \sqrt{\frac{2}{U}}\cos\frac{2\pi(j+\frac{1}{2})u}{U}, \\
g_j(u) &=& \sqrt{\frac{2}{U}}\sin\frac{2\pi(j+\frac{1}{2})u}{U},\nonumber
\eea
where the $j$ are non-negative integers, with eigenvalues
\be
\lambda_j = \left(\frac{2\pi(j+\frac{1}{2})}{U}\right)^2+\omega^2.
\ee
Before including the counter-term and regulating, we have
\be
{\rm Tr}\left[\exp(-\hat HU)\hat R\right] = \prod_{j=0}^{\infty}\frac{2\pi
U^2}{(2\pi(j+\frac{1}{2}))^2+(\omega U)^2}.
\ee
After including the counter-term action and dividing by the regulator, this
becomes
\bea
e^{-L_{\rm ct}U}\prod_{j=0}^{\infty}\frac{1+\left(\frac{\Omega
U}{2\pi(j+\frac{1}{2})}\right)^2}{1+\left(\frac{\omega
U}{2\pi(j+\frac{1}{2})}\right)^2} &=& e^{-L_{\rm
ct}U}\frac{\cosh\frac{1}{2}\Omega U}{\cosh\frac{1}{2}\omega U}\nonumber \\
&\sim& \frac{e^{(\Omega/2-L_{\rm ct})U}}{2\cosh\frac{1}{2}\omega U},
\eea
so the answer is
\be
{\rm Tr}\left[\exp(-\hat HU)\hat R\right] =
\frac{1}{2\cosh\frac{1}{2}\omega U}.
\ee
This result can easily be reproduced by summing over eigenstates of $\hat
H$ with weight $(-1)^R$, since the even $i$ eigenstates are also even under
reflection, and odd $i$ eigenstates odd under reflection:
\bea
{\rm Tr}\left[\exp(-\hat HU)\hat R\right] &=&
\sum_{i=0}^\infty(-1)^ie^{-(j+1/2)\omega U}\nonumber \\
&=& \frac{1}{2\cosh\frac{1}{2}\omega U}.
\eea

\subsection{Problem A.3}

The action can be written
\bea
S &=& \frac{1}{4\pi\alpha'}\int d^2\sigma
X(-\partial_1^2-\partial_2^2+m^2)X\nonumber \\
&=& \frac{1}{2}\int d^2\sigma X\Delta X,
\eea
\be
\Delta = \frac{1}{2\pi\alpha'}(-\partial_1^2-\partial_2^2+m^2).
\ee
The periodic eigenfunctions of $\Delta$ can be given in a basis of products
of periodic eigenfunction of $-\partial_1^2$ on $\sigma_1$ with periodic
eigenfunctions of $-\partial_2^2$ on $\sigma_2$:
\bea
F_{jk}(\sigma_1,\sigma_2) &=& f_j(\sigma_1)g_k(\sigma_2), \\
f_0(\sigma_1) &=& \frac{1}{\sqrt{2\pi}},\nonumber \\
f_j(\sigma_1) &=&
\frac{1}{\sqrt{\pi}}\left\{\begin{array}{c}\sin\\\cos\end{array}\right\}j\sigma_1,\quad
j=1,2,\dots,\nonumber \\
g_0(\sigma_2) &=& \frac{1}{\sqrt{T}},\nonumber \\
g_k(\sigma_2) &=&
\frac{1}{\sqrt{2T}}\left\{\begin{array}{c}\sin\\\cos\end{array}\right\}k\sigma_2,\quad
k=1,2,\dots.\nonumber
\eea
The eigenvalues are
\be
\lambda_{jk} = \frac{1}{2\pi\alpha'}\left(j^2+\left(\frac{2\pi
k}{T}\right)^2+m^2\right),
\ee
with multiplicity $n_jn_k$, where $n_0=1$ and $n_i=2$ ($i=1,2,\dots$). The
path integral is
\bea
\left(\det{}_{\rm P}\frac{\Delta}{2\pi}\right)^{-1/2} &=&
\prod_{j,k=0}^\infty\left(\frac{2\pi}{\lambda_{jk}}\right)^{n_jn_k/2}\nonumber
\\
&=&
\prod_{j=0}^\infty\left(\prod_{k=0}^{\infty}\left(\frac{2\pi}{\lambda_{jk}}\right)^{n_k/2}\right)^{n_j}\nonumber
\\
&=& 
\prod_{j=0}^\infty\left(\frac{1}{2\sinh\frac{1}{2}\sqrt{j^2+m^2}T}\right)^{n_j},
\eea 
where in the last step we have used the result of problem A.1. The infinite
product vanishes; it would have to be regulated and a counter-term
introduced to extract the finite part.

\subsection{Problem A.5}

We assume that the Hamiltonian for this system is
\be
H = m\chi\psi.
\ee
The periodic trace is
\bea
{\rm Tr}\left[(-1)^{\hat F}\exp(-\hat HU)\right] &=& \int
d\psi\langle\psi,U|\psi,0\rangle_{\rm E}\nonumber \\
&=&
\int[d\chi d\psi]\exp\left[\int_0^Udu(-\chi\partial_u\psi-H)\right]\nonumber
\\
&=& \int[d\chi d\psi]\exp\left[\int_0^Udu\chi\Delta\psi\right],
\eea
where 
\be
\Delta=-\partial_u-m.
\ee
The periodic eigenfunctions of $\Delta$ on $[0,U]$ are
\be
f_j(u) = \frac{1}{\sqrt{U}}e^{2\pi iju/U},
\ee
while the eigenfunctions of $\Delta^T=\partial_u-m$ are
\be
g_j(u) = \frac{1}{\sqrt{U}}e^{-2\pi iju/U},
\ee
with $j$ running over the integers. Their eigenvalues are
\be
\lambda_j = -\left(\frac{2\pi ij}{U}-m\right),
\ee
so the trace becomes
\bea
\int[d\chi d\psi]\exp\left[\int_0^Udu\chi\Delta\psi\right] &=&
\prod_{j=-\infty}^\infty\lambda_j\nonumber \\
&=& \prod_{j=-\infty}^\infty-\left(\frac{2\pi ij}{U}+m\right)\nonumber \\
&=& -m\prod_{j=1}^\infty\left(\left(\frac{2\pi j}{U}\right)^2+m^2\right).
\eea
This is essentially the inverse of the infinite product that was considered
in problem A.1(a) (eq. (3)). Regulating and renormalizing in the same
manner as in that problem yields:
\be
{\rm Tr}\left[(-1)^{\hat F}\exp(-\hat HU)\right] = 2\sinh\frac{1}{2}mU.
\ee
This answer can very easily be checked by explicit calculation of the
LHS. The Hamiltonian operator can be obtained from the classical
Hamiltonian by first antisymmetrizing on $\chi$ and $\psi$:
\be
H = m\chi\psi = \frac{1}{2}m(\chi\psi-\psi\chi)
\ee
\be
\longrightarrow \hat H = \frac{1}{2}m(\hat\chi\hat\psi-\hat\psi\hat\chi).
\ee
Now
\bea
\hat H|\uparrow\rangle &=& \frac{1}{2}|\uparrow\rangle, \\
\hat H|\downarrow\rangle &=& -\frac{1}{2}|\downarrow\rangle,\nonumber
\eea
so, using A.2.22,
\be
{\rm Tr}\left[(-1)^{\hat F}\exp(-\hat HU)\right] = e^{mU/2}-e^{-mU/2},
\ee
in agreement with (27).

The anti-periodic trace is calculated in the same way, the only difference
being that the index $j$ runs over the half-integers rather than the
integers in order to make the eigenfunctions (23) and (24)
anti-periodic. Eq. (26) becomes
\be
\int[d\chi d\psi]\exp\left[\int_0^Udu\chi\Delta\psi\right] =
\prod_{j=1/2,3/2,\dots}^\infty\left(\left(\frac{2\pi
j}{U}\right)^2+m^2\right).
\ee
When we regulate the product, it becomes
\bea
\prod_j\frac{\left(\frac{2\pi
j}{U}\right)^2+m^2}{\left(\frac{2\pi j}{U}\right)^2+M^2} &=&
\frac{\prod_j\left(1+\left(\frac{mU}{2\pi
j}\right)^2\right)}{\prod_j\left(1+\left(\frac{MU}{2\pi
j}\right)^2\right)}\nonumber \\
&=& \frac{\cosh\frac{1}{2}mU}{\cosh\frac{1}{2}MU}.
\eea
With the same counter-term Lagrangian as before to cancel the divergence in
the denominator as $M\rightarrow\infty$, we are simply left with
\be
{\rm Tr}\exp(-\hat HU) = 2\cosh\frac{1}{2}mU,
\ee
the same as would be found using (30).

\setcounter{equation}{0}
\newpage

\section{Chapter 10}

\subsection{Problem 10.1}

\paragraph{(a)}

The OPEs are:
\begin{equation}
T_F(z)X^\mu(0) \sim -i\sqrt\frac{\alpha'}{2}\frac{\psi^\mu(0)}z,
\qquad
T_F(z)\psi^\mu(0) \sim i\sqrt\frac{\alpha'}2\frac{\partial X^\mu(0)}z.
\end{equation}

\paragraph{(b)}

The result follows trivially from (2.3.11) and the above OPEs.

\subsection{Problem 10.2}

\paragraph{(a)}

We have
\begin{equation}
\delta_{\eta_1}\delta_{\eta_2}X =
\delta_{\eta_1}(\eta_2\psi+\eta_2^*\tilde\psi) =
-\eta_2\eta_1\partial X-\eta_2^*\eta_1^*\bar\partial X,
\end{equation}
so, using the anti-commutativity of the $\eta_i$,
\begin{equation}
[\delta_{\eta_1},\delta_{\eta_2}]X =
2\eta_1\eta_2\partial X+2\eta_1^*\eta_2^*\bar\partial X =
\delta_vX
\end{equation}
(see (2.4.7)). For $\psi$ we must apply the equation of motion
$\partial\tilde\psi=0$:
\begin{equation}
\delta_{\eta_1}\delta_{\eta_2}\psi =
\delta_{\eta_1}(-\eta_2\partial X) =
-\eta_2\partial(\eta_1\psi+\eta_1^*\tilde\psi) =
-\eta_2\eta_1\partial\psi-\eta_2\partial\eta_1\psi,
\end{equation}
so
\begin{equation}
[\delta_{\eta_1},\delta_{\eta_2}]\psi =
-v\partial \psi-\frac12\partial v\,\psi =
\delta_v\psi,
\end{equation}
the second term correctly reproducing the weight of $\psi$. The
$\tilde\psi$ transformation works out similarly.

\paragraph{(b)}

For $X$:
\begin{align}
\delta_\eta\delta_vX &=
-v\eta\partial\psi-v\partial\eta\,\psi
-v^*\eta^*\bar\partial\tilde\psi-v^*(\partial\eta)^*\,\tilde\psi, \\
\delta_v\delta_\eta X &=
-v\eta\partial\psi-\frac12\eta\partial v\psi
-v^*\eta^*\bar\partial\tilde\psi-\frac12\eta^*(\partial v)^*\tilde\psi,
\end{align}
so
\begin{equation}
[\delta_\eta,\delta_v]X = \delta_{\eta'}X,
\end{equation}
where
\begin{equation}
\eta' = -v\partial\eta+\frac12\partial v\,\eta.
\end{equation}
For $\psi$,
\begin{align}
\delta_\eta\delta_v\psi &=
v\partial\eta\partial X+v\eta\partial^2X+\frac12\eta\partial v\partial X, \\
\delta_v\delta_\eta\psi &=
\eta v\partial^2X+\eta\partial v\partial X,
\end{align}
so
\begin{equation}
[\delta_\eta,\delta_v]\psi = \delta_{\eta'}\psi,
\end{equation}
and similarly for $\tilde\psi$.

\subsection{Problem 10.3}

\paragraph{(a)}

Since the OPE of $T_B^X=-(1/\alpha')\partial X^\mu\partial X_\mu$ and
$T_B^\psi = -(1/2)\psi^\mu\partial\psi_\mu$ is non-singular,
\begin{equation}
T_B(z)T_B(0) \sim T_B^X(z)T_B^X(0)+T_B^\psi(z)T_B^\psi(z),
\end{equation}
which does indeed reproduce (10.1.13a). We then have
\begin{equation}
\begin{split}
\frac{T_B(z)T_F(0)}{i\sqrt{2/\alpha'}} &\sim
\frac1{z^2}\partial X^\mu(z)\psi_\mu(0)
+ \frac{1}{2z}\partial^\mu(z)\partial\psi^\mu(z)\partial X_\mu(0)
+ \frac1{2z^2}\psi^\mu(z)\partial X_\mu(0) \\
&\sim \frac3{2z^2}\psi^\mu\partial X_\mu(0) + \psi^\mu\partial^2X_\mu(0)
+ \frac1z\partial\psi^\mu\partial X_\mu(0), \\
T_F(z)T_F(0) &\sim
\frac D{z^3}-\frac2{\alpha'z}\partial X_\mu(z)\partial X^\mu(0)
+ \frac1{z^2}\psi^\mu(z)\psi_\mu(0).
\end{split}
\end{equation}

\paragraph{(b)}

Again, there is no need to check the $T_BT_B$ OPE, since that is simply the
sum of the $X$ part and the $\psi$ part. The new terms in $T_B$ and $T_F$
add two singular terms to their OPE, namely
\begin{equation}
\begin{split}
V_\mu\partial^2X^\mu(z)i\sqrt\frac2{\alpha'}\psi^\nu\partial X_\nu(0)
+
\frac12&\psi^\mu\partial\psi_\mu(z)i\sqrt{2\alpha'}V_\nu\partial\psi^\nu(0) \\
&\sim
\frac{i\sqrt{2\alpha'}}{z^3}V_\mu\psi^\mu(0)
-i\sqrt\frac2{\alpha}\frac1{z^2}V_\mu\partial\psi^\mu(z)
-\frac{i\sqrt{2\alpha}}{z^3}V_\mu\psi^\mu(z) \\
&\sim
-3i\sqrt\frac{\alpha'}2\frac1{z^2}V_\mu\partial\psi^\mu(0)
-\frac{i\sqrt{2\alpha'}}zV_\mu\partial^2\psi^\mu(0),
\end{split}
\end{equation}
which are precisely the extra terms we expect on the right hand side of
(10.1.1b). The new terms in the $T_FT_F$ OPE are
\begin{equation}
\begin{split}
2\psi^\mu\partial X_\mu(z)V_\nu\partial\psi^\nu(0)
+2V_\mu\partial\psi^\mu(z)\psi^\nu&\partial X_\nu(0)
-2\alpha'V_\mu\partial\psi^\mu(z)V_\nu\partial\psi^\nu(0) \\
&\sim
\frac2{z^2}V_\mu\partial X^\mu(z)-\frac2{z^2}V_\mu\partial X^\mu(0)
+\frac{4\alpha'V^2}{z^3} \\
&\sim \frac2zV_\mu\partial^2X^\mu(0)+\frac{4\alpha'V^2}{z^3}.
\end{split}
\end{equation}

\subsection{Problem 10.4}

The $[L_m,L_n]$ commutator (10.2.11a) is as in the bosonic case. The
current associated with the charge $\{G_r,G_s\}$ is, according to (2.6.14),
\begin{equation}\begin{split}
\te{Res}_{z_1\to z_2}\,z_1^{r+1/2}T_F(z_1)z_2^{s+1/2}T_F(z_2) 
&= 
\te{Res}_{z_{12}\to0}\,z_2^{r+s+1}\left(1+\frac{z_{12}}{z_2}\right)^{r+1/2} 
\left(\frac{2c}{3z_{12}^3}+\frac2{z_{12}}T_B(z_2)\right) \\
&= \frac{(4r^2-1)c}{12}z_2^{r+s-1} + 2z_2^{r+s+1}T_B(z_2).
\end{split}\end{equation}
$\{G_r,G_s\}$ is in turn the residue of this expression in $z_2$, which is
easily seen to equal the RHS of (10.2.11b). Similarly, for $[L_m,G_r]$ we
have
\begin{equation}\begin{split}
\te{Res}_{z_1\to z_2}\,z_1^{m+1}T_B(z_1)z_2^{r+1/2}&T_F(z_2) \\
&= 
\te{Res}_{z_{12}\to0}\,z_2^{r+m+3/2}\left(1+\frac{z_{12}}{z_2}\right)^{m+1}
\left(\frac3{2z_{12}^2}T_F(z_2)+\frac1{z_{12}}\partial T_F(z_2)\right) \\
&= \frac{3(m+1)}2z_2^{r+m+1/2}T_F(z_2) + z_2^{r+m+3/2}\partial T_F(z_2).
\end{split}\end{equation}
The residue in $z_2$ of this is
\begin{equation}
\frac{3(m+1)}2G_{r+m} - \left(r+m+\frac32\right)G_{r+m},
\end{equation}
in agreement with (10.2.11c).

\subsection{Problem 10.5}

Let us denote  by $c_B$ the central charge appearing in the $T_BT_B$ OPE,
and by $c_F$ that appearing in the $T_FT_F$ OPE. One of the Jacobi identies
for the superconformal generators is, using (10.2.11),
\begin{equation}
\begin{split}
0 &= [L_m,\{G_r,G_s\}] + \{G_r,[G_s,L_m]\} - \{G_s,[L_m,G_r]\} \\
&= \frac16\left(
c_B(m^3-m) + \frac{c_F}4\left((2s-m)(4r^2-1)+(2r-m)(4s^2-1)\right)
\right)\delta_{m+r+s,0} \\
&= \frac16(c_B-c_F)(m^3-m)\delta_{m+r+s,0}.
\end{split}
\end{equation}
Hence $c_B=c_F$.

\subsection{Problem 10.7}

Taking (for example) $z_1$ to infinity while holding the other $z_i$
fixed at finite values, the expectation value (10.3.7) goes like
$z_1^{-1}$. Since $e^{i\epsilon_1H(z_1)}$ is a tensor of weight $1/2$,
transforming to the $u=1/z_1$ frame this expectation value becomes
constant, $O(1)$, in the limit $u\to0$. This is the correct behavior---the
only poles and zeroes of the expectation value should be at the positions
of the other operators. If we now consider some other function, with
exactly the same poles and zeroes and behavior as $z_1\to\infty$, the ratio
between this function and the one given in (10.3.7) would have to be an
entire function which approaches a constant as $z_1\to\infty$. But the only
such function is a constant, so (applying the same argument to the
dependence on all the $z_i$) the expression in (10.3.7) is unique up to a
constant.

\subsection{Problem 10.10}

On the bosonic side, the energy eigenvalue in terms of the momentum $k_L$
and oscillator occupation numbers $N_n$ is
\begin{equation}
L_0 = \frac12k_L^2 + \sum_{n=1}^\infty nN_n.
\end{equation}
On the fermionic side, there are two sets of oscillators, generated by the
fields $\psi$ and $\bar\psi$, and the energy is
\begin{equation}
L_0 = \sum_{n=1}^\infty \left(n-\frac12\right)(N_n+\bar N_n).
\end{equation}
We will denote states on the bosonic side by $(k_L,N_1,N_2)$, and on the
fermionic side by $(N_1+\bar N_1,N_2+\bar N_2,N_3+\bar N_3)$. We won't need
any higher oscillators for this problem. On the fermionic side $N_n+\bar
N_n$ can take the values 0, 1, or 2, with degeneracy 1, 2, and 1
respectively. Here are the states with $L_0 = 0,1/2,\dots,5/2$ on each
side:
\begin{equation}
\begin{array}{c|c|c}
L_0 & (k_L,N_1,N_2) & (N_1+\bar N_1,N_2+\bar N_2,N_3+\bar N_3) \\
\hline
0 & (0,0,0) & (0,0,0) \\
1/2 & (\pm1,0,0) & (1,0,0) \\
1 & (0,1,0) & (2,0,0) \\
3/2 & (\pm1,1,0) & (0,1,0) \\
2 & (\pm2,0,0) & (1,1,0) \\
& (0,2,0) & \\
& (0,0,1) & \\
5/2 & (\pm1,2,0) & (2,1,0) \\
& (\pm1,0,1) & (0,0,1)
\end{array}
\end{equation}

\subsection{Problem 10.11}

We will use the first of the suggested methods. The OPE we need is
\begin{equation}
:e^{iH(z)}::e^{inH(0)}: =
z^n:e^{i(n+1)H(0)}: + O(z^{n+1}).
\end{equation}
Hence
\begin{equation}
\psi(z)F_n(0) = z^nF_{n+1}(0) + O(z^{n+1}).
\end{equation}
It's easy to see that this is satisfied by
\begin{equation}
F_n = :\prod_{i=0}^{n-1}\frac1{i!}\partial^i\psi:.
\end{equation}
The OPE of $\psi(z)$ with $F_n(0)$ is non-singular, but as we Taylor expand
$\psi(z)$, all the terms vanish until the $n$th one because $\psi$ is
fermionic. $F_n$ and $e^{inH}$ obviously have the same fermion number
$n$. Their dimensions work out nicely: for $e^{inH}$ we have $n^2/2$;
for $F_n$ we have $n$ $\psi$s and $n(n-1)/2$ derivatives, for a total of
$n/2+n(n-1)/2=n^2/2$.

For $e^{-inH}$ we obviously just replace $\psi$ with $\bar\psi$.

\subsection{Problem 10.14}

We start with the NS sector. The most general massless state is
\begin{equation}
|\psi\rangle = 
(e\cdot\psi_{-1/2} + f\beta_{-1/2} + g\gamma_{-1/2})|0;k\rangle_{\rm NS},
\end{equation}
where $k^2=0$ (by the $L_0$ condition) and $|0;k\rangle_{\rm NS}$ is
annihilated by $b_0$. The BRST charge acting on $|\psi\rangle$ is
\begin{align}
Q_{\rm B}|\psi\rangle &= 
(c_0L_0 + \gamma_{-1/2}G_{1/2}^{\rm m} + \gamma_{1/2}G_{-1/2}^{\rm m})
|\psi\rangle \notag \\
&=
\sqrt{2\alpha'}(e\cdot k\gamma_{-1/2} + fk\cdot\psi_{-1/2})
|0;k\rangle_{\rm NS}.
\end{align}
The $L_0$ term of course vanishes, along with many others we have not
indicated. For $|\psi\rangle$ to be closed requires $e\cdot
k=f=0$. Furthermore exactness of (28) implies $g \cong
g+\sqrt{2\alpha'}e'\cdot k$ for any $e'$ (so we might as well set $g=0$),
while $e \cong e+\sqrt{2\alpha'}f'k$ for any $f'$. We are left with the 8
transverse polarizations of a massless vector.

The R case is even easier: all of the work is done by the constraint
(10.5.26), and none by the BRST operator. The general massless state is
\begin{equation}
|\psi\rangle = |u;k\rangle_{\rm R},
\end{equation}
where $u$ is a 10-dimensional Dirac spinor, and $k^2=0$. $|\psi\rangle$ is
defined to be annihilated by $b_0$ and $\beta_0$, which implies that it is
annihilated by $G_0^{\rm g}$. According to (10.5.26), this in turn implies
that it is annihilated by $G^{\rm m}_0$, which is exactly the OCQ
condition. The only thing left to check is that all the states satisfying
these conditions are BRST closed, which follows more or less trivially from
all the above conditions. Finally, since they are all closed, none of them
can be exact.

\newcommand{\Rd}{{\rm R}}

\setcounter{equation}{0}
\newpage
\section{Chapter 11}

\subsection{Problem 11.1}

In order to establish the normalizations, we first calculate the
$T_F^+T_F^-$ OPE:
\begin{equation}
e^{+i\sqrt3H(z)}e^{-i\sqrt3H(0)} \sim
\frac1{z^3} + \frac{i\sqrt3\partial H(0)}{z^2} - 
\frac{3\partial H(0)\partial H(0)}{2z} + \frac{i2\sqrt3\partial^2H(0)}{z}.
\end{equation}
Since the third term is supposed to be $2T_B(0)/z$, where
\begin{equation}
T_B = -\frac12\partial H\partial H,
\end{equation}
we need
\begin{equation}
T_F^\pm = \sqrt\frac23e^{\pm i\sqrt3H}.
\end{equation}
It follows from the first term that $c=1$ (as we already knew), and from
the second that
\begin{equation}
j = \frac i{\sqrt3}\partial H.
\end{equation}

It is now straightforward to verify each of the OPEs in turn. The
$T_BT_F^\pm$ and $T_Bj$ OPEs are from Chapter 2, and show that $T_F^\pm$
and $j$ have weight $\frac32$ and 1 respectively. The fact that the
$T_F^\pm T_F^\pm$ OPE is non-singular was also shown in Chapter 2. For the
$jT_F^\pm$ OPE we have
\begin{equation}
j(z)T_F^\pm(0) = \pm\sqrt\frac23\frac{e^{\pm i\sqrt3H(0)}}z.
\end{equation}
Finally,
\begin{equation}
j(z)j(0) = \frac1{3z^2}.
\end{equation}

\subsection{Problem 11.3}

See the last paragraph of section 11.2.

\subsection{Problem 11.4}

\paragraph{(a)}

Dividing the 32 left-moving fermions into two groups of 16, the untwisted
theory contains 8 sectors:
\begin{equation}
\{(+++),(--+),(+--),(-+-)\}\times
\{(\N,\N,\N),(\Rd,\Rd,\Rd)\},
\end{equation}
where the first symbol in each triplet corresponds to the first 16
left-moving fermions, the second to the second 16, and the third to the 8
right-moving fermions. If we now twist by $\exp(\pi i F_1)$, we project out
$(--+)$ and $(-+-)$, but project in the twisted sectors $(\Rd,\N,\N)$ and
$(\N,\Rd,\Rd)$. We again have 8 sectors:
\begin{equation}
\{(+++),(+--)\}\times
\{(\N,\N,\N),(\Rd,\Rd,\Rd),(\Rd,\N,\N),(\N,\Rd,\Rd)\},
\end{equation}
Let us find the massless spacetime bosons first, to establish the gauge
group. These will have right-movers in the $\N+$ sector, so there are
two possibilities, $(\N+,\N+,\N+)$ and $(\Rd+,\N+,\N+)$, and the states are
easily enumerated (primes refer to the second set of left-moving fermions):
\begin{align}
\alpha^i_{-1}\tilde\psi^j_{-1/2}|0_\N,0_\N,0_\N\rangle
&\qquad\te{graviton, dilaton, antisymmetric tensor}; \notag \\
\lambda^{A'}_{-1/2}\lambda^{B'}_{-1/2}\tilde\psi^i_{-1/2}|0_\N,0_\N,0_\N\rangle
&\qquad\te{adjoint of $SO(16)'$}; \notag \\
\lambda^A_{-1/2}\lambda^B_{-1/2}\tilde\psi^i_{-1/2}|0_\N,0_\N,0_\N\rangle
&\qquad\te{adjoint of $SO(16)$}; \notag \\
\tilde\psi^i_{-1/2}|u_\Rd,0_\N,0_\N\rangle
&\qquad\te{{\bf 128} of $SO(16)$}.
\end{align}
The last two sets
combine to form an adjoint of $E_8$, so the gauge group is $E_8\times
SO(16)$. There are similarly two sectors containing massless spacetime
fermions, $(\N+,\Rd+,\Rd+)$ and $(\N+,\Rd-,\Rd-)$ (the $(\Rd,\Rd,\Rd)$ states are
all massive due to the positive normal-ordering constant for the
left-moving R fermions); these will give respectively $(\bf{8,1,128})$ and
$(\bf{8',1,128'})$ of $SO(16)_\te{spin}\times E_8\times SO(16)$. Finally,
the tachyon must be in $(\N+,\N-,\N-)$; the only states are
$\lambda_{-1/2}^{A'}|0_\N,0_\N,0_\N\rangle$, which transform as
$(\bf{1,1,8_v})$.

\paragraph{(b)}

Dividing the left-moving fermions into four groups of 8, if we twist the
above theory by the total fermion number of the first and third groups, we
get a total of 32 sectors:
\begin{multline}
\{(+++++),(----+),(+++--),(---+-)\}\times \\
\{(\N,\N,\N,\N,\N),(\N,\Rd,\Rd,\N,\N),(\Rd,\N,\Rd,\N,\N),(\Rd,\Rd,\N,\N,\N), \\
(\N,\N,\Rd,\Rd,\Rd),(\N,\Rd,\N,\Rd,\Rd),(\Rd,\N,\N,\Rd,\Rd),(\Rd,\Rd,\Rd,\Rd,\Rd)\}.
\end{multline}
Again we begin by listing the massless spacetime bosons, together with
their $SO(8)_\te{spin}\times SO(8)_1\times SO(8)_2\times SO(8)_3\times
SO(8)_4$ quantum numbers:
\begin{align}
(\N+,\N+,\N+,\N+,\N+): \qquad\qquad\qquad\qquad\qquad\qquad\quad& \notag \\
\alpha_{-1}^i\tilde\psi_{-1/2}^j|0_\N,0_\N,0_\N,0_\N,0_\N\rangle &\qquad
(\bf{1\oplus28\oplus35,1,1,1,1}) \notag \\
\lambda^{A_4}_{-1/2}\lambda^{B_4}_{-1/2}\tilde\psi_{-1/2}^i
|0_\N,0_\N,0_\N,0_\N,0_\N\rangle
&\qquad (\bf{8_v,1,1,1,28}) \notag \\
\lambda^{A_1}_{-1/2}\lambda^{B_1}_{-1/2}\tilde\psi_{-1/2}^i
|0_\N,0_\N,0_\N,0_\N,0_\N\rangle
&\qquad (\bf{8_v,28,1,1,1}) \notag \\
\lambda^{A_2}_{-1/2}\lambda^{B_2}_{-1/2}\tilde\psi_{-1/2}^i
|0_\N,0_\N,0_\N,0_\N,0_\N\rangle
&\qquad (\bf{8_v,1,28,1,1}) \notag \\
\lambda^{A_3}_{-1/2}\lambda^{B_3}_{-1/2}\tilde\psi_{-1/2}^i
|0_\N,0_\N,0_\N,0_\N,0_\N\rangle
&\qquad (\bf{8_v,1,1,28,1}) \notag \\
(\N+,\Rd+,\Rd+,\N+,\N+): \qquad
\tilde\psi_{-1/2}^i|0_\N,u_\Rd,v_\Rd,0_\N,0_\N\rangle
&\qquad (\bf{8_v,1,8,8,1}) \notag \\
(\Rd+,\N+,\Rd+,\N+,\N+): \qquad
\tilde\psi_{-1/2}^i|u_\Rd,0_\N,v_\Rd,0_\N,0_\N\rangle
&\qquad (\bf{8_v,8,1,8,1}) \notag \\
(\Rd+,\Rd+,\N+,\N+,\N+): \qquad
\tilde\psi_{-1/2}^i|u_\Rd,v_\Rd,0_\N,0_\N,0_\N\rangle
&\qquad (\bf{8_v,8,8,1,1}).
\end{align}
The first set of states gives the dilaton, antisymmetric tensor, and
graviton. The next set gives the $SO(8)_4$ gauge bosons. The rest combine
to give gauge bosons of $SO(24)$, once we perform triality rotations on
$SO(8)_{1,2,3}$ to turn the bispinors into bivectors.

The massless spacetime fermions are:
\begin{align}
(\Rd+,\N+,\N+,\Rd+,\Rd+): \qquad
|u_\Rd,0_\N,0_\N,v_\Rd,w_\Rd\rangle
&\qquad (\bf{8,8,1,1,8}) \notag \\
(\N+,\Rd+,\N+,\Rd+,\Rd+): \qquad
|0_\N,u_\Rd,0_\N,v_\Rd,w_\Rd\rangle
&\qquad (\bf{8,1,8,1,8}) \notag \\
(\N+,\N+,\Rd+,\Rd+,\Rd+): \qquad
|0_\N,0_\N,u_\Rd,v_\Rd,w_\Rd\rangle
&\qquad (\bf{8,1,1,8,8}).
\end{align}
The triality rotation again turns the $SO(8)_{1,2,3}$ spinors into vectors,
which combine into an $SO(24)$ vector.

Finally, the tachyon is an $SO(8)_4$ vector but is neutral under $SO(24)$:
\begin{equation}
(\N+,\N+,\N+,\N-,\N-):\qquad
\lambda^{A_4}_{-1/2}|0_\N,0_\N,0_\N,0_\N\rangle \qquad
(\bf{1,1,1,1,8_v}).
\end{equation}

\subsection{Problem 11.7}

We wish to show that the state $j^a_{-1}j^a_{-1}|0\rangle$ corresponds to
the operator $:jj(0):$. Since $j^a_{-1}|0\rangle$ clearly corresponds to
$j^a(0)$, and $j^a_{-1}=\oint dz/(2\pi i)j^a(z)/z$, we have
\begin{equation}
:jj(0):\, = 
\oint \frac{dz}{2\pi i}\frac{j^a(z)j^a(0)}{z},
\end{equation}
where the contour goes around the origin. The contour integral picks out
the $z^0$ term in the Laurent expansion of $j^a(z)j^a(0)$, which is
precisely (11.5.18).

We will now use this to prove the first line of (11.5.20), with $z_1=0$ and
$z_3=z$. Using the Laurent expansion (11.5.2) we have
\begin{equation}
:jj(0):j^c(z) \cong
\sum_{m=-\infty}^\infty\frac{1}{z^{m+1}}j^c_mj^a_{-1}j^a_{-1}|0\rangle.
\end{equation}
Only three terms in this sum are potentially interesting; the rest are
either non-singular (for $m<0$) or zero (for $m>2$, since the total level
of the state would be negative). In fact, the term with $m=2$ must also
vanish, since (being at level 0) it can only be proportional to the ground
state $|0\rangle$, leaving no room for the free Lie algebra index on (15);
this can also be checked explicitly. For the other two terms we have:
\begin{align}
j_0^cj^a_{-1}j^a_{-1}|0\rangle &= 
if^{cab}(j^a_{-1}j^b_{-1}+j^b_{-1}j^a_{-1})|0\rangle = 0, \\
j_1^cj^a_{-1}j^a_{-1}|0\rangle &= 
(\hat k\delta^{ca}+if^{cab}j^b_0+j_{-1}^aj_1^c)j^a_{-1}|0\rangle \notag \\
& = (2\hat k j_{-1}^c - f^{cab}f^{bad}j^d_{-1})|0\rangle \notag \\
& = (k+h(g))\psi^2j^c_{-1}|0\rangle.
\end{align}
The RHS of (17) clearly corresponds to the RHS of (11.5.20).

To check the $TT$ OPE (11.5.24), we employ the same strategy, using the
Laurent coefficients (11.5.26). The OPE will be the operator corresponding
to
\begin{equation}
\frac1{(k+h(g))\psi^2}\left(
\frac1{z^4}L_2 + \frac1{z^3}L_1 + \frac1{z^2}L_0 + \frac1zL_{-1}\right)
j_{-1}^aj_{-1}^a|0\rangle;
\end{equation}
terms with $L_m$ are non-singular for $m<-1$ and vanish for $m>2$. Life is
made much easier by the fact that $j_m^bj_{-1}^aj_{-1}^a|0\rangle = 0$
for $m=0$ and $m>1$, and the value for $m=1$ is given by (17) above. Thus:
\begin{align}
L_2j_{-1}^aj_{-1}^a|0\rangle &=
\frac1{(k+h(g))\psi^2}j_1^bj_1^bj_{-1}^aj_{-1}^a|0\rangle 
= j^b_1j^b_{-1}|0\rangle 
=\hat k {\rm dim}(g)|0\rangle, \notag \\
L_1j_{-1}^aj_{-1}^a|0\rangle &= 0, \notag \\
L_0j_{-1}^aj_{-1}^a|0\rangle &= 
\frac2{(k+h(g))\psi^2}j_{-1}^bj_1^bj_{-1}^aj_{-1}^a|0\rangle
= 2j^b_{-1}j^b_{-1}|0\rangle, \notag \\
L_{-1}j_{-1}^aj_{-1}^a|0\rangle &= 
\frac2{(k+h(g))\psi^2}j_{-2}^bj_1^bj_{-1}^aj_{-1}^a|0\rangle
= 2j^b_{-2}j^b_{-1}|0\rangle.
\end{align}
All of these states are easily translated back into operators, the only
slightly non-trivial one being the last. From the Laurent expansion we see
that the state corresponding to $\partial T^{\rm s}_B(0)$ is indeed
$L_{-3}|0\rangle = 2/((k+h(g))\psi^2)j^b_{-2}j^b_{-1}|0\rangle$. Thus we
have precisely the OPE (11.5.24).

\subsection{Problem 11.8}

The operator $:jj(0):$ is defined to be the $z^0$ term in the Laurent
expansion of $j^a(z)j^a(0)$. First let us calculate the contribution from a
single current $i\lambda^A\lambda^B$ ($A\neq B$):
\begin{multline}
i\lambda^A(z)\lambda^B(z)i\lambda^A(0)\lambda^B(0) =
\lambda^A(z)\lambda^A(0)\lambda^B(z)\lambda^B(0) \qquad \te{(no sum)}\\
= \,:\lambda^A(z)\lambda^A(0)\lambda^B(z)\lambda^B(0): +
\frac1z:\lambda^A(z)\lambda^A(0): +
\frac1z:\lambda^B(z)\lambda^B(0): +
\frac1{z^2}.
\end{multline}
Clearly the order $z^0$ term is $:\partial\lambda^A\lambda^A: +
:\partial\lambda^B\lambda^B:$. Summing over all $A$ and $B$ with $B\neq A$
double counts the currents, so we divide by 2:
\begin{equation}
:jj: \,= (n-1):\partial\lambda^A\lambda^A: \qquad \te{(sum)}.
\end{equation}
Finally, we have $k=1$, $h(SO(n))=n-2$, and $\psi^2=2$, so
\begin{equation}
T^{\rm s}_B =
\frac12:\partial\lambda^A\lambda^A: \qquad \te{(sum)}.
\end{equation}

\subsection{Problem 11.9}

In the notation of (11.6.5) and (11.6.6), the lattice $\Gamma$ is
$\Gamma_{22,6}$, i.e. the set of points of the form
\begin{align}
&(n_1,\dots,n_{28})\quad \te{or}\quad 
(n_1+\frac12,\dots,n_{28}+\frac12), \notag \\
&\sum_in_i\in2Z
\end{align}
for any integers $n_i$. It will be convenient to divide $\Gamma$ into two
sublattices, $\Gamma=\Gamma_1\cup\Gamma_2$ where
\begin{align}
\Gamma_1 &= \{(n_1,\dots,n_{28}):\sum n_i\in2Z\}, \notag \\
\Gamma_2 &= \Gamma_1+l_0, \qquad l_0 \equiv (\frac12,\dots,\frac12).
\end{align}
Evenness of $l\in\Gamma_1$ follows from the fact that the number of odd
$n_i$ must be even, implying
\begin{equation}
l\circ l=\sum_{i=1}^{22}n_i^2-\sum_{i=23}^{28}n_i^2\in2Z.
\end{equation}
Evenness of $l+l_0\in\Gamma_2$ follows from the same fact:
\begin{equation}
(l+l_0)\circ(l+l_0) = l\circ l +2l_0\circ l + l_0\circ l_0
= l\circ l+\sum_{i=1}^{22}n_i-\sum_{i=23}^{28}n_i+4\in2Z.
\end{equation}
Evenness implies integrality, so $\Gamma\subset\Gamma^*$. It's easy to see
that the dual lattice to $\Gamma_1$ is
$\Gamma_1^*=Z^{28}\cup(Z^{28}+l_0)\supset\Gamma$. But to be dual for
example to $l_0$ requires a vector to have an even number of odd $n_i$s, so
$\Gamma^*=\Gamma$.

To find the gauge bosons we need to find the lattice vectors satisfying
(11.6.15). But these are obviously the root vectors of $SO(44)$. In
addition there are the 22 gauge bosons with vertex operators $\partial
X^m\tilde\psi^\mu$, providing the Cartan generators to fill out the adjoint
representation of $SO(44)$, and the 6 gauge bosons with vertex operators
$\partial X^\mu\tilde\psi^m$, generating $U(1)^6$.

\subsection{Problem 11.12}

It seems to me that both the statement of the problem and the derivation of
the Hagedorn temperature for the bosonic string (Vol. I, pp. 320-21) are
misleading. Equation (7.3.20) is not the correct one to use to find the
asymptotic density of states of a string theory, since it does not take
into account the level matching constraint in the physical
spectrum. It's essentially a matter of luck that Polchinski ends up
with the right Hagedorn temperature, (9.8.13).

Taking into account level matching we have $n(m) = n_{\rm L}(m)n_{\rm
R}(m)$. To find $n_{\rm L}$ and $n_{\rm R}$ we treat the left-moving and
right-moving CFTs separately, and include only the physical parts of the
spectrum, that is, neglect the ghosts and the timelike and longitudinal
oscillators. Then we have, as in (9.8.11),
\begin{equation}\label{cardy}
\sum_{m^2}n_{\rm L}(m)e^{-\pi\alpha'm^2l/2}
\sim e^{\pi c/(12l)},
\end{equation}
implying
\begin{equation}
n_{\rm L}(m) \sim e^{\pi m\sqrt{\alpha'c/6}}.
\end{equation}
Similarly for the right-movers, giving
\begin{equation}
n(m) \sim e^{\pi m\sqrt{\alpha'/6}(\sqrt{c}+\sqrt{\tilde c})}.
\end{equation}
The Hagedorn temperature is then given by
\begin{equation}
T_{\rm H}^{-1} = 
\pi\sqrt{\alpha'}\left(\sqrt\frac{c}6+\sqrt\frac{\tilde c}6\right).
\end{equation}
For the type I and II strings, this gives
\begin{equation}
\label{typeII}
T_{\rm H} = \frac1{2\pi\sqrt{2\alpha'}},
\end{equation}
while for the heterotic theories we have
\begin{equation}
T_{\rm H} = \frac1{\pi\sqrt{\alpha'}}(1-\frac1{\sqrt2}).
\end{equation}
The Hagedorn temperature for an open type I string is the same,
\eqref{typeII}, as for the closed string, since $n_{\rm open}(m) = n_{\rm
L}(2m)$.

There is an interesting point that we glossed over above. The
RHS of \eqref{cardy} is obtained by doing a modular transformation
$l\to1/l$ on the torus partition function with $\tau=il$. Then for small
$l$ only the lowest-lying state in the theory contributes. We implicitly
took the lowest-lying state to be the vacuum, corresponding to the unit
operator. However, that state is projected out by the GSO projection in all
of the above theories, else it would give rise to a tachyon. So should we
really consider it? To see that we should, let us derive \eqref{cardy} more
carefully, for example in the case of the left-movers of the type II
string. The GSO projection there is $(-1)^F=-1$, where $F$ is the
worldsheet fermion number of the transverse fermions ({\it not} including
the ghosts). The partition function from which we will extract $n_{\rm
L}(m)$, the number of projected-in states, is
\begin{equation}\label{careful}
Z(il) = \sum_{i\in\te{R,NS}}q^{h_i-c/24}\frac12(1-(-1)^{F_i}) =
\sum_{m^2}n_{\rm L}(m)e^{-\pi\alpha'm^2l/2}.
\end{equation}
This partition function corresponds to a path integral on the torus in
which we sum over all four spin structures, with minus signs when the
fermions are periodic in the $\sigma_2$ (``time'') direction. Upon doing
the modular integral, this minus sign corresponds to giving a minus sign to
R sector states. But the sum on R and NS sectors in \eqref{careful} means
that we now project onto states with $(-1)^F=1$:
\begin{equation}
Z(il) = 
\sum_{i\in\te{R,NS}}e^{-2\pi/l(h_i-c/24)}(-1)^{\alpha_i}\frac12(1+(-1)^{F_i})
\sim e^{\pi c/(12l)}.
\end{equation}
We see that the ground state, with $h=0$, is indeed projected in, and
therefore dominates in the limit $l\to0$.

\setcounter{equation}{0}
\newpage
\section{Chapter 13}

\subsection{Problem 13.2}

An open string with ends attached to D$p$-branes is T-dual to an open type I
string. An open string with both ends attached to the same D$p$-brane and
zero winding number is T-dual to an open
type I string with zero momentum in the $9-p$ dualized directions, and
Chan-Paton factor of the form
\be
t^a =
\frac{1}{\sqrt{2}}\left[\begin{array}{cc}0&i\\-i&0\end{array}\right]\otimes{\rm
diag}(1,0,\dots,0).
\ee
Four open strings attached to the same D$p$-brane are T-dual to four open
type I strings with zero momentum in the $9-p$ dualized directions and the
same Chan-Paton factor (1). The scattering amplitude for four gauge boson
open string states was calculated in section 12.4. Using
\be
{\rm Tr}(t^a)^4 = \frac{1}{2},
\ee
the result (12.4.22) becomes in this case
\bea
\lefteqn{S(k_i,e_i) =} \\
&& -8ig_{\rm
YM}^2\alpha'^2(2\pi)^{10}\delta^{10}(\sum_ik_i)K(k_i,e_i)\left(\frac{\Gamma(-\alpha's)\Gamma(-\alpha'u)}{\Gamma(1-\alpha's-\alpha'u)}+\mbox{2
permutations}\right).\nonumber
\eea
The kinematic factor $K$ is written in three different ways in (12.4.25)
and (12.4.26), and we won't bother to reproduce it here. Since the momenta
$k_i$ all have vanishing components in the $9-p$ dualized directions, the
amplitude becomes,
\bea
\lefteqn{-8ig_{(p+1),\rm
YM}^2\alpha'^2V_{9-p}^2(2\pi)^{p+1}\delta^{p+1}(\sum_ik_i)}\qquad\qquad\qquad\qquad\qquad\quad \\
&&\times K(k_i,e_i)\left(\frac{\Gamma(-\alpha's)\Gamma(-\alpha'u)}{\Gamma(1-\alpha's-\alpha'u)}+\mbox{2
permutations}\right),\nonumber
\eea
where $V_{9-p}$ is the volume of the transverse space, and we have used
(13.3.29). But in order to get the proper $(p+1)$-dimensional scattering
amplitude, we must renormalize the wave function of each string (which is
spread out uniformly in the transverse space) by a factor of
$\sqrt{V_{9-p}}$:
\bea
\lefteqn{S'(k_i,e_i) = -8ig_{(p+1),\rm YM}^2\alpha'^2(2\pi)^{p+1}\delta^{p+1}(\sum_ik_i)}\qquad\qquad\qquad\qquad\qquad\quad \\
&& \times K(k_i,e_i)\left(\frac{\Gamma(-\alpha's)\Gamma(-\alpha'u)}{\Gamma(1-\alpha's-\alpha'u)}+\mbox{2
permutations}\right).\nonumber
\eea
Finally, using (13.3.30) and (13.3.28), we can write the
dimensionally reduced type I Yang-Mills coupling $g_{(p+1),\rm YM}$, in terms
of the coupling $g_{{\rm D}p}$ on the brane:
\be
g_{(p+1),\rm YM}^2 = g_{{\rm D}p,{\rm SO}(32)}^2 = 2g_{{\rm D}p}^2,
\ee
so
\bea
\lefteqn{S'(k_i,e_i) = -16ig_{{\rm D}p}^2\alpha'^2(2\pi)^{p+1}\delta^{p+1}(\sum_ik_i)}\qquad\qquad\qquad\qquad\qquad\quad \\
&& \times K(k_i,e_i)\left(\frac{\Gamma(-\alpha's)\Gamma(-\alpha'u)}{\Gamma(1-\alpha's-\alpha'u)}+\mbox{2
permutations}\right).\nonumber
\eea
(One can also use (13.3.25) to write $g_{{\rm D}p}$ in terms of the string
coupling $g$.)

\subsection{Problem 13.3}

\paragraph{(a)}
By equations (B.1.8) and (B.1.10),
\be
\Gamma^{2a}\Gamma^{2a+1} = -2iS_a,
\ee
where $a=1,2,3,4$, so
\be
\beta^{2a}\beta^{2a+1} = 2iS_a.
\ee
If we define
\be
\beta \equiv \beta^1\beta^2\beta^3,
\ee
and label the D4-branes extended in the (6,7,8,9), (4,5,8,9), and (4,5,6,7)
directions by the subscripts 2, 3, and 4 respectively, then
\be
\beta^\perp_2 = \beta^1\beta^2\beta^3\beta^4\beta^5 = 2iS_2\beta
\ee
and similarly
\be
\beta^\perp_3 = 2iS_3\beta,\qquad\beta^\perp_4 = 2iS_4\beta.
\ee
The supersymmetries preserved by brane $a$ ($a=2,3,4$) are
\be
Q_{\bf s}+(\beta_a^\perp\tilde{Q})_{\bf s} = Q_{\bf
s}+2is_a(\beta\tilde{Q})_{\bf s},
\ee
so for a supersymmetry to be unbroken by all three branes simply requires
$s_2=s_3=s_4$. Taking into account the chirality condition $\Gamma=+1$ on
$Q_{\bf s}$, there are four unbroken supersymmetries:
\bea
Q_{(+++++)}+i(\beta\tilde{Q})_{(+++++)},\nonumber \\
Q_{(--+++)}+i(\beta\tilde{Q})_{(--+++)},\nonumber \\
Q_{(+----)}-i(\beta\tilde{Q})_{(+----)},\nonumber \\
Q_{(-+---)}-i(\beta\tilde{Q})_{(-+---)}.
\eea

\paragraph{(b)}
For the D0-brane,
\be
\beta_{\rm D0}^\perp =
\beta^1\beta^2\beta^3\beta^4\beta^5\beta^6\beta^7\beta^8\beta^9 =
-8iS_2S_3S_4\beta,
\ee
so the unbroken supersymmetries are of the form,
\be
Q_{\bf s}+(\beta_{\rm D0}^\perp\tilde{Q})_{\bf s} = Q_{\bf
s}-8is_2s_3s_4(\beta\tilde{Q})_{\bf s}.
\ee
The signs in all four previously unbroken supersymmetries (14) are just
wrong to remain unbroken by the D0-brane, so that this configuration preserves
no supersymmetry.

\paragraph{(c)}
Let us make a brane scan of the original configuration:

\begin{tabular}{c|ccccccccc}
  &1&2&3&4&5&6&7&8&9 \\
\hline
D4${}_2$&D&D&D&D&D&N&N&N&N \\
D4${}_3$&D&D&D&N&N&D&D&N&N \\
D4${}_4$&D&D&D&N&N&N&N&D&D \\
D0      &D&D&D&D&D&D&D&D&D
\end{tabular}
\newline
There are nine distinct T-dualities that can be performed in the 4, 5, 6,
7, 8, and 9 directions, up to the symmetries $4\leftrightarrow5$,
$6\leftrightarrow7$, $8\leftrightarrow9$, and
$(45)\leftrightarrow(67)\leftrightarrow(89)$. They result in the following
brane content:

\begin{tabular}{r|l}
T-dualized directions&$(p_1,p_2,p_3,p_4)$ \\
\hline
4&$(5,3,3,1)$ \\
4, 5&$(6,2,2,2)$ \\
4, 6&$(4,4,2,2)$ \\
4, 5, 6&$(5,3,1,3)$ \\
4, 5, 6, 7&$(4,4,0,4)$ \\
4, 6, 8&$(3,3,3,3)$ \\
4, 5, 6, 8&$(4,2,2,4)$ \\
4, 5, 6, 7, 8&$(3,3,1,5)$ \\
4, 5, 6, 7, 8, 9&$(2,2,2,6)$
\end{tabular}
\newline
Further T-duality in one, two, or all three of the 1, 2, and 3 directions will
turn any of these configurations into $(p_1+1,p_2+1,p_3+1,p_4+1)$,
$(p_1+2,p_2+2,p_3+2,p_4+2)$, and $(p_1+3,p_2+3,p_3+3,p_4+3)$ respectively.

T-dualizing at general angles to the coordinate axes will result in
combinations of the above configurations for the directions involved, with
the smaller-dimensional brane in each column being replaced by a magnetic
field on the larger-dimensional brane.

\subsection{Problem 13.4}

\paragraph{(a)}
Let the D2-brane be extended in the 8 and 9 directions, and let it be
separated from the D0-brane in the 1 direction by a distance
$y$. T-dualizing this configuration in the 2, 4, 6, and 8 directions yields
2 D4-branes, the first (from the D2-brane) extended in the 2, 4, 6, and 9
directions, and the second (from the D0-brane) in the 2, 4, 6, and 8
directions. This is in the class of D4-brane configurations studied in
section 13.4; in our case the angles defined there take the values
\be
\phi_1 = \phi_2 = \phi_3 = 0,\qquad\phi_4 = \frac{\pi}{2},
\ee
and therefore (according to (13.4.22)),
\be
\phi_1' = \phi_4' = \frac{\pi}{4},\qquad\phi_2' = \phi_3' = -\frac{\pi}{4}.
\ee
For the three directions in which the D4-branes are parallel, we must make
the substitution (13.4.25) (without the exponential factor, since we have
chosen the separation between the D0- and D2-branes to vanish in those
directions, but with a factor $-i$, to make the potential real and
attractive). The result is
\be
V(y) =
-\int_0^\infty\frac{dt}{t}\,(8\pi^2\alpha't)^{-1/2}\exp\left(-\frac{ty^2}{2\pi\alpha'}\right)\frac{i\vartheta_{11}^4(it/4,it)}{\vartheta_{11}(it/2,it)\eta^9(it)}.
\ee
Alternatively, using the modular transformations (7.4.44b) and (13.4.18b),
\be
V(y) = -\frac{1}{\sqrt{8\pi^2\alpha'}}\int_0^\infty
dt\,t^{3/2}\exp\left(-\frac{ty^2}{2\pi\alpha'}\right)\frac{\vartheta_{11}^4(1/4,i/t)}{\vartheta_{11}(1/2,i/t)\eta^9(i/t)}.
\ee

\paragraph{(b)}
For the field theory calculation we lean heavily on the similar calculation
done in section 8.7, adapting it to $D=10$. Polchinski employs the shifted
dilaton 
\be
\tilde{\Phi} = \Phi-\Phi_0,
\ee
whose expectation value vanishes, and the Einstein metric 
\be
\tilde{G} = e^{-\tilde{\Phi}/2}G;
\ee
their propagators are given in (8.7.23):
\bea
\langle\tilde{\Phi}\tilde{\Phi}(k)\rangle &=& -\frac{2i\kappa^2}{k^2}, \\
\langle h_{\mu\nu}h_{\sigma\rho}(k)\rangle &=&
-\frac{2i\kappa^2}{k^2}\left(\eta_{\mu\sigma}\eta_{\nu\rho}+\eta_{\mu\rho}\eta_{\nu\sigma}-\frac{1}{4}\eta_{\mu\nu}\eta_{\sigma\rho}\right),
\eea
where $h=\tilde{G}-\eta$. The D-brane action (13.3.14) expanded for small
values of $\tilde{\Phi}$ and $h$ is
\be
S_p = -\tau_p\int
d^{p+1}\!\xi\,\left(\frac{p-3}{4}\tilde{\Phi}+\frac{1}{2}h_a{}^a\right),
\ee
where the trace on $h$ is taken over directions tangent to the
brane. From the point of view of the supergravity, the D-brane is thus a
source for $\tilde{\Phi}$,
\be
J_{\tilde{\Phi},p}(X) = \frac{3-p}{4}\tau_p\delta^{9-p}(X_\perp-X_\perp'),
\ee
and for $h$,
\be
J_{h,p}^{\mu\nu}(X) =
-\frac{1}{2}\tau_pe_p^{\mu\nu}\delta^{9-p}(X_\perp-X'_\perp),
\ee
where $X'_\perp$ is the position of the brane in the transverse
coordinates, and $e_p^{\mu\nu}$ is $\eta^{\mu\nu}$ in the directions parallel
to the brane and 0 otherwise. In momentum space the sources are
\bea
\tilde{J}_{\tilde{\Phi},p}(k) &=&
\frac{3-p}{4}\tau_p(2\pi)^{p+1}\delta^{p+1}(k_\parallel)e^{ik_\perp\cdot
X'_\perp}, \\
\tilde{J}_{h,p}^{\mu\nu}(k) &=&
-\frac{1}{2}\tau_pe_p^{\mu\nu}(2\pi)^{p+1}\delta^{p+1}(k_\parallel)e^{ik_\perp\cdot
X'_\perp}.
\eea
Between the D0-brane, located at the
origin of space, and the D2-brane, extended in the 8 and 9 directions and
located in the other directions at the point
\be
(X'_1,X'_2,X'_3,X'_4,X'_5,X'_6,X'_7)=(y,0,0,0,0,0,0),
\ee
the amplitude for dilaton exchange is
\bea
\mathcal{A}_{\tilde{\Phi}} &=&
-\int\frac{d^{10}\!k}{(2\pi)^{10}}\,\tilde{J}_{\tilde{\Phi},0}(k)\langle\tilde{\Phi}\tilde{\Phi}(k)\rangle\tilde{J}_{\tilde{\Phi},2}(-k)
\nonumber \\
&=&
i\frac{3}{8}\tau_0\tau_2\kappa^2\int\frac{d^{10}\!k}{(2\pi)^{10}}\,2\pi\delta(k_0)\frac{1}{k^2}(2\pi)^3\delta^3(k_0,k_8,k_9)e^{ik_1y}
\nonumber \\
&=&
iT\frac{3}{8}\tau_0\tau_2\kappa^2\int\frac{d^7\!k}{(2\pi)^7}\,\frac{e^{iky}}{k^2}
\nonumber \\
&=& iT\frac{3}{8}\tau_0\tau_2\kappa^2G_7(y),
\eea
where $G_7$ is the 7-dimensional massless scalar Green function. We divide
the amplitude by $-iT$ to obtain the static potential due to dilaton
exchange:
\be
V_{\tilde{\Phi}}(y) = -\frac{3}{8}\tau_0\tau_2\kappa^2G_7(y).
\ee
The calculation for the graviton exchange is similar, the only difference
being that the numerical factor $(1/4)2(3/4)$ is replaced by
\be
\frac{1}{2}e_0^{\mu\nu}2\left(\eta_{\mu\sigma}\eta_{\nu\rho}+\eta_{\mu\rho}\eta_{\nu\sigma}-\frac{1}{4}\eta_{\mu\nu}\eta_{\sigma\rho}\right)\frac{1}{2}e_2^{\sigma\rho}
= \frac{5}{8}.
\ee
The total potential between the D0-brane and D2-brane is therefore
\be
V(y) = \tau_0\tau_2\kappa^2G_7(y) = -\pi(4\pi^2\alpha')^2G_7(y),
\ee
where we have applied (13.3.4). As expected, gravitation and dilaton
exchange are both attractive forces.

In the large-$y$ limit of (20), the integrand becomes very small except
where $t$ is very small. The ratio of modular functions involved in the
integrand is in fact finite in the limit $t\to0$,
\be
\lim_{t\to0}\frac{\vartheta_{11}^4(1/4,i/t)}{\vartheta_{11}(1/2,i/t)\eta^9(i/t)}
= 2,
\ee
(according to {\sl Mathematica}), so that the first term in the asymptotic
expansion of the potential in $1/y$ is
\bea
V(r) &\approx& -\frac{1}{\sqrt{2\pi^2\alpha'}}\int_0^\infty
dt\,t^{3/2}\exp\left(-\frac{ty^2}{2\pi\alpha'}\right)\nonumber \\
&=& -\pi^{-1/2}(2\pi\alpha')^2\Gamma(\frac{5}{2})y^{-5} \nonumber \\
&=& -\pi(4\pi^2\alpha')^2G_7(y),
\eea
in agreement with (34).

\subsection{Problem 13.12}

The tension of the $(p_i,q_i)$-string is (13.6.3)
\be
\tau_{(p_i,q_i)} =\frac{\sqrt{p_i^2+q_i^2/g^2}}{2\pi\alpha'}.
\ee
Let the three strings sit in the $(X^1,X^2)$ plane. If the angle string
$i$ makes with the $X^1$ axis is $\theta_i$, then the force it exerts on
the junction point is
\be
(F_i^1,F_i^2) =
\frac{1}{2\pi\alpha'}(\cos\theta_i\sqrt{p_i^2+q_i^2/g^2},\sin\theta_i\sqrt{p_i^2+q_i^2/g^2}).
\ee
If we orient each string at the angle
\be
\cos\theta_i = \frac{p_i}{\sqrt{p_i^2+q_i^2/g^2}},\qquad\sin\theta_i =
\frac{q_i/g}{\sqrt{p_i^2+q_i^2/g^2}},
\ee
then the total force exerted on the junction point is
\be
\frac{1}{2\pi\alpha'}\sum_{i=1}^3(p_i,q_i/g),
\ee
which vanishes if $\sum p_i=\sum q_i=0$. This is the unique stable
configuration, up to rotations and reflections.

The supersymmetry algebra for a static $(p,q)$ string extended in the $X^i$
direction is (13.6.1)
\be
\frac{1}{2L}\left\{\left[\begin{array}{c}Q_\alpha\\\tilde{Q}_\alpha\end{array}\right],\left[\begin{array}{cc}Q^\dag_\beta&\tilde{Q}^\dag_\beta\end{array}\right]\right\}
= \tau_{(p,q)}\delta_{\alpha\beta}\left[\begin{array}{cc}1\,&0\\0\,&1\end{array}\right]+\frac{(\Gamma^0\Gamma^i)_{\alpha\beta}}{2\pi\alpha'}\left[\begin{array}{cc}p\,&q/g\\q/g\,&-p\end{array}\right].
\ee
Defining
\be
u \equiv \frac{p}{\sqrt{p^2+q^2/g^2}},\qquad U \equiv
\frac{1}{\sqrt{2}}\left[\begin{array}{cc}\sqrt{1+u}\,&\sqrt{1-u}\\-\sqrt{1-u}\,&\sqrt{1+u}\end{array}\right],
\ee
we can use $U$ to diagonalize the matrix on the RHS of (41):
\be
\frac{1}{2L\tau_{(p,q)}}
\left\{U\left[\begin{array}{c}Q_\alpha\\\tilde{Q}_\alpha\end{array}\right],
\left[\begin{array}{cc}Q_\beta^\dag&\tilde{Q}\beta^\dag\end{array}\right]
U^T\right\}
=
\left[\begin{array}{cc}(I_{16}+\Gamma^0\Gamma^i)_{\alpha\beta}\,&0\\0\,&(I_{16}-\Gamma^0\Gamma^i)_{\alpha\beta}\end{array}\right].
\ee
The top row of this $2\times2$ matrix equation tells us that, in a basis in
spinor space in which $\Gamma^0\Gamma^i$ is diagonal, the supersymmetry
generator
\be
\sqrt{1+u}Q_\alpha+\sqrt{1-u}\tilde{Q}_\alpha
\ee
annihilates this state if $(I_{16}+\Gamma^0\Gamma^i)_{\alpha\alpha}=0$. We
can use $(I_{16}-\Gamma^0\Gamma^i)$ to project onto this eight-dimensional
subspace, yielding eight supersymmetries that leave this state invariant:
\be
\left[(I_{16}-\Gamma^0\Gamma^i)(\sqrt{1+u}Q+\sqrt{1-u}\tilde{Q})\right]_\alpha.
\ee
The other eight unbroken supersymmetries are given by the bottom row of (43),
after projecting onto the subspace annihilated by
$(I_{16}-\Gamma^0\Gamma^i)$: 
\be
\left[(I_{16}+\Gamma^0\Gamma^i)(-\sqrt{1-u}Q+\sqrt{1+u}\tilde{Q})\right]_\alpha.
\ee

Now let us suppose that the string is aligned in the direction (39), which
depends on $p$ and $q$. We will show that eight of the sixteen unbroken
supersymmetries do not depend on $p$ or $q$, and therefore any
configuration of $(p,q)$ strings that all obey (39) will leave these eight
unbroken. If the string is aligned in the direction (39),
then
\be
\Gamma^i =
\frac{p}{\sqrt{p^2+q^2/g^2}}\Gamma^1+\frac{q/g}{\sqrt{p^2+q^2/g^2}}\Gamma^2
= u\Gamma^1+\sqrt{1-u^2}\Gamma^2.
\ee
Our first set of unbroken supersymmetries (45) becomes
\be
\left[(I_{16}-u\Gamma^0\Gamma^1-\sqrt{1-u^2}\Gamma^0\Gamma^2)(\sqrt{1+u}Q+\sqrt{1-u}\tilde{Q})\right]_\alpha.
\ee
We work in a basis of eigenspinors of the operators $S_a$ defined in
(B.1.10). In this basis $\Gamma^0\Gamma^1=2S_0$, while
\be
\Gamma^0\Gamma^2 = \left[\begin{array}{cc}0\,&1\\-1\,&0\end{array}\right]
\otimes\left[\begin{array}{cc}0\,&-1\\1\,&0\end{array}\right]\otimes
I_2\otimes I_2\otimes I_2.
\ee
We can divide the sixteen values of the spinor index $\alpha$ into four
groups of four according to the eigenvalues of $S_0$ and $S_1$:
\bea
\lefteqn{\left[(I_{16}-u\Gamma^0\Gamma^1-\sqrt{1-u^2}\Gamma^0\Gamma^2)(\sqrt{1+u}Q+\sqrt{1-u}\tilde{Q})\right]_{(++s_2s_3s_4)}}\qquad\qquad\qquad\qquad\qquad\qquad \\
&=&
(1-u)(\sqrt{1+u}Q+\sqrt{1-u}\tilde{Q})_{(++s_2s_3s_4)}\nonumber \\
&&\quad{}+\sqrt{1-u^2}(\sqrt{1+u}Q+\sqrt{1-u}\tilde{Q})_{(--s_2s_3s_4)},
\nonumber \\
\lefteqn{\left[(I_{16}-u\Gamma^0\Gamma^1-\sqrt{1-u^2}\Gamma^0\Gamma^2)(\sqrt{1+u}Q+\sqrt{1-u}\tilde{Q})\right]_{(++s_2s_3s_4)}}\qquad\qquad\qquad\qquad\qquad\qquad \\
&=&
(1+u)(\sqrt{1+u}Q+\sqrt{1-u}\tilde{Q})_{(--s_2s_3s_4)}\nonumber \\
&&\quad{}+\sqrt{1-u^2}(\sqrt{1+u}Q+\sqrt{1-u}\tilde{Q})_{(++s_2s_3s_4)},
\nonumber \\
\lefteqn{\left[(I_{16}-u\Gamma^0\Gamma^1-\sqrt{1-u^2}\Gamma^0\Gamma^2)(\sqrt{1+u}Q+\sqrt{1-u}\tilde{Q})\right]_{(+-s_2s_3s_4)}}\qquad\qquad\qquad\qquad\qquad\qquad \\
&=&
(1-u)(\sqrt{1+u}Q+\sqrt{1-u}\tilde{Q})_{(+-s_2s_3s_4)}\nonumber \\
&&\quad{}-\sqrt{1-u^2}(\sqrt{1+u}Q+\sqrt{1-u}\tilde{Q})_{(-+s_2s_3s_4)},
\nonumber \\
\lefteqn{\left[(I_{16}-u\Gamma^0\Gamma^1-\sqrt{1-u^2}\Gamma^0\Gamma^2)(\sqrt{1+u}Q+\sqrt{1-u}\tilde{Q})\right]_{(-+s_2s_3s_4)}}\qquad\qquad\qquad\qquad\qquad\qquad \\
&=&
(1+u)(\sqrt{1+u}Q+\sqrt{1-u}\tilde{Q})_{(-+s_2s_3s_4)}\nonumber \\
&&\quad{}-\sqrt{1-u^2}(\sqrt{1+u}Q+\sqrt{1-u}\tilde{Q})_{(+-s_2s_3s_4)}.
\nonumber
\eea
(The
indexing by $s_2,s_3,s_4$ is somewhat redundant, since the chirality
condition on both $Q$ and $\tilde{Q}$ implies the restriction
$8s_2s_3s_4=1$ in the case of (50) and (51), and $8s_2s_3s_4=-1$ in the
case of (52) and (53).) It easy to see that (50) and (51) differ only by a
factor of $\sqrt{(1-u)/(1+u)}$, and (52) and (53) similarly by a factor of
$-\sqrt{(1-u)/(1+u)}$, so (50) and (52) alone are sufficient to describe
the eight independent supersymmetry generators in this sector. In the other
sector, given by (46), there is a similar repetition of generators, and the
eight independent generators are
\bea
\lefteqn{\left[(I_{16}+u\Gamma^0\Gamma^1+\sqrt{1-u^2}\Gamma^0\Gamma^2)
(-\sqrt{1-u}Q+\sqrt{1+u}\tilde{Q})\right]_{(++s_2s_3s_4)}}\qquad\qquad\quad\qquad\qquad\qquad
\\ 
&=&
(1+u)(-\sqrt{1-u}Q+\sqrt{1+u}\tilde{Q})_{(++s_2s_3s_4)} \nonumber \\
&& \quad{}-\sqrt{1-u^2}(-\sqrt{1-u}Q+\sqrt{1+u}\tilde{Q})_{(--s_2s_3s_4)},
\nonumber \\
\lefteqn{\left[(I_{16}+u\Gamma^0\Gamma^1+\sqrt{1-u^2}\Gamma^0\Gamma^2)
(-\sqrt{1-u}Q+\sqrt{1+u}\tilde{Q})\right]_{(+-s_2s_3s_4)}}\qquad\qquad\quad\qquad\qquad\qquad
\\ 
&=&
(1+u)(-\sqrt{1-u}Q+\sqrt{1+u}\tilde{Q})_{(+-s_2s_3s_4)} \nonumber \\
&& \quad{}+\sqrt{1-u^2}(-\sqrt{1-u}Q+\sqrt{1+u}\tilde{Q})_{(-+s_2s_3s_4)}.
\nonumber
\eea
Are there linear combinations of the generators (50), (52), (54), and (55)
that are independent of $u$, and therefore unbroken no matter what the
values of $p$ and $q$? Indeed, by dividing (50) by $2\sqrt{1-u}$ and (54) by
$2\sqrt{1+u}$ and adding them, we come up with four such generators:
\be
\tilde{Q}_{(++s_2s_3s_4)}+Q_{(--s_2s_3s_4)}.
\ee
Four more are found by dividing (52) by $2\sqrt{1-u}$ and (55) by
$\sqrt{1+u}$:
\be
\tilde{Q}_{(+-s_2s_3s_4)}-Q_{(-+s_2s_3s_4)}.
\ee
As promised, one quarter of the original supersymmetries leave the entire
configuration described in the first paragraph of this solution invariant.

\setcounter{equation}{0}
\newpage\section{Chapter 14}

\subsection{Problem 14.1}

The excitation on the F-string will carry some energy (per unit length)
$p_0$, and momentum (per unit length) in the 1-direction $p_1$. Since the
string excitations move at the speed of light, left-moving excitation have
$p_0=-p_1$, while right-moving excitations have $p_0=p_1$. The
supersymmetry algebra (13.2.9) for this string is similar to (13.6.1), with
additional terms for the excitation:
\bea
\lefteqn{\frac{1}{2L}
\left\{\left[\begin{array}{c}Q_\alpha\\\tilde{Q}_\alpha\end{array}\right],
\left[\begin{array}{cc}
Q^\dag_\beta&\tilde{Q}^\dag_\beta
\end{array}\right]\right\}}\qquad \\
&& = \frac{1}{2\pi\alpha'}
\left[\begin{array}{cc}(\delta+\Gamma^0\Gamma^1)_{\alpha\beta}\,&0\\
0\,&(\delta-\Gamma^0\Gamma^1)_{\alpha\beta}\end{array}\right]
+\left(p_0\delta_{\alpha\beta}+p_1(\Gamma^0\Gamma^1)_{\alpha\beta}\right)
\left[\begin{array}{cc}1\,&0\\0\,&1\end{array}\right]\nonumber.
\eea
The first term on the RHS vanishes for those supersymmetries preserved by
the unexcited F-string, namely $Q$s for which $\Gamma^0\Gamma^1=-1$ and
$\tilde Q$s for which $\Gamma^0\Gamma^1=1$. The second term thus also
vanishes (making the state BPS) for the $Q$s if the excitation is
left-moving, and for the $\tilde Q$s if the excitation is right-moving.

For the D-string the story is almost the same, except that the first term
above is different:
\bea
\lefteqn{\frac{1}{2L}
\left\{\left[\begin{array}{c}Q_\alpha\\\tilde{Q}_\alpha\end{array}\right],
\left[\begin{array}{cc}
Q^\dag_\beta&\tilde{Q}^\dag_\beta
\end{array}\right]\right\}}\qquad \\
&& = \frac{1}{2\pi\alpha'g}
\left[
\begin{array}{cc}\delta_{\alpha\beta}\,&(\Gamma^0\Gamma^1)_{\alpha\beta} \\
(\Gamma^0\Gamma^1)_{\alpha\beta}\,&\delta_{\alpha\beta}\end{array}\right]
+\left(p_0\delta_{\alpha\beta}+p_1(\Gamma^0\Gamma^1)_{\alpha\beta}\right)
\left[\begin{array}{cc}1\,&0\\0\,&1\end{array}\right]\nonumber.
\eea
When diagonalized, the first term yields the usual preserved
supersymmetries, of the form $Q_\alpha+(\beta^\perp\tilde Q)_\alpha$. When
$1-(\Gamma^0\Gamma^1)_{\alpha\alpha}=0$ this supersymmetry is also
preserved by the second term if the excitation is left-moving; when
$1+(\Gamma^0\Gamma^1)_{\alpha\alpha}=0$ it is preserved if the
excitation is right-moving. Either way, the state is BPS.

\subsection{Problem 14.2}

The supergravity solution for two static parallel NS5-branes is given in
(14.1.15) and (14.1.17):
\bea
e^{2\Phi} &=&
g^2+\frac{Q_1}{2\pi^2(x^m-x_1^m)^2}+\frac{Q_2}{2\pi^2(x^m-x_2^m)^2},
\nonumber
\\
G_{mn} &=& g^{-1}e^{2\Phi}\delta_{mn},\qquad G_{\mu\nu} = g\eta_{\mu\nu},
\nonumber \\
H_{mnp} &=& -\epsilon_{mnp}{}^q\partial_q\Phi,
\eea
where $\mu,\nu=0,\dots,5$ and $m,n=6,\dots,9$ are the parallel and
transverse directions respectively, and the branes are located in the
transverse space at $x^m_1$ and $x_2^m$. (We have altered (14.1.15a)
slightly in order to make (3) S-dual to the D-brane solution (14.8.1).) A
D-string stretched between the two branes at any given excitation level is
a point particle with respect to the 5+1 dimensional Poincar\'e symmetry of
the parallel dimensions. In other words, if we make an ansatz for the
solution of the form
\be
X^\mu = X^\mu(\tau),\qquad X^m=X^m(\sigma),
\ee
then, after performing the integral over $\sigma$ in the D-string action,
we should obtain the point-particle action (1.2.2) in 5+1 dimensions,
\be
S_{\rm pp} = -m\int d\tau\,\sqrt{-\partial_\tau X^\mu\partial_\tau X_\mu},
\ee
where $m$ is the mass of the solution $X^m(\sigma)$  with respect to the
5+1 dimensional Poincar\'e symmetry.

Assuming that the gauge field is not excited, with this ansatz the D-string
action (13.3.14) factorizes:
\bea
S_{\rm D1} &=& -\frac{1}{2\pi\alpha'}\int d\tau
d\sigma\,e^{-\Phi}\sqrt{-\det(G_{ab}+B_{ab})} \nonumber \\
&=& -\frac{1}{2\pi\alpha'}\int d\tau
d\sigma\,e^{-\Phi} \nonumber \\
&&\qquad\qquad\times\sqrt{-\left|\begin{array}{cc}G_{\mu\nu}\partial_\tau
X^\mu\partial_\tau X^\nu\, & (G_{\mu n}+B_{\mu n})\partial_\tau
X^\mu\partial_\sigma X^n \\ (G_{m\nu}+B_{m\nu})\partial_\sigma
X^m\partial_\tau X^\nu\, & G_{mn}\partial_\sigma
X^m\partial_\sigma X^n\end{array}\right|} \nonumber \\
&=& -\frac{1}{2\pi\alpha'}\int d\sigma\,e^{-\Phi}\sqrt{\partial_\sigma
X^m\partial_\sigma X_m}\int d\tau\,\sqrt{-\partial_\tau
X^\mu\partial_\tau X_\mu}.
\eea
In the last equality we have used the fact that neither the metric nor the
two-form potential in the solution (3) have mixed $\mu n$ components.
Comparison with (5) shows that
\be
m = \frac{g^{-1/2}}{2\pi\alpha'}\int d\sigma\,|\partial_\sigma X^m|,
\ee
where the integrand is the coordinate (not the proper) line element in this
coordinate system. The ground state is therefore a straight line connecting
the two branes:
\be
m = \frac{g^{-1/2}|x^m_2-x^m_1|}{2\pi\alpha'}.
\ee

As explained above, this mass is defined with respect the geometry of the
parallel directions, and it is only in string frame that the parallel
metric $G_{\mu\nu}$ is independent of the transverse position. We can
nonetheless define an Einstein-frame mass $m_{\rm E}$ with respect to
$G_{\mu\nu}$ at $|x^m|=\infty$, and it is this mass that transforms simply
under S-duality. (Here we are using the definition (14.1.7) of the Einstein
frame, $G_{\rm E}=e^{-\Phi/2}G$, which is slightly different from the one
used in volume I and in Problem 14.6 below, where $G_{\rm
E}=e^{-\tilde\Phi/2}G$.) From the definition (5) of the mass,
\be
m_{\rm E} = g^{1/4}m,
\ee
so (7) becomes
\be
m_{\rm E} = \frac{g^{-1/4}}{2\pi\alpha'}\int d\sigma\,|\partial_\sigma
X^m|.
\ee

We can calculate the mass of an F-string stretched between two D5-branes in
two different pictures: we can use the black 5-brane supergravity solution
(14.8.1) and do a calculation similar to the one above, or we can consider
the F-string to be stretched between two elementary D5-branes embedded in
flat spacetime. In the first calculation, the S-duality is manifest at
every step, since the NS5-brane and the black 5-brane solutions are related
by S-duality, as are the D-string and F-string actions. The second
calculation yields the same answer, and is much easier: since the
tension of the F-string is $1/2\pi\alpha'$, and in flat spacetime
($G_{\mu\nu}=\eta_{\mu\nu}$) its proper length and coordinate length are
the same, its total energy is
\be
m = \frac{1}{2\pi\alpha'}\int d\sigma\,|\partial_\sigma X^m|.
\ee
Its Einstein-frame mass is then
\be
m_{\rm E} = \frac{g^{1/4}}{2\pi\alpha'}\int d\sigma\,|\partial_\sigma X^m|,
\ee
which indeed agrees with (10) under $g\to1/g$.

\subsection{Problem 14.6}

To find the expectation values of the dilaton and graviton in the low
energy field theory, we add to the action a source term
\be
S' = \int
d^{10}\!X\,\left(K_{\tilde\Phi}\tilde\Phi+K_h^{\mu\nu}h_{\mu\nu}\right),
\ee
and take functional derivatives of the partition function
$Z[K_{\tilde\Phi},K_h]$ with respect to $K_{\tilde\Phi}$ and $K_h$.
For the D-brane, which is a real, physical source for the fields, we also
include the sources $J_{\tilde\Phi}$ and $J_h$, calculated in problem
13.4(b) (see (26) and (27) of that solution):
\bea
J_{\tilde\Phi}(X) &=& \frac{3-p}{4}\tau_p\delta^{9-p}(X_\perp), \\
J_h^{\mu\nu}(X) &=& -\frac{1}{2}\tau_pe^{\mu\nu}_p\delta^{9-p}(X_\perp).
\eea
Recall that $h$ is the perturbation in the Einstein-frame metric,
$\tilde{G} = \eta+h$, and that $e^{\mu\nu}$ equals $\eta^{\mu\nu}$ for
$\mu,\nu$ parallel to the brane and zero otherwise. (We have put the brane
at the origin, so that $X'_\perp=0$.) Since the dilaton decouples from the
Einstein-frame graviton, we can calculate the partition functions
$Z[K_{\tilde\Phi}]$ and $Z[K_h]$ separately. Using the
propagator (23),
\bea
Z[K_{\tilde\Phi}] &=&
-Z[0]\int\frac{d^{10\!}k}{(2\pi)^{10}}\,\tilde
J_{\tilde\Phi}(-k)\langle\tilde\Phi\tilde\Phi(k)\rangle\tilde K_{\tilde\Phi}(k)
\nonumber \\
&=&
Z[0]\frac{3-p}{2}i\kappa^2\tau_p\int\frac{d^{9-p}\!k_\perp}{(2\pi)^{9-p}}\,\frac{1}{k_\perp^2}\tilde
K_{\tilde\Phi}(k_\perp,k_\parallel=0)
\nonumber \\
&=& Z[0]\frac{3-p}{2}i\kappa^2\tau_p\int
d^{9-p}\!X_\perp\,G_{9-p}(X_\perp)\int
d^{p+1}\!X_\parallel\,K_{\tilde\Phi}(X),
\eea
so that
\bea
\langle\tilde\Phi(X)\rangle &=& \frac{1}{iZ[0]}\frac{\delta
Z[K_{\tilde\Phi}]}{\delta K_{\tilde\Phi}(X)}
\nonumber \\
&=& \frac{3-p}{2}\kappa^2\tau_pG_{9-p}(X_\perp).
\eea
Using (13.3.22), (13.3.23), and the position-space expression for $G_d$,
this becomes
\bea
\langle\tilde\Phi(X)\rangle &=&
\frac{3-p}{4}(4\pi)^{(5-p)/2}\Gamma(\frac{7-p}{2})g\alpha'^{(7-p)/2}r^{p-7}
\nonumber \\
&=& \frac{3-p}{4}\frac{\rho^{7-p}}{r^{7-p}},
\eea
where $\rho^{7-p}$ is as defined in (14.8.2b) with $Q=1$. Hence
\bea
e^{2\langle\Phi\rangle} &\approx& g^2(1+2\langle\tilde\Phi\rangle)
\nonumber \\
&\approx&g^2\left(1+\frac{\rho^{7-p}}{r^{7-p}}\right)^{(3-p)/2},
\eea
in agreement with (14.8.1b) (corrected by a factor of $g^2$). 

The graviton calculation is very similar. Using the propagator (24),
\bea
Z[K_h] &=& -Z[0]\int\frac{d^{10}\!X}{(2\pi)^{10}}\,\tilde
J_h^{\mu\nu}(-k)\langle h_{\mu\nu}h_{\rho\sigma}(k)\rangle\tilde
K^{\rho\sigma}_h(k) \nonumber
\\
&=&
Z[0]\left(\frac{p+1}{8}\eta_{\mu\nu}-e_{\mu\nu}\right)2i\kappa^2\tau_p\int\frac{d^{9-p}\!k_\perp}{(2\pi)^{10}}\,\frac{1}{k_\perp^2}\tilde
K_h^{\mu\nu}(k_\perp,k_\parallel=0) \nonumber \\
&=&
Z[0]\left(\frac{p+1}{8}\eta_{\mu\nu}-e_{\mu\nu}\right)2i\kappa^2\tau_p \\
&&\qquad\qquad\qquad\qquad\qquad\quad\times\int
d^{9-p}\!X_\perp\,G_{9-p}(X_\perp)\int
d^{p+1}\!X_\parallel\,K_h^{\mu\nu}(X), \nonumber
\eea
so
\bea
\langle h_{\mu\nu}(X)\rangle &=&
\left(\frac{p+1}{8}\eta_{\mu\nu}-e_{\mu\nu}\right)2\kappa^2\tau_pG_{9-p}(X_\perp)
\nonumber \\
&=&
\left(\frac{p+1}{8}\eta_{\mu\nu}-e_{\mu\nu}\right)\frac{\rho^{7-p}}{r^{7-p}}.
\eea
Hence for $\mu,\nu$ aligned along the brane,
\be
\langle\tilde G_{\mu\nu}\rangle \approx
\left(1+\frac{\rho^{7-p}}{r^{7-p}}\right)^{(p-7)/8}\eta_{\mu\nu},
\ee
while for $m,n$ transverse to the brane,
\be
\langle\tilde G_{mn}\rangle \approx
\left(1+\frac{\rho^{7-p}}{r^{7-p}}\right)^{(p+1)/8}\delta_{mn}.
\ee
The string frame metric, $G=e^{\tilde\Phi/2}\tilde G$, is therefore
\bea
\langle G_{\mu\nu}\rangle &\approx&
\left(1+\frac{\rho^{7-p}}{r^{7-p}}\right)^{-1/2}\eta_{\mu\nu}, \\
\langle G_{mn}\rangle &\approx&
\left(1+\frac{\rho^{7-p}}{r^{7-p}}\right)^{1/2}\delta_{mn},
\eea
in agreement with (14.8.1).

\setcounter{equation}{0}
\newpage\section{Chapter 15}

\subsection{Problem 15.1}

The matrix of inner products is
\begin{align}
{\cal M}^3 &= 
\langle h|
\begin{bmatrix}L_1^3 \\ L_1L_2 \\ L_3 \end{bmatrix}
\begin{bmatrix}L_{-1}^3 & L_{-2}L_{-1} & L_{-3} \end{bmatrix}
|h\rangle \\
&=
\begin{bmatrix} 
24h(h+1)(2h+1) & 12h(3h+1) & 24h \\
12h(3h+1) & h(8h+8+c) & 10h \\
24h & 10h & 6h+2c
\end{bmatrix}.
\end{align}
This matches the Kac formula, with
\begin{equation}
K_3 = 2304.
\end{equation}

\subsection{Problem 15.3}

Let's begin by recording some useful symmetry relations of the
operator product coefficient with lower indices, derived from the
definition (6.7.13) and (6.7.14),
\begin{equation}
c_{ijk} = 
\left\langle
\mathcal{A}'_i(\infty,\infty)\mathcal{A}_j(1,1)\mathcal{A}_k(0,0)
\right\rangle_{S_2}.
\end{equation}
The following relations then hold, with the sign of the coefficient
depending on the statistics of the operators:
\begin{align}
c_{ijk} &= \pm(-1)^{h_j+\tilde h_j}c_{kji}\qquad
\te{if $\mathcal{A}_j$ is primary} \\
c_{ijk} &= 
\pm(-1)^{h_i+h_j+h_k+\tilde h_i+\tilde h_j+\tilde h_k}c_{ikj}\qquad
\te{if $\mathcal{A}_i$ is primary} \\
c_{ijk} &= \pm(-1)^{h_k+\tilde h_k}c_{jik}\qquad
\te{if $\mathcal{A}_i,\mathcal{A}_j,\mathcal{A}_k$ are primary}.
\end{align}
Actually, in the above ``primary'' may be weakened to
``quasi-primary'' (meaning annihilated by $L_1$, rather than $L_n$ for
all $n>0$, and therefore transforming as a tensor under $PSL(2,C)$
rather than general local conformal transformations), but Polchinski
does not seem to find the notion of quasi-primary operator interesting
or useful.

Armed with these symmetries (and in particular relation (5)), but
glibly ignoring phase factors as Polchinski does, we would like to
claim that the correct form for (15.2.7) should be as follows (we
haven't bothered to raise the index):
\begin{equation}
c_{i\{k,\tilde k\},mn} =
\lim_{\substack{z_n\to\infty\\z_m\to1}}
z_n^{2h_n}\bar z_n^{2\tilde h_n}
\mathcal{L}_{-\{k\}}\tilde{\mathcal{L}}_{-\{\tilde k\}}
\left\langle
\mathcal{O}_n(z_n,\bar z_n)
\mathcal{O}_m(z_m,\bar z_m)
\mathcal{O}_i(0,0)
\right\rangle_{S_2}.
\end{equation}
We would also like to claim that the LHS of (15.2.9) should read
\begin{equation}
\left\langle
\mathcal{O}'_l(\infty,\infty)
\mathcal{O}_j(1,1)
\mathcal{O}_m(z,\bar z)
\mathcal{O}_n(0,0)
\right\rangle_{S_2}.
\end{equation}

We are now ready to solve the problem. The case $N=0$ is trivial,
since by the definition (15.2.8),
\begin{equation}
\beta^{i\{\}}_{mn}=1,
\end{equation}
so the coefficient of $z^{-h_m-h_n+h_i}$ in
$\mathcal{F}^{jl}_{mn}(i|z)$ is 1. For $N=1$ there is again only one
operator, $L_{-1}\cdot\mathcal{O}_i$. We have
\begin{equation}
\beta^{i\{1\}}_{mn} = \frac1{2h_i}(h_i+h_m-h_n).
\end{equation}
Thus the coefficient of $z^{-h_m-h_n+h_i+1}$ is
\begin{equation}
\frac1{2h_i}(h_i+h_m-h_n)(h_i+h_j-h_l).
\end{equation}

\setcounter{equation}{0}
\newpage
\section{Appendix B}

\subsection{Problem B.1}

Under a change of spinor representation basis, $\Gamma^\mu\to U\Gamma^\mu
U^{-1}$, $B_1$, $B_2$, and $C$, all transform the same way:
\be
B_1\to U^*B_1U^{-1},\qquad B_2\to U^*B_2U^{-1},\qquad C\to U^*CU^{-1}.
\ee
The invariance of the following equations under this change of basis is
more or less trivial:

(B.1.17) (the definition of $B_1$ and $B_2$):
\bea
U^*B_1U^{-1}U\Gamma^\mu U^{-1}(U^*B_1U^{-1})^{-1} &=& U^*B_1\Gamma^\mu
B_1^{-1}U^T\nonumber \\
&=& (-1)^kU^*\Gamma^{\mu*}U^T\nonumber \\
&=& (-1)^k(U\Gamma^\mu U^{-1})^*, \\
U^*B_2U^{-1}U\Gamma^\mu U^{-1}(U^*B_2U^{-1})^{-1} &=& U^*B_2\Gamma^\mu
B_2^{-1}U^T\nonumber \\
&=& (-1)^{k+1}U^*\Gamma^{\mu*}U^T\nonumber \\
&=& (-1)^{k+1}(U\Gamma^\mu U^{-1})^*.
\eea

(B.1.18), using the fact that $\Sigma^{\mu\nu}$ transforms the same way as
$\Gamma^\mu$:
\bea
U^*BU^{-1}U\Sigma^{\mu\nu}U^{-1}UB^{-1}U^T &=&
U^*B\Sigma^{\mu\nu}B^{-1}U^T\nonumber \\
&=& -U^*\Sigma^{\mu\nu*}U^T\nonumber \\
&=& -(U\Sigma^{\mu\nu}U^{-1})^*.
\eea

The invariance of (B.1.19) is the same as that of (B.1.17).

(B.1.21):
\bea
UB_1^*U^TU^*B_1U^{-1} = UB_1^*B_1U^{-1} = (-1)^{k(k+1)/2}UU^{-1} =
(-1)^{k(k+1)/2}, \\
UB_2^*U^TU^*B_2U^{-1} = UB_2^*B_2U^{-1} = (-1)^{k(k-1)/2}UU^{-1} =
(-1)^{k(k-1)/2}.
\eea

(B.1.24) (the definition of C):
\bea
U^*CU^{-1}U\Gamma^\mu U^{-1}UC^{-1}U^T &=& U^*C\Gamma^\mu
C^{-1}U^T\nonumber \\ 
&=& -U^*\Gamma^{\mu T}U^T\nonumber \\
&=& -(U\Gamma^\mu U^{-1})^T.
\eea

(B.1.25): all three sides clearly transform by multiplying on the left by
$U$ and on the right by $U^{-1}$.

(B.1.27):
\be
U^*BU^{-1}U\Gamma^0U^{-1} = U^*B\Gamma^0U^{-1} = U^*CU^{-1}.
\ee

We will determine the relation between $B$ and $B^T$ in the $\zeta^{(\bf
s)}$ basis, where (for $d=2k+2$) $B_1$ and $B_2$ are defined by equation
(B.1.16):
\be
B_1 = \Gamma^3\Gamma^5\cdots\Gamma^{d-1},\qquad B_2 = \Gamma B_1.
\ee
Since all of the $\Gamma$'s that enter into this product are antisymmetric
in this basis (since they are Hermitian and imaginary), we have
\bea
B_1^T &=& (\Gamma^3\Gamma^5\cdots\Gamma^{d-1})^T\nonumber \\
&=& (-1)^k\Gamma^{d-1}\Gamma^{d-3}\cdots\Gamma^3\nonumber \\
&=& (-1)^{k(k+1)/2}\Gamma^3\Gamma^5\cdots\Gamma^{d-1}\nonumber \\
&=& (-1)^{k(k+1)/2}B_1.
\eea
Using (10), the fact that $\Gamma$ is real and symmetric in this basis, and
(B.1.19), we find
\bea
B_2^T &=& (\Gamma B_1)^T\nonumber \\
&=& B_1^T\Gamma^T\nonumber \\
&=& (-1)^{k(k+1)/2}B_1\Gamma\nonumber \\
&=& (-1)^{k(k+1)/2+k}\Gamma^*B_1\nonumber \\
&=& (-1)^{k(k-1)/2}B_2.
\eea
Since $B_1$ is used when $k=0,3$ (mod 4), and $B_2$ is used when $k=0,1$
(mod 4), so in any dimension in which one can impose a Majorana condition
we have
\bea
B^T = B.
\eea

When $k$ is even, $C=B_1\Gamma^0$, so, using the fact that in this basis
$\Gamma^0$ is real and antisymmetric, and (B.1.17), we find
\bea
C^T &=& \Gamma^{0T}B_1^T\nonumber \\
&=& (-1)^{k(k+1)/2+1}\Gamma^{0*}B_1\nonumber \\
&=& (-1)^{k/2+1}B_1\Gamma^0\nonumber \\
&=& (-1)^{k/2+1}C.
\eea
On the other hand, if $k$ is odd, then $C=B_2\Gamma^0$, and
\bea
C^T &=& \Gamma^{0T}B_2^T\nonumber \\
&=& (-1)^{k(k-1)/2+1}\Gamma^{0*}B_2\nonumber \\
&=& (-1)^{(k+1)/2}B_2\Gamma^0\nonumber \\
&=& (-1)^{(k+1)/2}C.
\eea

That all of these relations are invariant under change of basis follows
directly from the transformation law (1), since $B^T$ and $C^T$ transform
the same way as $B$ and $C$.

\subsection{Problem B.3}

The decomposition of SO(1,3) spinor representations under the
subgroup SO(1,1)$\times$SO(2) 
is described most simply in terms of Weyl representations: one
positive chirality spinor, $\zeta^{++}$, transforms as a positive chirality
Weyl spinor under both SO(1,1) and SO(2), while the other,
$\zeta^{--}$, transforms as a negative chirality Weyl spinor under both
(see (B.1.44a)). Let us therefore use the Weyl-spinor description 
of the 4 real supercharges of $d=4$, $\mathcal{N}=1$ supersymmetry. The
supersymmetry algebra is (B.2.1a):
\bea
\{Q_{++},Q_{++}^\dag\} &=& 2(P^0-P^1), \nonumber \\
\{Q_{--},Q_{--}^\dag\} &=& 2(P^0+P^1), \\
\{Q_{++},Q_{--}^\dag\} &=& 2(P^2+iP^3). \nonumber
\eea
We must now decompose these Weyl spinors into Majorana-Weyl spinors. Define
\bea
Q_L^1 = \frac{1}{2}(Q_{++}+Q_{++}^\dag), &\qquad& Q_L^2 =
\frac{1}{2i}(Q_{++}-Q_{++}^\dag), \\
Q_R^1 = \frac{1}{2}(Q_{--}+Q_{--}^\dag), &\qquad& Q_R^2 =
\frac{1}{2i}(Q_{--}-Q_{--}^\dag).
\eea
(The subscripts $L$ and $R$ signify that the respective supercharges have
positive and negative SO(1,1) chirality.) Hence, for instance,
\be
\{Q^1_L,Q^1_L\} =
\frac{1}{4}\left(\{Q_{++},Q_{++}\}+\{Q_{++}^\dag,Q_{++}^\dag\}
+2\{Q_{++},Q^\dag_{++}\}\right).
\ee
The last term is given by the algebra (14), but what do we do with the
first two terms? The $d=4$ algebra has a U(1) R-symmetry under which
$Q_{++}$ and $Q_{--}$ 
are both multiplied by the same phase. In order for the $d=2$ algebra to
inherit that symmetry, we must assume that 
\be
Q_{++}^2 = Q_{--}^2 = \{Q_{++},Q_{--}\} = 0.
\ee
The $d=2$ algebra is then
\bea
\{Q_L^A,Q_L^B\} &=& \delta^{AB}(P^0-P^1), \\
\{Q_R^A,Q_R^B\} &=& \delta^{AB}(P^0+P^1), \\
\{Q_L^A,Q_R^B\} &=& Z^{AB},
\eea
where
\be
Z = \left[\begin{array}{cc} P^2 & \,-P^3 \\ P^3 & \,P^2 \end{array}\right].
\ee
The central charges are thus the Kaluza-Klein momenta associated with the
reduced dimensions. If these momenta are 0 (as in dimensional
reduction in the strict sense), then the algebra possesses a
further R-symmetry, namely the SO(2) of rotations in the 2-3 plane. We saw
at the beginning that $Q_{++}$ is positively charged and $Q_{--}$
negatively charged under this symmetry. This symmetry and the
original U(1) R-symmetry of the $d=4$ algebra can be recombined into two
independent SO(2) R-symmetry groups of the $Q_L^A$ and $Q_R^A$ pairs of
supercharges.

\subsection{Problem B.5}

Unfortunately, it appears that some kind of fudge will be necessary to get
this to work out correctly. There may be an error lurking in the book. The
candidate fudges are: (1) The vector multiplet is
$\mathbf{8}_v+\mathbf{8'}$, not $\mathbf{8}_v+\mathbf{8}$ (this is
suggested by the second sentence of Section B.6). (2) The supercharges of
the $N=1$ theory are in the $\mathbf{16'}$, not $\mathbf{16}$, of
SO(9,1). (3) The frame is one in which $k^0=k^1$, not $k_0=k_1$ as
purportedly used in the book. We will arbitrarily choose fudge \#1, although
it's hard to see how this can fit into the analysis of the type I spectrum
in Chapter 10.

The $\mathbf{8}_v$ states have helicities $(\pm1,0,0,0)$, $(0,\pm1,0,0)$,
$(0,0,\pm1,0)$, and $(0,0,0,\pm1)$. The $\mathbf{8'}$ states have
helicities $\pm(-\frac12,+\frac12,+\frac12,+\frac12)$, 
$\pm(+\frac12,-\frac12,+\frac12,+\frac12)$,
$\pm(+\frac12,+\frac12,-\frac12,+\frac12)$, and
$\pm(+\frac12,+\frac12,+\frac12,-\frac12)$.

In a frame in which $k_0=k_1$, the supersymmetry algebra is
\bea
\{Q_\alpha,Q_\beta^\dag\} = 2P_\mu(\Gamma^\mu\Gamma^0)_{\alpha\beta}
= -2k_0(1+2S_0)_{\alpha\beta},
\eea
so that supercharges with $s_0=-\frac12$ annihilate all states. The
supercharges with $s_0=+\frac12$ form an $\mathbf{8}$ representation of the
SO(8) little group. Since the operator $B$ switches the sign of all the
helicities $s_1,\dots,s_4$, the Majorana condition pairs these eight
supercharges into four independent sets of fermionic raising and lowering
operators. Let $Q_{(+\frac12,+\frac12,+\frac12,+\frac12,+\frac12)}$, 
$Q_{(+\frac12,+\frac12,+\frac12,-\frac12,-\frac12)}$,
$Q_{(+\frac12,+\frac12,-\frac12,+\frac12,-\frac12)}$, and
$Q_{(+\frac12,+\frac12,-\frac12,-\frac12,+\frac12)}$ be the raising operators.
We can obtain all sixteen of the states in $\mathbf{8}_v+\mathbf{8}'$ by
starting with the state $(-1,0,0,0)$ and acting on it with all possible
combinations of these four operators. The four states in the $\mathbf{8}'$ with
$s_1=-\frac12$ are obtained by acting with a single operator. Acting
with a second operator yields the six states in the $\mathbf{8}_v$ with
$s_1=0$. A third operator gives $s_1=+\frac12$, the other four states of
the $\mathbf{8}'$. Finally, acting with all four operators yields the last
state of the $\mathbf{8}_v$, $(+1,0,0,0)$.

\newpage

\bibliography{ref}
\bibliographystyle{JHEP}

\end{document}